\documentclass[12pt, a4paper]{article}
\usepackage{amsfonts, amssymb}
\usepackage{graphicx}
\usepackage{latexsym,amsmath,color,array}

\usepackage{natbib}
\usepackage{enumitem}
\usepackage{bbding}
\usepackage{amssymb}
\usepackage{amsmath}
\usepackage{graphicx}
\usepackage{amsmath}
\usepackage{bbm}
\usepackage{mathtools}
 \usepackage{color}
 \usepackage{array}
 \usepackage{subcaption}
 \usepackage{multirow}
 \usepackage[dvipsnames]{xcolor}
 \usepackage{makecell}
 \usepackage{colortbl}
 \usepackage{algorithm}
 \usepackage{rotating}
 \usepackage{siunitx}
\usepackage[hyphens]{url}
\usepackage{tikz}
\usepackage{bm}

\usepackage{booktabs}
\usepackage{threeparttable}
\usepackage{longtable}
\def\1{\mathbb{I}}

\DeclareMathOperator*{\argmax}{arg\,max}

\newcounter{appen}[section]

\begin{document}

\title{Investigating Statistical Inference and Covariate Effects in Shallow Neural Networks}
\author{Andrew McInerney\footnote{School of Medicine, University of Limerick; andrew.mcinerney@ul.ie} \hspace{3cm}
Kevin Burke\footnote{School of Medicine, University of Limerick; kevin.burke@ul.ie}}
\date{\today}

\maketitle

\begin{abstract}
Feedforward neural networks (FNNs) are typically viewed as pure prediction algorithms, and their strong predictive performance has led to their use in many machine-learning applications. 
However, their flexibility comes with an interpretability trade-off; thus, FNNs have been historically less popular among statisticians. 
Nevertheless, for suitably parsimonious shallow FNNs, classical statistical theory, such as significance testing and uncertainty quantification, may still provide useful regression-style summaries.
Supplementing FNNs with methods of statistical inference, and covariate-effect visualisations, can shift the focus away from black-box prediction and move FNNs towards traditional statistical models. 
This can allow for more inferential analysis, and, hence, make FNNs more accessible within the statistical-modelling context.
We investigate covariate-level Wald testing in the context of penalised FNNs, and also propose covariate-effect plots that emulate regression coefficients.
Simulation studies are used to extensively investigate the performance of Wald-based inference, with particular emphasis on when this approach performs well.
This statistical-based approach to neural networks is demonstrated through an application to insurance data.

\smallskip


\end{abstract}

\qquad

\newpage
\section{Introduction}
Many applications require the estimation of a complex underlying relationship between a response variable and a set of explanatory variables. 
Neural networks are often used to approximate such relationships, particularly when predictive performance is the primary objective \citep{lecun2015deep}.
However, while neural networks can capture nonlinearities and interactions, they are most often treated as prediction algorithms rather than statistical models. This limits their use in settings where analysts require not only accurate predictions, but also uncertainty quantification, covariate-level summaries, and interpretable descriptions of the fitted relationship. This is particularly relevant for small-to-moderate-sized tabular data problems, where a shallow neural network may provide useful flexibility while remaining close enough to classical regression models to permit statistical investigation.
Shallow architectures also remain popular in applied tabular-data settings. 
In a recent systematic review of 430 data analyses comparing neural networks with generalised additive models, feedforward neural networks accounted for 89\% of the neural networks considered and, among studies reporting network depth, 70\% used a single hidden layer \citep{doohan2026comparison}.

In the current paper, the \texttt{insurance} data \citep{lantz2019machine} provide a simple motivating example.
The data contain information relating to an insurance plan's primary beneficiaries and the total amount of medical expenses charged to each account in the United States (measured in thousands of dollars). 
Whilst a shallow neural network has strong performance in the prediction of the amount billed to each account, it does not immediately offer any insights into the underlying relationship. (Note: a neural network with all covariates and two hidden nodes achieves a five-fold cross-validation root mean squared error of 4.634 ($\pm 0.144$); see Section \ref{sec: app_to_data} for more detail.)
Indeed, most neural network software offers very little information beyond predictive performance metrics; summaries beyond out-of-sample have apparently been less of a focus in this context \citep{efron2020prediction}.
By contrast, a linear regression model fitted to the \texttt{insurance} data results in a worse root mean squared error ($6.107 \pm 0.146$) compared to the neural network, but it does provide a model summary, shown in Table \ref{tab: insurance_lm}, which is useful and familiar to statisticians.
This table effectively answers two questions that naturally arise when modelling the relationship between a covariate and a response.
First, is there any relationship?
This is covered by the p-values, which test for statistically significant associations.
Second, if there is a relationship, what is the nature of this relationship?
This is covered by the regression coefficients, which give us an idea of the size and direction of the relationship.
From Table \ref{tab: insurance_lm}, we can see that \texttt{age}, \texttt{bmi}, \texttt{children} and \texttt{smoker} are all highly statistically significant, and increased values for each variable are associated with higher total medical expenses charged, on average (for the case of \texttt{smoker}, an increased value corresponds to going from non-smoker to smoker).

\begin{table}[t!]
    \centering
    \caption{Summary of linear model for the \texttt{insurance} data.}
    \begin{tabular}[c]{lrr}
      \toprule
       &  \multicolumn{1}{c}{$\hat{\beta}_j$ (SE)}  & p-value  \\ 
       \cmidrule(lr){2-2} \cmidrule(lr){3-3} 
  \texttt{intercept}  & -11.939 (0.988)  & $<0.001$ \\
  \texttt{age}        & ~\space0.257 (0.012)  & $<0.001$ \\
  \texttt{sex.male}   & -0.131 (0.333)  & ~~0.693 \\
  \texttt{bmi}        & ~0.339 (0.029)  & $<0.001$ \\
  \texttt{children}   & ~0.476 (0.138)  & $<0.001$ \\
  \texttt{smoker}     & ~23.849 (0.413)  & $<0.001$\\
  \texttt{region.nw}  & -0.353 (0.476)  & ~~0.459 \\
  \texttt{region.se}  & -1.035 (0.479)  & ~~0.031 \\
  \texttt{region.sw}  & -0.960 (0.478)  & ~~0.045 \\
       \bottomrule
    \end{tabular}\label{tab: insurance_lm}
  \end{table}

The motivating example highlights the trade-off between model interpretability and model predictivity and also distinguishes between the traditional approaches of statistical modelling (focused on explanation) and machine learning (focused on prediction) \citep{shmueli2010explain}.
These two fields have not developed in unison and the majority of neural network research has been conducted outside of the field of statistics \citep{breiman2001statistical, efron_hastie_2021, hooker2021bridging}.
As a result of this, other avenues toward explainability exist (see \citet{guidotti2018survey} for a comprehensive review).
Model-agnostic methods, i.e., methods that can be used on any black-box model, are a popular alternative \citep{ribeiro2016model}.
This includes methods that perturb inputs such as Shapley values \citep{strumbelj2010efficient} and local surrogate models such as Locally Interpretable Model Explanations (LIME) \citep{ribeiro2016lime}.
These approaches are valuable, but they are not the focus of this paper. 
Instead, we focus on shallow FNNs used within a statistical-modelling framework, with the aim of obtaining covariate-level inferential summaries that are closer in spirit to those produced by more traditional statistical regression models.

In this paper, we investigate statistical inference and interpretation in shallow neural networks used as flexible non-linear regression models.
We tackle the applied problem of interpreting neural networks using likelihood-based inferential techniques, and demonstrate that they are useful beyond pure prediction problems.
The presence of a relationship between a covariate and the response is primarily addressed through the multiple-parameter Wald test, since multiple network weights link each input to the hidden nodes.
To the best of our knowledge, covariate-level Wald testing is not currently used for fitted neural networks in practice, has not been extensively investigated empirically, nor has it been considered in the context of penalised FNNs.
We therefore consider Wald-based testing under ridge-penalised estimation, which provides more stable estimation and improves its practical usability.
Here, we carry out an empirical investigation to assess the estimation and inferential performance of Wald-based testing in shallow neural networks.
In particular, we investigate when testing performs as expected, and when performance deteriorates; we use this to provide practical recommendations related to model size and regularisation.
We also provide a convenient visualisation of the results from these tests, which are overlaid on the neural network architecture.
Furthermore, the nature of statistically significant relationships is addressed through covariate-effect plots that emulate regression coefficients from statistical models, yielding interpretable model outputs that are more familiar in the statistical-modelling context.

The remainder of this paper is structured as follows.
In Section \ref{sec: FNN}, we introduce the feedforward neural network (FNN) model, and summarise the relevant likelihood theory.
Section \ref{sec: hypo_test} discusses significance testing for input nodes (covariates) and individual weights (parameters).
In Section \ref{sec: cov_eff}, we propose covariate-effect plots that play a similar role to regression coefficients obtained from classical statistical models.
The performance of the aforementioned methods is thoroughly investigated using simulation studies in Section \ref{sec: sim}.
The motivating \texttt{insurance} data example is revisited throughout the paper, but will be explored in more detail in Section \ref{sec: app_to_data}.
Finally, we conclude with a discussion in Section \ref{sec: disc}.

\section{Feedforward Neural Networks}\label{sec: FNN}

Let $y_i$ be the response variable of interest, for $i = 1, \dotsc, n$.
We observe a vector of covariates, $x_i = (x_{i0}, x_{i1}, \dotsc, x_{ip})^T$, for the $i$th observation with covariates indexed by $j = 0, 1, \dotsc, p$, and $x_{i0} \equiv 1$.
The general form of an FNN model can be written as
\begin{equation*}
    \mathbb{E}(y_i | x_i) = \text{NN}(x_i, \theta) ,
\end{equation*}
where 
\begin{equation}\label{eq: model}
  \text{NN}(x_i, \theta) = \phi_o \left[ \gamma_0+\sum_{k=1}^q \gamma_k \phi_h \left( \sum_{j=0}^p \omega_{jk}x_{ij}\right) \right],
\end{equation}
and $\theta$ denotes the neural network parameters.
The parameters are: $\omega_{jk}$, the weight that connects the $j$th covariate (input node) to the $k$th hidden node, where $\omega_{0k}$ is an intercept term; $\gamma_k$, the weight that connects the $k$th hidden node to the output node; and $\gamma_0$, the intercept term associated with the output node.
Let $\omega_j = (\omega_{j1},\omega_{j2},\dotsc,\omega_{jq})^T$ be the vector representing all of the connection weights from input node $j$ to the hidden layer, and let $\gamma=(\gamma_0,\gamma_1,\dotsc,\gamma_q)^T$ be the vector representing all of the connection weights between the hidden layer and the output layer.
Finally, $\theta=(\omega_0^T,\omega_1^T,\dotsc,\omega_p^T,\gamma^T)^T$ is the vector of all parameters in the neural network model, which is of dimension $r = (p + 2)q + 1$.
The function $\phi_h(\cdot)$ is the activation function for the hidden layer, which is often a logistic function, and the function $\phi_o(\cdot)$ is the activation function for the output layer.
In the case where $y_i$ is a continuous outcome variable, $\phi_o(\cdot)$ is typically the identity function where one might assume $y_i \sim N(\text{NN}(x_i, \theta), \sigma^2)$; and in the case where $y_i$ is a binary outcome variable, $\phi_o(\cdot)$ is typically the logistic function with $y_i \sim \text{Bernoulli}(\text{NN}(x_i, \theta))$.

Given that we assume the response variable $y_i$ has an associated underlying distribution, penalised maximum likelihood can be used to estimate the parameters, i.e., we aim to maximise 
\begin{equation}\label{eq: loglike}
  \ell(\theta) = \sum_{i=1}^n \log f(y_i | \theta)
 - \lambda\lVert \Tilde\theta \rVert_2^2,
\end{equation}
where $f(y_i | \theta)$ is the assumed density function for $y_i$, $\lambda$ is the size of the penalty, $\Tilde\theta = (\omega_1^T, \omega_2^T, \dotsc, \omega_p^T, \gamma_1, \gamma_2, \dotsc, \gamma_q)^T$ is the vector of all weights in the neural network, $\theta$, with the intercept terms omitted, and  $\lVert \cdot \rVert_2^2$ is the squared $L_2$ norm.
We then define the penalised maximum likelihood estimate (MLE), 
\begin{equation}\label{eq: thetahat}
    \hat{\theta} = \argmax_\theta \ell(\theta).
\end{equation}
Note that, with $\lambda = 0$, maximising $\ell(\theta)= -\frac{n}{2}\log(2\pi\sigma^2)-\frac{1}{2\sigma^2} \sum_{i=1}^n (y_i - \text{NN}(x_i, \theta))^2$ for continuous (normal) outputs is equivalent to minimising the residual sum of squares; and maximising $\ell(\theta)= \sum_{i=1}^n y_i\log(\text{NN}(x_i, \theta)) + (1 - y_i)\log(1 - \text{NN}(x_i, \theta))$ for binary outputs is equivalent to minimising the logistic (cross-entropy) loss.
These are commonly used objective functions in neural network optimisers, which makes the implementation of maximum likelihood in existing software straightforward \citep{geron2022hands, goodfellow2016deep}. 
Building upon this, the use of maximum likelihood theory further allows us to quantify uncertainty about the estimated parameters, i.e., we have that $\hat{\theta} \sim N(\theta, \Sigma)$ as $n \rightarrow \infty$, where $\Sigma$ is the variance-covariance matrix of the neural network parameters (once symmetries in the parameter space are accounted for (see Appendix \ref{app: practicalities})).
The variance-covariance matrix for the MLE of the neural network parameters can then be estimated using the sandwich formula \citep{fan2001sandwich},
\begin{equation}\label{eq: sandwich}
 \hat\Sigma = (I_o(\hat\theta) + 2\lambda I)^{-1} I_o(\hat\theta) (I_o(\hat\theta) + 2\lambda I)^{-1},
\end{equation}
where $I_o(\theta) = - \nabla_\theta\nabla_\theta^T [\ell(\theta)\rvert_{\lambda=0}]$.
However, the presence of redundant hidden nodes can inhibit the computation of $I_o(\hat\theta)$; this issue is considered further in simulation studies presented in Appendices \ref{app: practicalities} and \ref{app: sim_valid}.
Overall, the estimation of $\theta$ and $I_o(\theta)$ has improved stability in a penalisation framework (also known as regularization and shrinkage), and
Equation \ref{eq: loglike} makes use of ridge penalisation, which is popular in context of neural networks and more commonly referred to as weight decay \citep{hinton1989connectionist, hoerl1970ridge, krogh1991weightdecay}.
For the ridge penalty, \citet{ripley1994neuralflexible} recommended $\lambda \in [10^{-4}, 10^{-2}]$ when minimising the sum of squares, and $\lambda \in [10^{-2}, 10^{-1}]$ when minimising the cross entropy, and small values in line with this are typically used in practice in the neural network context \citep{james2013introduction, smith2018disciplined}.
(Of course, $\lambda$ could also be tuned for predictive performance, but our primary aim here is to improve estimation stability using a small ridge penalty.)

\section{Hypothesis Testing}\label{sec: hypo_test}
Hypothesis tests can be used to determine the statistical significance of individual parameters (i.e., neural network weights), and they can be used to determine the statistical significance of groups of parameters (i.e., neural network nodes).
Due to the asymptotic normality of the MLE, the single-parameter Wald test can be used to determine if a given weight in an FNN is statistically different from zero.
For a single parameter, $\theta_j$, we can test the null hypothesis $H_0: \theta_{j} = 0$ using the fact that
$\hat{\theta}_{j}^2 / \hat\Sigma_{jj} \sim \chi^2_1$,
where $\hat\Sigma_{jj}$ is the $j$th diagonal element of $\hat\Sigma$, and an associated p-value can be obtained.

While testing the significance of single parameters has some practical value in neural networks, it will usually be of greater interest to test groups of parameters in this setting.
As each input node has multiple weights associated with it, we therefore suggest using a multiple-parameter Wald test to test a single hypothesis on each of these parameters, i.e., test the overall significance of the $j$th covariate by testing $H_0: \omega_{j} = 0_q$, where $\omega_j = (\omega_{j1},\omega_{j2},\dotsc,\omega_{jq})^T$ is the vector of weights connecting that covariate to the hidden layer, and $0_q$ is a zero vector of length $q$.
First, let $A = (I_o(\hat\theta) + 2\lambda I)^{-1} I_o(\hat\theta)$, where $\text{tr}(A)$ is the overall degrees of freedom of the penalised FNN and $\text{tr}(\cdot)$ denotes the trace operator \citep{hastie2009elements}.
Using this, and the fact that (asymptotically) $\hat{\omega}_j \sim N(\omega_j, \Sigma_{\omega_{j}})$, we have that
\begin{equation}\label{eq: mp_wald}
    (\hat{\omega}_{j} - \omega_j)^T\hat\Sigma_{{\omega}_{j}}^{-1}(\hat{\omega}_{j} - \omega_j) \sim \chi^2_{\tilde q},
\end{equation}where $\Sigma_{{\omega}_{j}} = S\Sigma S^T$ is the relevant $q \times q$ sub-matrix of $\Sigma$ with $S$ being a $q \times r$ selection matrix, $r$ is the total number of parameters in the model, and
\begin{equation*}
    \tilde q = \text{tr}(SAS^T)
\end{equation*}
is the effective number of parameters associated with the test statistic, which is approximately equal to $q$ when $\lambda$ is small.
To the best of our knowledge, this result, with modified degrees of freedom accounting for the penalty, has not been considered in the context of FNNs; an associated p-value can then be obtained by setting $\omega_j = 0_q$ and comparing this statistic to the $\chi^2_{\tilde q}$ distribution.

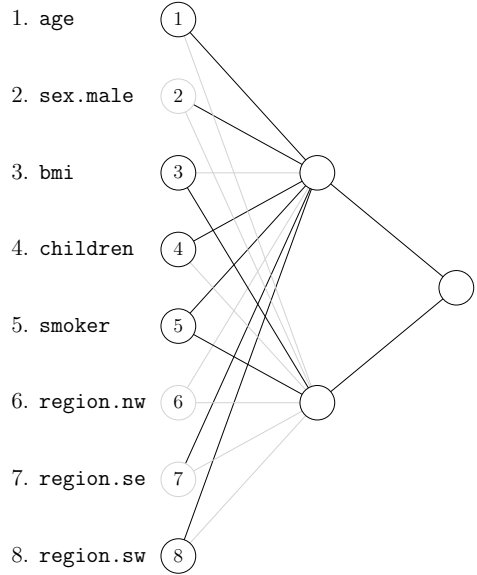
\begin{figure}[t]
		\centering
		\centerline{\resizebox{0.35\textwidth}{!}{% Created by tikzDevice version 0.12.4 on 2023-10-06 21:50:31
% !TEX encoding = UTF-8 Unicode
\begin{tikzpicture}[x=1pt,y=1pt]
\definecolor{fillColor}{RGB}{255,255,255}
\path[use as bounding box,fill=fillColor,fill opacity=0.00] (100,0) rectangle (405.89,505.89);
\begin{scope}
\path[clip] (  0.00,  0.00) rectangle (505.89,505.89);
\definecolor{drawColor}{RGB}{211,211,211}

\path[draw=drawColor,line width= 0.4pt,line join=round,line cap=round] (151.77, 56.21) --
	(252.94,168.63);
\definecolor{drawColor}{RGB}{0,0,0}

\path[draw=drawColor,line width= 0.4pt,line join=round,line cap=round] (151.77, 56.21) --
	(252.94,337.26);

\node[text=drawColor,anchor=base west,inner sep=0pt, outer sep=0pt, scale=  1.40] at ( 30.35, 52.02) {8. \texttt{region.sw}       };
\end{scope}
\begin{scope}
\path[clip] (  0.00,  0.00) rectangle (505.89,505.89);
\definecolor{drawColor}{RGB}{0,0,0}
\definecolor{fillColor}{RGB}{255,255,255}

\path[draw=drawColor,line width= 0.4pt,line join=round,line cap=round,fill=fillColor] (151.77, 56.21) circle ( 12.65);

\node[text=drawColor,anchor=base,inner sep=0pt, outer sep=0pt, scale=  1.20] at (151.77, 52.08) {8};
\definecolor{drawColor}{RGB}{211,211,211}

\path[draw=drawColor,line width= 0.4pt,line join=round,line cap=round] (151.77,112.42) --
	(252.94,168.63);
\definecolor{drawColor}{RGB}{0,0,0}

\path[draw=drawColor,line width= 0.4pt,line join=round,line cap=round] (151.77,112.42) --
	(252.94,337.26);

\node[text=drawColor,anchor=base west,inner sep=0pt, outer sep=0pt, scale=  1.40] at ( 30.35,108.23) {7. \texttt{region.se}       };
\end{scope}
\begin{scope}
\path[clip] (  0.00,  0.00) rectangle (505.89,505.89);
\definecolor{drawColor}{RGB}{211,211,211}
\definecolor{fillColor}{RGB}{255,255,255}

\path[draw=drawColor,line width= 0.4pt,line join=round,line cap=round,fill=fillColor] (151.77,112.42) circle ( 12.65);
\definecolor{drawColor}{RGB}{0,0,0}

\node[text=drawColor,anchor=base,inner sep=0pt, outer sep=0pt, scale=  1.20] at (151.77,108.29) {7};
\definecolor{drawColor}{RGB}{211,211,211}

\path[draw=drawColor,line width= 0.4pt,line join=round,line cap=round] (151.77,168.63) --
	(252.94,168.63);

\path[draw=drawColor,line width= 0.4pt,line join=round,line cap=round] (151.77,168.63) --
	(252.94,337.26);
\definecolor{drawColor}{RGB}{0,0,0}

\node[text=drawColor,anchor=base west,inner sep=0pt, outer sep=0pt, scale=  1.40] at ( 30.35,164.44) {6. \texttt{region.nw}       };
\end{scope}
\begin{scope}
\path[clip] (  0.00,  0.00) rectangle (505.89,505.89);
\definecolor{drawColor}{RGB}{211,211,211}
\definecolor{fillColor}{RGB}{255,255,255}

\path[draw=drawColor,line width= 0.4pt,line join=round,line cap=round,fill=fillColor] (151.77,168.63) circle ( 12.65);
\definecolor{drawColor}{RGB}{0,0,0}

\node[text=drawColor,anchor=base,inner sep=0pt, outer sep=0pt, scale=  1.20] at (151.77,164.50) {6};

\path[draw=drawColor,line width= 0.4pt,line join=round,line cap=round] (151.77,224.84) --
	(252.94,168.63);

\path[draw=drawColor,line width= 0.4pt,line join=round,line cap=round] (151.77,224.84) --
	(252.94,337.26);

\node[text=drawColor,anchor=base west,inner sep=0pt, outer sep=0pt, scale=  1.40] at ( 30.35,220.65) {5. \texttt{smoker}       };
\end{scope}
\begin{scope}
\path[clip] (  0.00,  0.00) rectangle (505.89,505.89);
\definecolor{drawColor}{RGB}{0,0,0}
\definecolor{fillColor}{RGB}{255,255,255}

\path[draw=drawColor,line width= 0.4pt,line join=round,line cap=round,fill=fillColor] (151.77,224.84) circle ( 12.65);

\node[text=drawColor,anchor=base,inner sep=0pt, outer sep=0pt, scale=  1.20] at (151.77,220.71) {5};
\definecolor{drawColor}{RGB}{211,211,211}

\path[draw=drawColor,line width= 0.4pt,line join=round,line cap=round] (151.77,281.05) --
	(252.94,168.63);
\definecolor{drawColor}{RGB}{0,0,0}

\path[draw=drawColor,line width= 0.4pt,line join=round,line cap=round] (151.77,281.05) --
	(252.94,337.26);

\node[text=drawColor,anchor=base west,inner sep=0pt, outer sep=0pt, scale=  1.40] at ( 30.35,276.86) {4. \texttt{children}       };
\end{scope}
\begin{scope}
\path[clip] (  0.00,  0.00) rectangle (505.89,505.89);
\definecolor{drawColor}{RGB}{0,0,0}
\definecolor{fillColor}{RGB}{255,255,255}

\path[draw=drawColor,line width= 0.4pt,line join=round,line cap=round,fill=fillColor] (151.77,281.05) circle ( 12.65);

\node[text=drawColor,anchor=base,inner sep=0pt, outer sep=0pt, scale=  1.20] at (151.77,276.92) {4};

\path[draw=drawColor,line width= 0.4pt,line join=round,line cap=round] (151.77,337.26) --
	(252.94,168.63);
\definecolor{drawColor}{RGB}{211,211,211}

\path[draw=drawColor,line width= 0.4pt,line join=round,line cap=round] (151.77,337.26) --
	(252.94,337.26);
\definecolor{drawColor}{RGB}{0,0,0}

\node[text=drawColor,anchor=base west,inner sep=0pt, outer sep=0pt, scale=  1.40] at ( 30.35,333.07) {3. \texttt{bmi}       };
\end{scope}
\begin{scope}
\path[clip] (  0.00,  0.00) rectangle (505.89,505.89);
\definecolor{drawColor}{RGB}{0,0,0}
\definecolor{fillColor}{RGB}{255,255,255}

\path[draw=drawColor,line width= 0.4pt,line join=round,line cap=round,fill=fillColor] (151.77,337.26) circle ( 12.65);

\node[text=drawColor,anchor=base,inner sep=0pt, outer sep=0pt, scale=  1.20] at (151.77,333.13) {3};
\definecolor{drawColor}{RGB}{211,211,211}

\path[draw=drawColor,line width= 0.4pt,line join=round,line cap=round] (151.77,393.47) --
	(252.94,168.63);
\definecolor{drawColor}{RGB}{0,0,0}

\path[draw=drawColor,line width= 0.4pt,line join=round,line cap=round] (151.77,393.47) --
	(252.94,337.26);

\node[text=drawColor,anchor=base west,inner sep=0pt, outer sep=0pt, scale=  1.40] at ( 30.35,389.28) {2. \texttt{sex.male}       };
\end{scope}
\begin{scope}
\path[clip] (  0.00,  0.00) rectangle (505.89,505.89);
\definecolor{drawColor}{RGB}{211,211,211}
\definecolor{fillColor}{RGB}{255,255,255}

\path[draw=drawColor,line width= 0.4pt,line join=round,line cap=round,fill=fillColor] (151.77,393.47) circle ( 12.65);
\definecolor{drawColor}{RGB}{0,0,0}

\node[text=drawColor,anchor=base,inner sep=0pt, outer sep=0pt, scale=  1.20] at (151.77,389.34) {2};
\definecolor{drawColor}{RGB}{211,211,211}

\path[draw=drawColor,line width= 0.4pt,line join=round,line cap=round] (151.77,449.68) --
	(252.94,168.63);
\definecolor{drawColor}{RGB}{0,0,0}

\path[draw=drawColor,line width= 0.4pt,line join=round,line cap=round] (151.77,449.68) --
	(252.94,337.26);

\node[text=drawColor,anchor=base west,inner sep=0pt, outer sep=0pt, scale=  1.40] at ( 30.35,445.49) {1. \texttt{age}       };
\end{scope}
\begin{scope}
\path[clip] (  0.00,  0.00) rectangle (505.89,505.89);
\definecolor{drawColor}{RGB}{0,0,0}
\definecolor{fillColor}{RGB}{255,255,255}

\path[draw=drawColor,line width= 0.4pt,line join=round,line cap=round,fill=fillColor] (151.77,449.68) circle ( 12.65);

\node[text=drawColor,anchor=base,inner sep=0pt, outer sep=0pt, scale=  1.20] at (151.77,445.55) {1};

\path[draw=drawColor,line width= 0.4pt,line join=round,line cap=round] (252.94,168.63) --
	(354.12,252.94);

\path[draw=drawColor,line width= 0.4pt,line join=round,line cap=round,fill=fillColor] (252.94,168.63) circle ( 12.65);

\path[draw=drawColor,line width= 0.4pt,line join=round,line cap=round] (252.94,337.26) --
	(354.12,252.94);

\path[draw=drawColor,line width= 0.4pt,line join=round,line cap=round,fill=fillColor] (252.94,337.26) circle ( 12.65);
\end{scope}
\begin{scope}
\path[clip] (  0.00,  0.00) rectangle (505.89,505.89);
\definecolor{drawColor}{RGB}{0,0,0}
\definecolor{fillColor}{RGB}{255,255,255}

\path[draw=drawColor,line width= 0.4pt,line join=round,line cap=round,fill=fillColor] (354.12,252.94) circle ( 12.65);
\end{scope}
\end{tikzpicture}}}
		\caption{Results of the single- and multiple-parameter Wald tests overlaid on the neural network architecture for the \texttt{insurance} data. 
    Weights coloured in black are statistically significant at the 5\% level from the Wald single-parameter test.
    Nodes coloured in black are statistically significant at the 5\% level from the Wald multiple-parameter test.
    The intercept terms are omitted for conciseness.}
		\label{fig: isurance_nn}
	\end{figure}

Returning to the motivating example, Table \ref{tab: insurance_mpwald} contains the results from both the single-parameter and multiple-parameter Wald tests.
For the single-parameter test, the weights that connect each input to the hidden layer are reported, along with an indication of their statistical significance.
For the multiple-parameter test, the p-value associated with each input is reported.
From the multiple-parameter Wald test, \texttt{age}, \texttt{bmi}, \texttt{children}, \texttt{smoker} and \texttt{region.sw} are all statistically significant at the 5\% significance level; these results align broadly with those previously found using linear regression (in Table \ref{tab: insurance_lm}).
A diagram of the corresponding neural network architecture that highlights the statistically significant weights and inputs is given in Figure \ref{fig: isurance_nn}.
This diagram provides a useful and convenient representation of the inferential results for neural networks.

\begin{table}\centering
 \begin{threeparttable}
    \caption{Single- and multiple-parameter Wald test results for the \texttt{insurance} data.
    }
    \begin{tabular}[c]{lllr}
      \toprule
      & \multicolumn{2}{c}{SP} & \multicolumn{1}{c}{MP} \\
        & \multicolumn{1}{c}{$\hat{\omega}_{j1}$} &  \multicolumn{1}{c}{$\hat{\omega}_{j2}$}  & \multicolumn{1}{c}{p-value}  \\ 
       \cmidrule(lr){2-3} \cmidrule(lr){4-4} 
  \texttt{age}        & ~0.65$^{***}$   & ~~0.06         & $<0.001$ \\
  \texttt{sex.male}   & -0.11$^*$       & ~~0.06         & ~~0.101 \\
  \texttt{bmi}        & ~0.02           & ~~3.49$^{***}$ & $<0.001$ \\
  \texttt{children}   & ~0.12$^{***}$   & ~-0.22         & ~~0.001 \\
  \texttt{smoker} & ~1.78$^{***}$   & ~14.54$^{***}$ & $<0.001$\\
  \texttt{region.nw}  & -0.08           & ~~0.19         & ~~0.415 \\
  \texttt{region.se}  & -0.16$^*$       & ~~0.10         & ~~0.080 \\
  \texttt{region.sw}  & -0.22$^{**}$    & ~~0.13         & ~~0.016 \\
       \bottomrule
    \end{tabular}\label{tab: insurance_mpwald}
    {\footnotesize\begin{tablenotes}[para, flushleft]
	\item[]{Significance codes: 0 $^{***}$ 0.001 $^{**}$ 0.01 $^{*}$ 0.05}
 \item[] For the single-parameter test, $\hat{\omega}_{jk}$ is the weight that connects the $j$th covariate to the $k$th hidden node.
	\end{tablenotes}}
 \end{threeparttable}
\end{table}

\section{Covariate Effects}\label{sec: cov_eff}
Hypothesis testing informs us whether or not there is a significant relationship between a covariate and the response, but, when a relationship is present, we also need to determine the nature of this relationship.
Linear regression models naturally provide this information in the form of regression coefficients, which are point estimates, and are easily interpreted, e.g., a unit increase in $x$ results in a change of $\beta$ units in $y$.
For neural networks, it is common to use a graphical approach to understand the (potentially) complex covariate effects that are captured by the model.
A popular approach to assess the relationship between a covariate and the response is using partial dependence (PD) plots \citep{friedman2001greedy}.
The ``partial dependence" of the response on the $j$th covariate can be estimated from the data using 
\begin{equation}\label{eq: pd}
 \overline{\text{NN}}(x^{(j)}) =  \frac{1}{n}\sum_{i=1}^n \text{NN}(x_i^{(j)}, \hat\theta),
\end{equation}
where $x^{(j)}$ is a scalar value to which the $j$th covariate is set,
$x_i^{(j)} = (x_{i0}, x_{i1}, \dotsc, x_{ij} = x^{(j)}, \dotsc, x_{ip})^T$, and $\hat\theta$ is estimated from the data as in Equation \ref{eq: thetahat}.
Equation \ref{eq: pd} can be computed for a sequence of $x^{(j)}$ values, and the pairs of points, $(x^{(j)}, \overline{\text{NN}}(x^{(j)}))$, can then be used to construct a PD plot.

Although the PD plot provides the change in the average predicted response as a covariate varies, it is also useful to consider the \emph{difference} in the average predicted response for a $d$-unit increase the covariate. 
The motivation behind this is to provide an analogous interpretation to that of a regression coefficient ($\beta$) obtained from a classical statistical model.
Thus, adapting the PD plot, we define the effect of a $d$-unit increase in the $j$th covariate on the response as 
\begin{equation}\label{eq: PCE_est}
 \hat{\beta}(x^{(j)},d) = \overline{\text{NN}}(x^{(j)} + d) - \overline{\text{NN}}(x^{(j)}),
\end{equation}
where we suggest setting $d$ to either one or the standard deviation of the $j$th covariate (but, of course, any value can be used).
Again, the $(x^{(j)}, \hat{\beta}(x^{(j)},d))$ pairs can be used to construct a plot, which we term a Partial Covariate-Effect (PCE) plot.
Furthermore, since we take a statistical approach throughout our work, we also estimate (and plot) the associated uncertainty of the PCE function using the delta method (which builds on top of the likelihood-based inference outlined in Sections \ref{sec: FNN} and \ref{sec: hypo_test}).
Note that, if the PCE does not differ significantly from a straight line, the effect of the covariate is linear. 
Moreover, when the covariate of interest is binary, the PCE plot visualises the average difference in prediction when the covariate is changed from the $x^{(j)} = 0$ to the $x^{(j)} = 1$ level.

Due to the ability of neural networks to capture interactions between covariates, PCE plots can be extended to visualise interaction effects.
Following \citet{goldstein2015peeking}, we note that, when the $j$th covariate does not interact with any other covariates, this implies 
\begin{equation}\label{eq: pce_noint}
 \text{NN}(x_i, \theta) = g(x_{ij}) + h(x_i^{(-j)})
\end{equation}
for some functions $g(\cdot)$ and $h(\cdot)$, where $x_i^{(-j)} = (x_{i0}, x_{i1}, \dotsc, x_{ij} = 0,  \dotsc, x_{ip})^T$ is the vector of covariates for the $i$th observation with the $j$th covariate set to zero (which is equivalent to removing it).
Therefore, from Equations \ref{eq: pd} and \ref{eq: PCE_est}, in the case where the $j$th covariate does not interact with other covariates, we have that
\begin{equation*}
 \hat{\beta}(x^{(j)}, d) =  g(x^{(j)} + d) - g(x^{(j)}),
\end{equation*}
i.e., the PCE only depends on the value of the $j$th covariate. 
Therefore, the PCE plots will be equivalent when plotted for different values of other covariates contained in $x_i^{(-j)}$.
However, when an interaction is present, Equation \ref{eq: pce_noint} does not hold, and the PCE plots will vary with respect to $x_i^{(-j)}$.
A two-way interaction between the $j$th covariate and, say, the $k$th covariate can be visualised by fixing the $k$th covariate in Equation \ref{eq: PCE_est} to a (small) set of values; this captures how the effect of the $j$th covariate varies with respect to the $k$th covariate.
This ultimately leads to different PCE plots for each value of the $k$th covariate, and we suggest using two values (generating two plots) via: the mean $\pm$ one standard deviation for continuous variables, and the zero and one levels for binary variables.

\begin{figure}[t!]
\centering
\includegraphics[width=.49\textwidth]{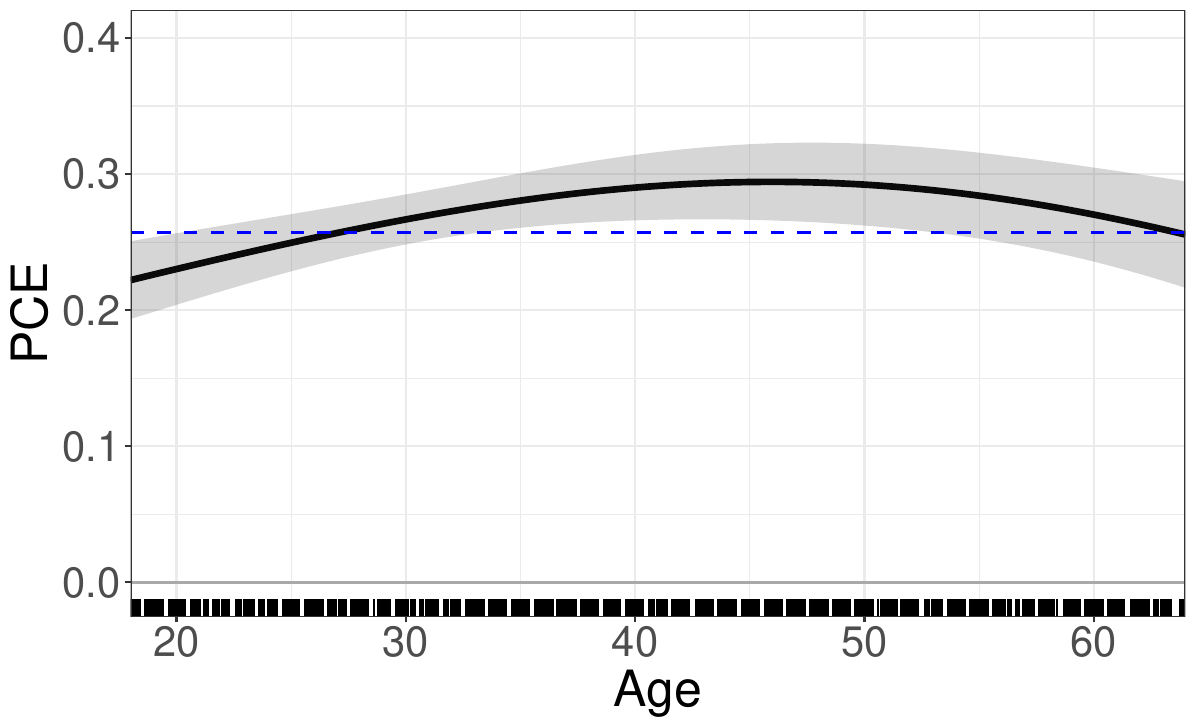}
\caption{PCE plot and its associated 95\% confidence bands for \texttt{age} (with $d$ set to the standard deviation of \texttt{age}). The blue dashed line denotes the corresponding effect from the linear model in Table \ref{tab: insurance_lm}.}
\label{fig: PCE}
\end{figure}

\begin{figure}[t!]
\centering
\includegraphics[width=.49\textwidth]{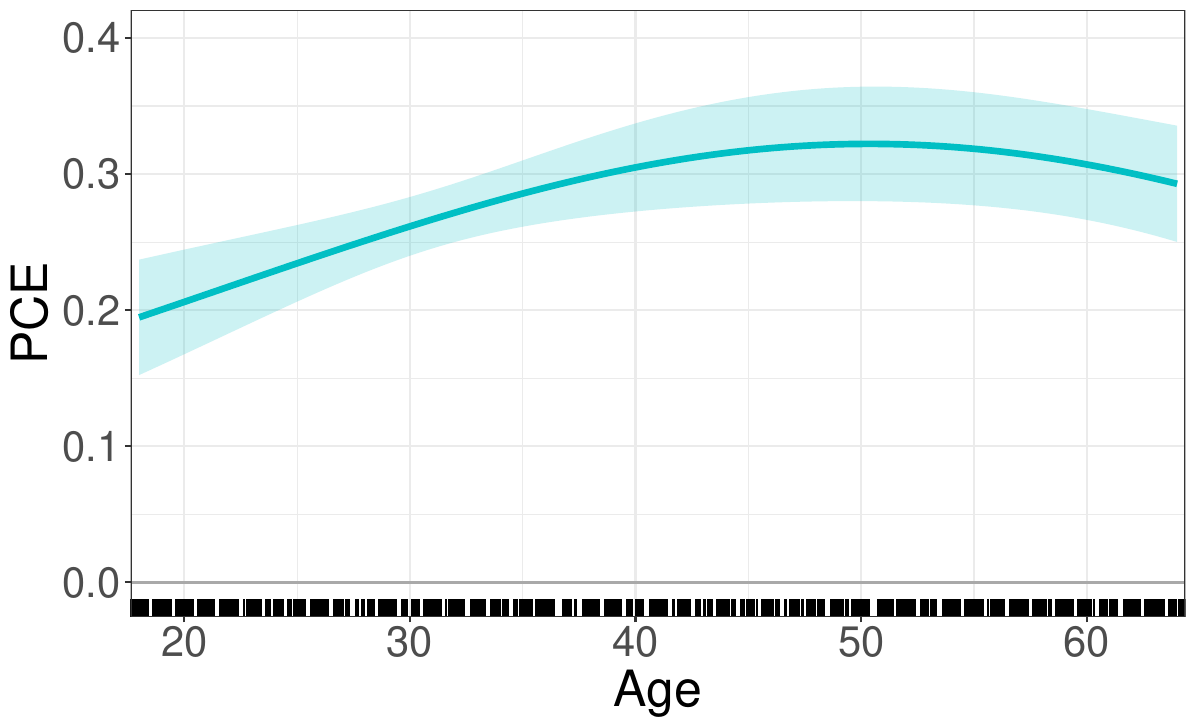}
\includegraphics[width=.49\textwidth]{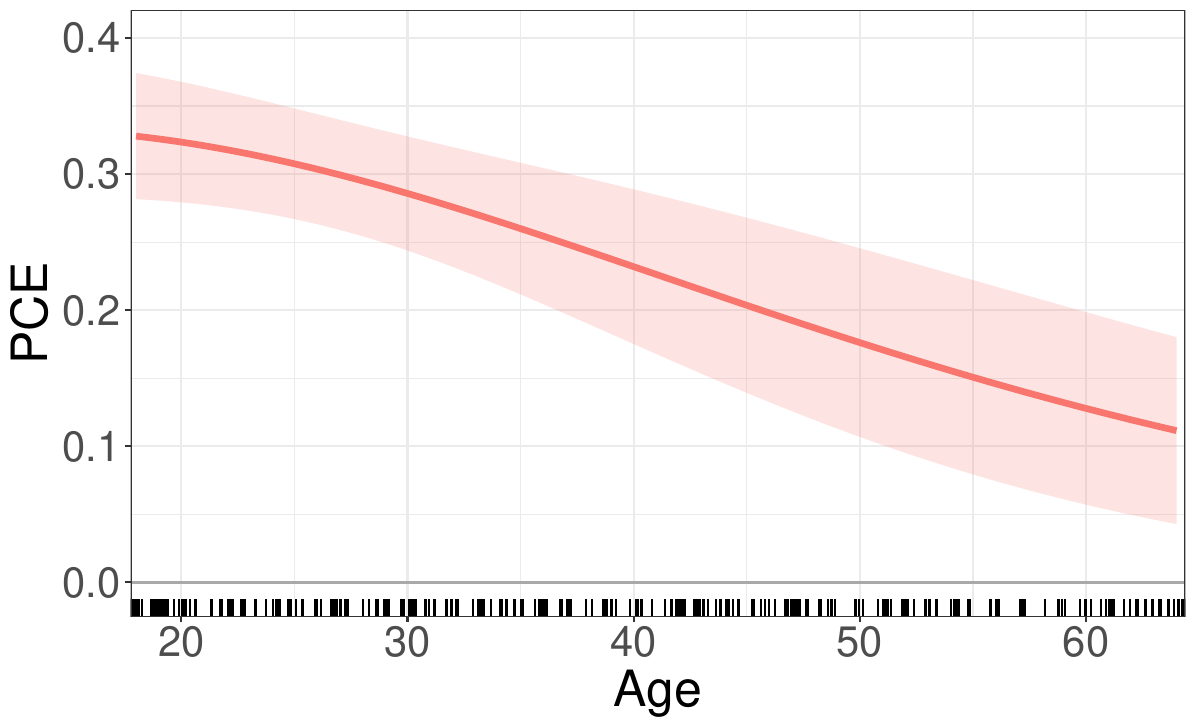}
\includegraphics[width=.7\textwidth]{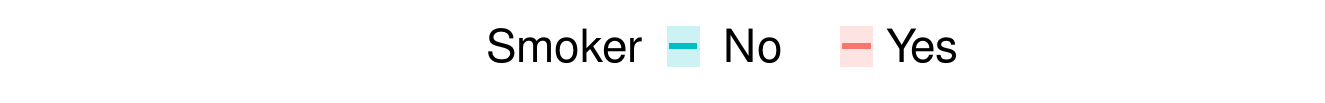}
\caption{PCE plot for \texttt{age} when \texttt{smoker} = 0 (left) and when \texttt{smoker} = 1 (right) (with $d$ set to the standard deviation of \texttt{age})}.
\label{fig: PCE_bmi}
\end{figure}

Revisiting the \texttt{insurance} data, the PCE plot for \texttt{age}, along with its associated uncertainty, is shown in Figure \ref{fig: PCE}.
It is clear that the covariate effect is significantly different from zero (as previously found in Table \ref{tab: insurance_mpwald}).
The effect for \texttt{age} is positive suggesting that older individuals charge higher expenses to their insurance, on average. 
Moreover, the effect increases in magnitude as \texttt{age} increases up to about 45 years, then it begins to reduce.
Investigating possible interactions, Figure \ref{fig: PCE_bmi} displays the partial covariate effect for \texttt{age} for the two levels of \texttt{smoker}.
That the two curves differ suggests the presence of an interaction between the two variables.
When \texttt{smoker} is zero (the individual is not a smoker), the effect for \texttt{age} is similar to the one in Figure \ref{fig: PCE} (and note that most of the individuals are not smokers).
However, when \texttt{smoker} is one (the individual is a smoker), the shape of the effect is quite different.
The effect is still positive, but it is much higher for younger individuals, i.e., young smokers accrue higher expenses.
Furthermore, the PCE decreases with \texttt{age} suggesting that older smokers charge expenses to a lesser extent.

\section{Simulation Study}\label{sec: sim}
The simulation study aims to determine the performance of both the single-parameter and the multiple-parameter Wald test for FNNs (described in Section~\ref{sec: hypo_test}).
The response is generated from an FNN with known ``true" architecture with $p = 6$ input nodes, where $x_1, x_2, \ldots, x_{6}$ are generated from a standard normal distribution.
The ratio of non-zero-to-zero covariates (N-Z) is varied (a zero covariate has $\omega_j = 0_q$) to allow for ratios of 5-to-1 (denoted 5-1), and 1-to-1 (denoted 3-3). 
More specifically, in the 5-1 scenario, $x_1$ is the zero covariate; and in the 3-3 scenario, $x_1, x_3, x_4$ are the zero covariates.
(The specific parameter values can be found in Appendix \ref{app: full_est_results}.)
Also, the hidden-layer structure is varied to contain $q \in \{2, 4, 6\}$ hidden nodes.
This allows for a number of different architectures to be explored, and, in particular, cases where the number of non-zero covariates is less than and greater than the number of hidden nodes.
Sample sizes of size  500, 1000 and 2000 are investigated, and each scenario is repeated for 1000 replicates.
Due to the presence of local maxima in the parameter space, and the requirement for initial weights to begin neural network learning, each model is allowed ten random initialisations to improve its chances of finding a global solution.

Table \ref{tab: sim_wald} contains the results from both the single-parameter and multiple-parameter Wald tests.
For each scenario of the single-parameter test, the empirical type-I error for a significance test with $\alpha = 0.05$ for the true-zero parameter $\omega_{11}$ (Type-I) and the empirical power of the significance test for the non-zero parameter $\omega_{21}$ (Power) are reported.
Similarly, for the multiple-parameter test, the empirical type-I error for a significance test with $\alpha = 0.05$ for the true-zero covariate $x_1$, via $\omega_1$, (Type-I), and the empirical power of the significance test for the non-zero covariate $x_2$, via $\omega_2$, (Power) are reported.
In order to determine the performance of the single-parameter test, the symmetries in the weight space need to be considered.
Therefore, for each replicate, $\hat\theta$ is compared to the true $\theta$ vector, and the necessary node reshuffling and sign flipping is performed to make each $\hat\theta$ comparable across simulation replicates (see Appendix \ref{app: practicalities} for more information).
A preliminary simulation study, provided in Appendix \ref{app: sim_valid}, highlights the improved numerical stability obtained by including the ridge penalty; this is included in all numerical studies presented in the main paper where we fix $\lambda = 0.01$.

\begin{table}
\caption{Simulation: estimates of rejection rates for both the single-parameter and multiple-parameter Wald tests.}
\label{tab: sim_wald}
\resizebox{\textwidth}{!}{
\begin{tabular}{@{}l@{~~} c@{~~} c@{~~}c@{~~}  c@{~~}c@{~~}c@{~~}c@{~~}c@{~~}c@{~~}  c@{~~} c@{~~}c@{~~}c@{~~}c@{~~}c@{~~}c@{~~}}
\toprule
{} & {} & {} & {} & \multicolumn{6}{c}{Single Parameter} && \multicolumn{6}{c}{Multiple Parameter}  \\
{} & {} & {} & {} & \multicolumn{2}{c}{$n = 500$} & \multicolumn{2}{c}{$n = 1000$} & \multicolumn{2}{c}{$n = 2000$} && \multicolumn{2}{c}{$n = 500$} & \multicolumn{2}{c}{$n = 1000$} & \multicolumn{2}{c}{$n = 2000$}  \\
\cmidrule(r){5-10} \cmidrule(r){12-17}
{$q$} & {} & {N-Z} & $K^*$ & Type-I   &  Power & Type-I   &  Power & Type-I   &  Power && Type-I   &  Power & Type-I   &  Power & Type-I   &  Power \\

  \midrule

2 && 5-1 & 15 & 0.058 & 0.596 & 0.066 & 0.687  &  0.068  & 0.818 &&
0.050 &  0.999 & 0.054 & 1.000  &  0.053  & 1.000  \\
  && 3-3 & 11 & 0.068 &  0.650 & 0.044 & 0.747  &  0.041  & 0.889 &&
0.055 &  1.000 & 0.039 & 1.000  &  0.042  & 1.000   \\[0.2cm]

4 && 5-1 & 29 & 0.054 & 0.535 & 0.054 & 0.767  &  0.061  & 0.835 &&
0.038 &  0.767 & 0.048 & 1.000  &  0.057  & 1.000   \\ 
  && 3-3 & 21 & 0.055 &  0.619 & 0.046 & 0.806 &  0.060  & 0.925   &&
0.035 &  0.851 & 0.035 & 0.996  &  0.037  & 1.000   \\ [0.2cm]

6 && 5-1 & 43 & 0.056 & 0.502 & 0.053 & 0.762  &  0.054  & 0.859 &&
0.053 &  0.990 & 0.064 & 1.000  &  0.035 & 1.000   \\ 
  && 3-3 & 31 & 0.194 &  0.433 & 0.147 & 0.547  &  0.128  & 0.749   &&
0.489 &  0.979 & 0.320 & 0.998  &  0.299  & 1.000   \\  
  \bottomrule
  \multicolumn{17}{p{1.2\textwidth}}{\footnotesize  $n$, sample size; $q$, number of nodes in the hidden layer; N-Z, ratio of non-zero-to-zero covariates; $K^*$, number of non-zero parameters in the neural network; Type-I, type-I error for $\omega_{11}$ for the single-parameter test and type-I error for $\omega_{1}$ for the multiple-parameter test; Power, statistical power for $\omega_{21}$ for the single-parameter test and statistical power for $\omega_{2}$ for the multiple-parameter test.}\\
  \end{tabular}}
\end{table}

For both the single- and multiple-parameter Wald tests, when the number of non-zero covariates (N) is greater than the number of hidden nodes ($N > q$), the test performs well both in terms of type-I error and statistical power.
However, when the number of non-zero covariates is less than the number of hidden nodes ($N < q$), neither test performs well in terms of the type-I error (albeit the multiple-parameter test still appears to be powerful). 
This highlights the importance of the neural network architecture, and indicates a clear trade-off between model flexibility and the ability to perform statistical inference.
Here, we have focused on $\omega_{11}$, $\omega_{21}$, $x_1$ via $\omega_{1}$, and $x_2$ via $\omega_{2}$, but the results are similar for the other parameters and covariates, and these results can be found in the Appendix \ref{app: full_est_results}.

\begin{figure}[t]
    \centering
    \includegraphics[width = 0.49\textwidth]{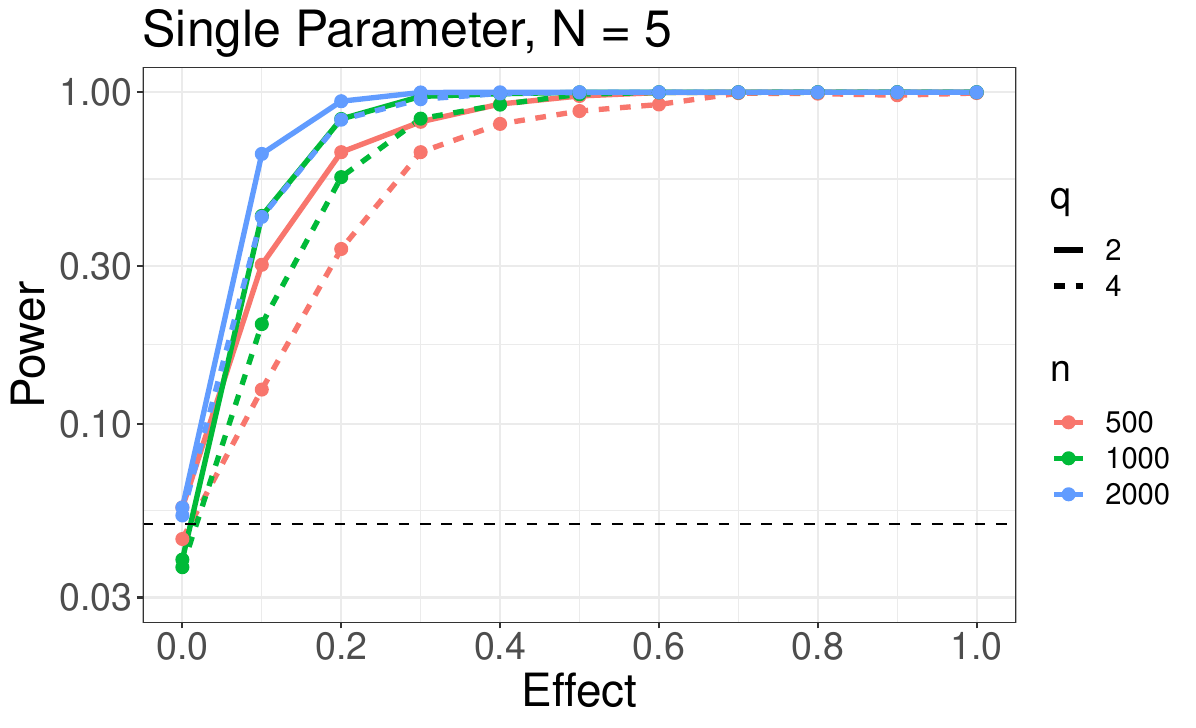}
    \includegraphics[width = 0.49\textwidth]{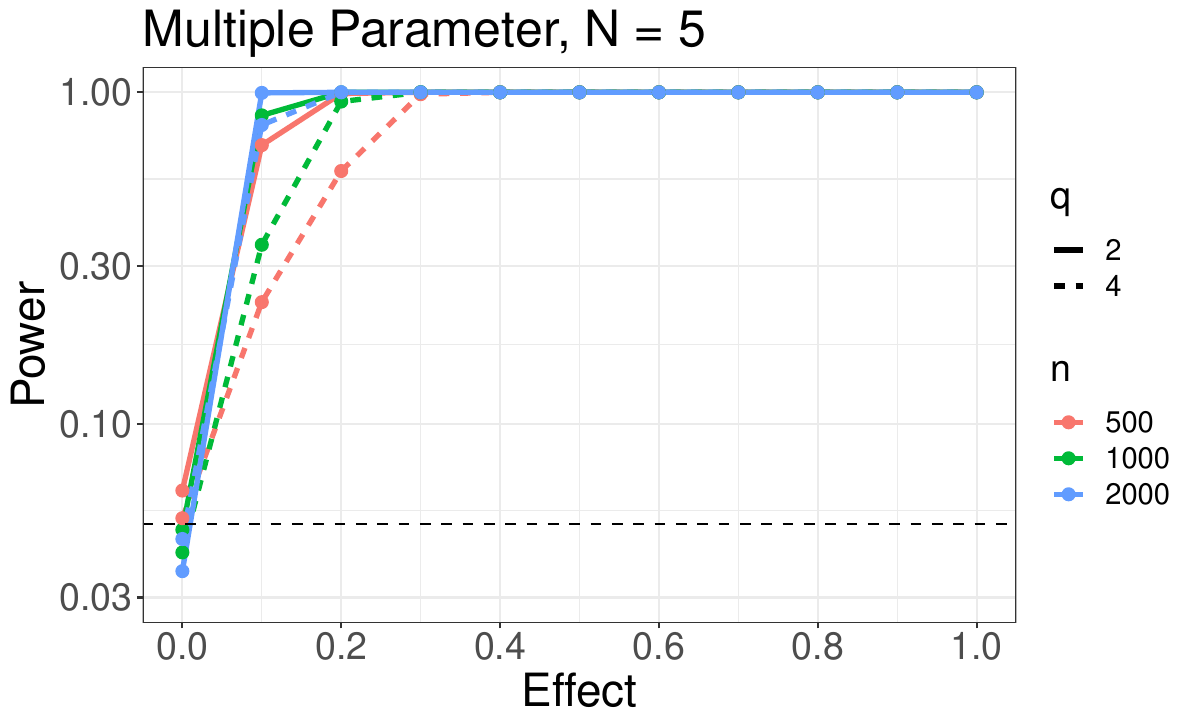}
     \includegraphics[width = 0.49\textwidth]{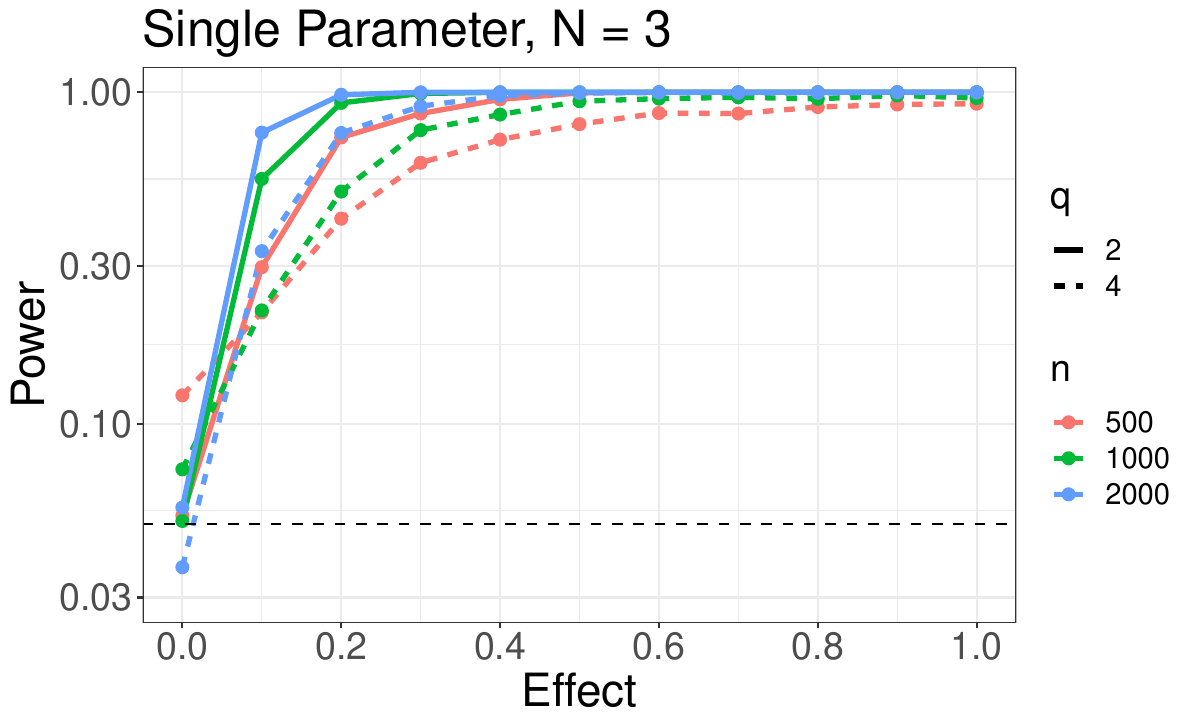}
    \includegraphics[width = 0.49\textwidth]{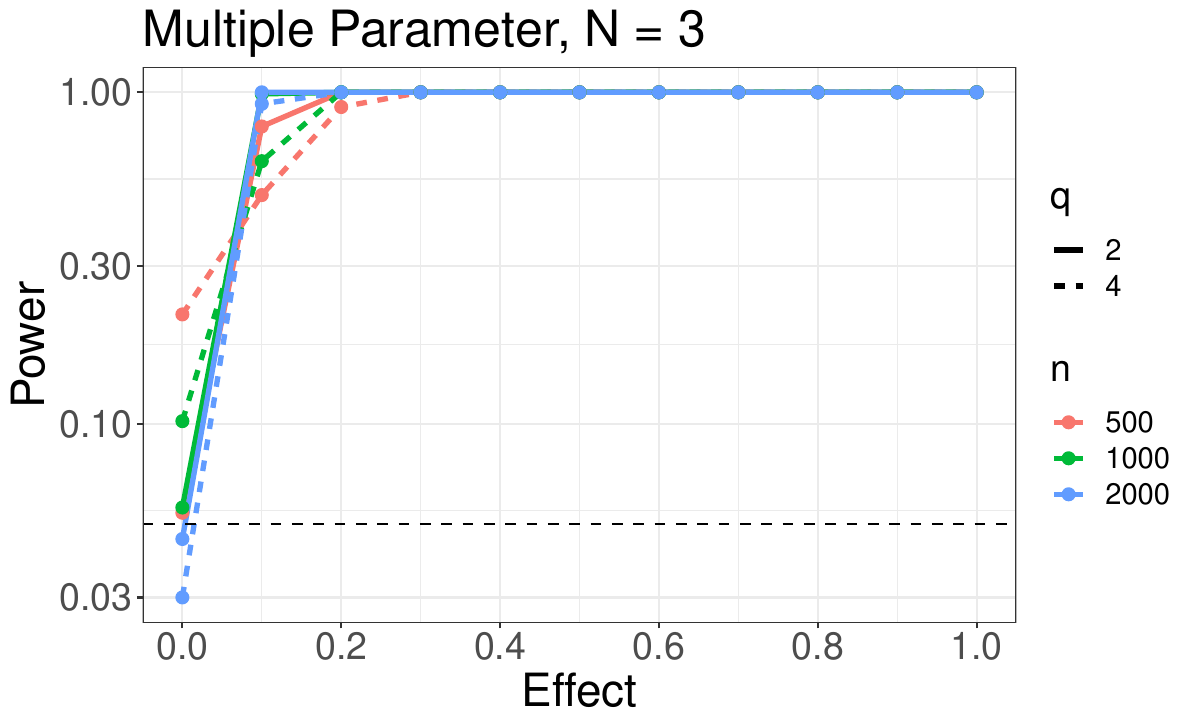}
    \caption{Plot of the statistical power (on a logarithmic scale) for the single- and multiple-parameter Wald test as the size of the effect varies for different sample sizes for $N = 5$ (top) and $N = 3$ (bottom).}
    \label{fig: power}
\end{figure}

To further investigate the performance of the Wald tests and, in particular, the statistical power for increasing effect sizes, we varied the effect of covariate $x_2$ by varying the values of $\omega_2 = (\omega_{21}, \dotsc, \omega_{2q})^T$ such that each $\omega_{2k}$ has the same weight value for values $0, 0.1, 0.2, \dotsc, 1.0$ (i.e., in the first scenario $\omega_2 = (0, \dotsc, 0)^T$, in the second scenario $\omega_2 = (0.1, \dotsc, 0.1)^T$ and so on).
This ensures that weights do not oppose each other (e.g., a positive weight cancelling with a negative weight) and the effect grows at a similar rate between each scenario.
Figure \ref{fig: power} contains a plot of the power versus effect size (the value of $\omega_{2k}$) for both the single- and multiple-parameter tests (the single-parameter test is evaluated using $\omega_{21}$); results are shown for the number of hidden nodes $q \in \{2, 4\}$.
The tests perform as expected: the power (i) is approximately at the nominal significance level ($\alpha = 0.05$) when there is no effect, except when $N < q$ and $n$ is relatively small (as per Table \ref{tab: sim_wald}), (ii) approaches one as the effect size increases, and (iii) converges to one more quickly in larger samples and more slowly in more complex models (larger $q$).
It is important to note that the power of the multiple-parameter Wald test increases far more quickly than for the single-parameter test.
In other words, it is easier to detect an overall covariate effect (which is the combination of multiple weights) than it is to detect the effect of one specific weight; indeed, the former will typically be of greater interest in practice.

\begin{figure}
    \centering
    \includegraphics[width = 0.59\textwidth]{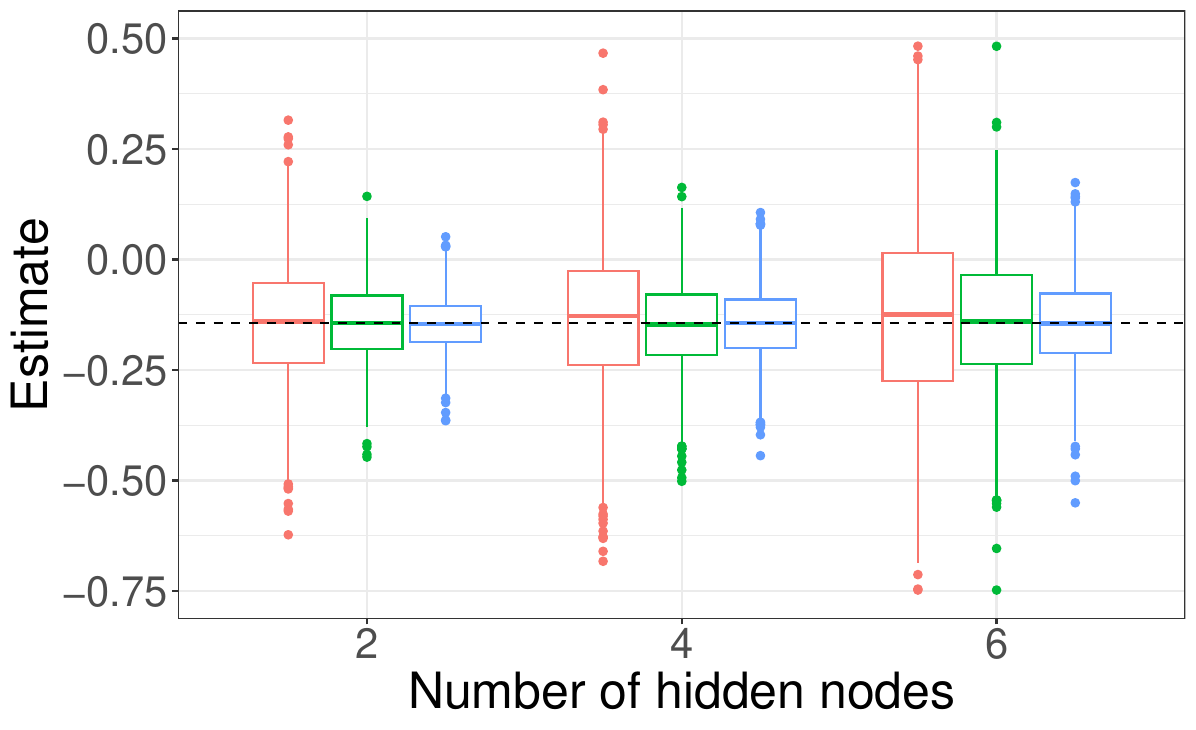}
    \includegraphics[width = 0.09\textwidth]{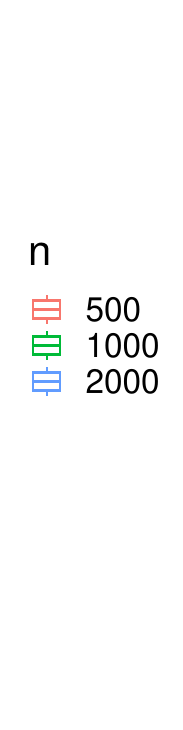}
    \caption{Boxplots for $\hat\omega_{21}$ for different values of $n$ and $q$. Dashed line denotes the true value ($\omega_{21} = -0.14$).}
    \label{fig: theta_hat}
\end{figure}

For the 5-1 scenario, additional estimation and inferential performance metrics are examined in Table \ref{tab: sim_sp_coverage}.
More specifically, we display the average estimate over simulation replicates ($\hat{\theta}$), the true standard error (SE) (i.e., the standard deviation of the estimates over the replicates), the average estimated standard error (SEE) (computed using Equation \ref{eq: sandwich} in a given replicate), and the empirical coverage probability (CP) for a nominal 95\% confidence interval.
In the interest of brevity, only the results for one input-to-hidden-layer weight vector, $\omega_{2}$, are displayed.
(The results for all parameters are available in Appendix \ref{app: full_est_results}.)
Boxplots of the estimates for $\omega_{21}$ for each value of $n$ and $q$ are displayed in Figure \ref{fig: theta_hat}.
We find that estimation bias is relatively low, uncertainty quantification is acceptable in the sense that the SEs are well estimated by the SEEs and the confidence interval coverage is near the nominal 95\% level; moreover, the estimation performance and uncertainty quantification improve with the sample size as expected.
The results from the 3-3 scenario can be found in Appendix \ref{app: full_est_results}.
As expected from our earlier results, the performance is good when $N = 3$ with $q = 2$ (more non-zero covariates than hidden nodes); it is slightly poorer but still acceptable when $N = 3$ with $q = 4$ (similar number of non-zero covariates and hidden nodes); and it is unacceptable when $N = 3$ and $q = 6$ (fewer non-zero covariates than hidden nodes).

\begin{table}
\caption{Simulation: estimation and inference metrics for the 5-1 non-zero-to-zero (N-Z) scenario.}
\label{tab: sim_sp_coverage}
\resizebox{\textwidth}{!}{
\begin{tabular}{@{}l@{~~} c@{~~} c@{~~}  c@{~~}c@{~~}c@{~~}c@{~~}  c@{~~}  c@{~~}c@{~~}c@{~~}c@{~~}  c@{~~}  c@{~~}c@{~~}c@{~~}c@{~~}}
\toprule
{N-Z}  & {5-1} & {} & \multicolumn{4}{c}{$n = 500$} && \multicolumn{4}{c}{$n = 1000$} && \multicolumn{4}{c}{$n = 2000$} \\
\cmidrule(){4-7} \cmidrule(){9-12} \cmidrule(){14-17}
{$q$} &  {} & $\theta$ & $\hat{\theta}$ & SE & SEE & CP && $\hat{\theta}$ & SE & SEE & CP && $\hat{\theta}$ & SE & SEE & CP \\
  \midrule
  2       & $\omega_{21}$ & -0.14 & -0.14 & 0.14 & 0.13 & 0.95  &  & -0.14 & 0.09 & 0.09 & 0.95  &  & -0.15 & 0.06 & 0.06 & 0.94 \\
  {}  & $\omega_{22}$ & -0.27 & -0.27 & 0.06 & 0.06 & 0.95  &  & -0.27 & 0.04 & 0.04 & 0.94  &  & -0.27 & 0.03 & 0.03 & 0.95 \\[0.2cm]
  4        & $\omega_{21}$ & -0.14 & -0.14 & 0.17 & 0.17 & 0.95 &  & -0.15 & 0.10 & 0.10 & 0.95  &  & -0.15 & 0.09 & 0.08 & 0.95 \\
    {}  & $\omega_{22}$ & -0.27 & -0.28 & 0.11 & 0.10 & 0.94 &  & -0.27 & 0.06 & 0.06 & 0.95 &  & -0.27 & 0.05 & 0.04 & 0.94 \\
      {} & $\omega_{23}$ & -0.20 & -0.22 & 0.14 & 0.12 & 0.95  &  & -0.21 & 0.08 & 0.08 & 0.95  &  & -0.20 & 0.06 & 0.06 & 0.93 \\
        {} & $\omega_{24}$ & -0.29 & -0.30 & 0.12 & 0.11 & 0.96 &  & -0.29 & 0.07 & 0.07 & 0.94  &  & -0.29 & 0.05 & 0.05 & 0.95 \\[0.2cm]
  6       & $\omega_{21}$ & -0.14 & -0.14 & 0.24 & 0.20 & 0.93  &  & -0.14 & 0.15 & 0.14 & 0.94  &  & -0.15 & 0.10 & 0.10 & 0.95 \\
    {} & $\omega_{22}$ & -0.27 & -0.28 & 0.14 & 0.12 & 0.93 &  & -0.28 & 0.09 & 0.08 & 0.94 &  & -0.27 & 0.05 & 0.06 & 0.94 \\
      {}  & $\omega_{23}$ & -0.20 & -0.22 & 0.19 & 0.15 & 0.92  &  & -0.21 & 0.11 & 0.10 & 0.93  &  & -0.20 & 0.07 & 0.07 & 0.95 \\
        {}  & $\omega_{24}$ & -0.29 & -0.30 & 0.13 & 0.12 & 0.92  &  & -0.30 & 0.09 & 0.08 & 0.94  &  & -0.29 & 0.06 & 0.06 & 0.95 \\
          {} & $\omega_{25}$ & ~0.27 & ~0.26 & 0.11 & 0.10 & 0.91  &  & ~0.27 & 0.07 & 0.06 & 0.94  &  & ~0.27 & 0.05 & 0.04 & 0.95 \\
            {}  & $\omega_{26}$ & ~0.20 & ~0.20 & 0.09 & 0.08 & 0.91  &  & ~0.20 & 0.06 & 0.06 & 0.93  &  & ~0.20 & 0.04 & 0.04 & 0.94 \\
  \bottomrule
  \multicolumn{17}{p{0.95\textwidth}}{\footnotesize Full results for all parameters available in Appendix \ref{app: full_est_results}; SE, standard deviation of estimates over 1000 replications; SEE, average of estimated standard errors over 1000 replications; CP, the empirical coverage probability of a nominal 95\% confidence interval.}\\
  \end{tabular}}
\end{table}

\clearpage
\section{Application to Data}\label{sec: app_to_data}
The application to the \texttt{insurance} data, which has been revisited throughout the paper, is now explored in more detail.
There are 1,338 observations in total, and the explanatory variables are: the age of the primary beneficiary (\texttt{age}), their gender (\texttt{sex}), their body mass index (\texttt{bmi}), the number of dependents covered by the insurance plan (\texttt{children}), an indicator of the insured's smoking status (\texttt{smoker}), and their region of residence (\texttt{region}), which is divided into north-east (\texttt{region.ne}), south-east (\texttt{region.se}), south-west (\texttt{region.sw}), and north-west (\texttt{region.nw}).
The response variable is the total medical expenses charged to the plan for the calendar year in \$1000s of dollars (\texttt{charges}).
All numeric covariates are standardised to have zero mean and unit variance, and all categorical covariates are dummy encoded.
The variables that are dummy encoded are named \texttt{variable.level} to clarify which level of the variable is represented by a dummy value of one (e.g. \texttt{sex.male}).
The response variable is also standardised to have zero mean and unit variance.
Note that standardisation (of continuous covariates and the response) is done for the purpose of optimisation stability. 
However, when producing the PCE plots, we transform both back to their original scales.

\begin{figure}[b!]
    \centering
    \includegraphics[width = 0.55\textwidth]{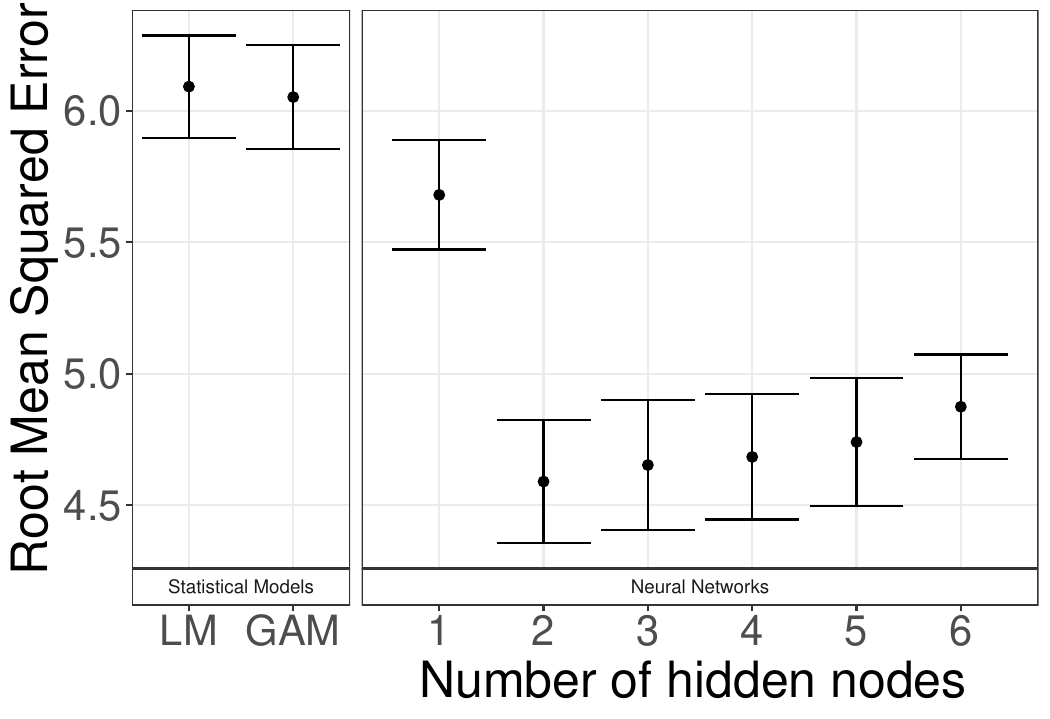}
    \caption{Cross-validated RMSE $\pm$ one standard error for a normal linear regression model (LM), a normal generalised additive model (GAM) and neural networks with varying hidden layer size fit to the \texttt{insurance} data.}
    \label{fig: insurance_mse}
\end{figure}

While model selection is not the focus of this paper, in order to apply the proposed statistical-modelling approach for neural networks, the choice of the number of hidden nodes, $q$, is required.
Here, we select $q$ by considering both out-of-sample performance using five-fold cross-validation.
Figure \ref{fig: insurance_mse} displays the average cross-validated root mean squared error (RMSE) $\pm$ one standard error for a normal linear regression model, a normal generalised additive model (GAM) and neural networks with varying hidden layer size, with $\lambda$ set to 0.01 for all neural networks.
Based on the out-of-sample predictive performance of the models, it is clear that the additional flexibility of neural networks improves upon the linear model and the GAM.
However, there is no further improvement in performance once the hidden layer size goes beyond two hidden nodes.

\begin{figure}
		\centering
		\centerline{\resizebox{0.95\textwidth}{!}{% Created by tikzDevice version 0.12.4 on 2023-10-06 15:59:22
% !TEX encoding = UTF-8 Unicode
\begin{tikzpicture}[x=1pt,y=1pt]
\definecolor{fillColor}{RGB}{255,255,255}
\path[use as bounding box,fill=fillColor,fill opacity=0.00] (00,0) rectangle (375.89,505.89);
\begin{scope}
\path[clip] (  0.00,  0.00) rectangle (505.89,505.89);
\definecolor{drawColor}{RGB}{211,211,211}

\path[draw=drawColor,line width= 0.4pt,line join=round,line cap=round] (151.77, 56.21) --
	(252.94,168.63);
\definecolor{drawColor}{RGB}{0,0,0}

\path[draw=drawColor,line width= 0.4pt,line join=round,line cap=round] (151.77, 56.21) --
	(252.94,337.26);

 \node[text=drawColor,anchor=base west,inner sep=0pt, outer sep=0pt, scale=  1.80] at ( 00.35, 52.02) {8. \texttt{region.sw}       };
\end{scope}
\begin{scope}
\path[clip] (  0.00,  0.00) rectangle (505.89,505.89);
\definecolor{drawColor}{RGB}{0,0,0}
\definecolor{fillColor}{RGB}{255,255,255}

\path[draw=drawColor,line width= 0.4pt,line join=round,line cap=round,fill=fillColor] (151.77, 56.21) circle ( 12.65);

\node[text=drawColor,anchor=base,inner sep=0pt, outer sep=0pt, scale=  1.80] at (151.77, 50) {8};
\definecolor{drawColor}{RGB}{211,211,211}

\path[draw=drawColor,line width= 0.4pt,line join=round,line cap=round] (151.77,112.42) --
	(252.94,168.63);
\definecolor{drawColor}{RGB}{0,0,0}

\path[draw=drawColor,line width= 0.4pt,line join=round,line cap=round] (151.77,112.42) --
	(252.94,337.26);

 \node[text=drawColor,anchor=base west,inner sep=0pt, outer sep=0pt, scale=  1.80] at ( 00.35,108.23) {7. \texttt{region.se}       };
\end{scope}
\begin{scope}
\path[clip] (  0.00,  0.00) rectangle (505.89,505.89);
\definecolor{drawColor}{RGB}{211,211,211}
\definecolor{fillColor}{RGB}{255,255,255}

\path[draw=drawColor,line width= 0.4pt,line join=round,line cap=round,fill=fillColor] (151.77,112.42) circle ( 12.65);
\definecolor{drawColor}{RGB}{0,0,0}

\node[text=drawColor,anchor=base,inner sep=0pt, outer sep=0pt, scale=  1.8] at (151.77,107) {7};
\definecolor{drawColor}{RGB}{211,211,211}

\path[draw=drawColor,line width= 0.4pt,line join=round,line cap=round] (151.77,168.63) --
	(252.94,168.63);

\path[draw=drawColor,line width= 0.4pt,line join=round,line cap=round] (151.77,168.63) --
	(252.94,337.26);

\definecolor{drawColor}{RGB}{0,0,0}
 \node[text=drawColor,anchor=base west,inner sep=0pt, outer sep=0pt, scale=  1.80] at ( 00.35,164.44) {6. \texttt{region.nw}       };
\end{scope}
\begin{scope}
\path[clip] (  0.00,  0.00) rectangle (505.89,505.89);
\definecolor{drawColor}{RGB}{211,211,211}
\definecolor{fillColor}{RGB}{255,255,255}

\path[draw=drawColor,line width= 0.4pt,line join=round,line cap=round,fill=fillColor] (151.77,168.63) circle ( 12.65);
\definecolor{drawColor}{RGB}{0,0,0}

\node[text=drawColor,anchor=base,inner sep=0pt, outer sep=0pt, scale=  1.80] at (151.77,163) {6};

\path[draw=drawColor,line width= 0.4pt,line join=round,line cap=round] (151.77,224.84) --
	(252.94,168.63);

\path[draw=drawColor,line width= 0.4pt,line join=round,line cap=round] (151.77,224.84) --
	(252.94,337.26);

 \node[text=drawColor,anchor=base west,inner sep=0pt, outer sep=0pt, scale=  1.80] at ( 00.35,220.65) {5. \texttt{smoker}       };
\end{scope}
\begin{scope}
\path[clip] (  0.00,  0.00) rectangle (505.89,505.89);
\definecolor{drawColor}{RGB}{0,0,0}
\definecolor{fillColor}{RGB}{255,255,255}

\path[draw=drawColor,line width= 0.4pt,line join=round,line cap=round,fill=fillColor] (151.77,224.84) circle ( 12.65);

\node[text=drawColor,anchor=base,inner sep=0pt, outer sep=0pt, scale=  1.80] at (151.77,219.2) {5};
\definecolor{drawColor}{RGB}{211,211,211}

\path[draw=drawColor,line width= 0.4pt,line join=round,line cap=round] (151.77,281.05) --
	(252.94,168.63);
\definecolor{drawColor}{RGB}{0,0,0}

\path[draw=drawColor,line width= 0.4pt,line join=round,line cap=round] (151.77,281.05) --
	(252.94,337.26);

 \node[text=drawColor,anchor=base west,inner sep=0pt, outer sep=0pt, scale=  1.80] at ( 00.35,276.86) {4. \texttt{children}       };
\end{scope}
\begin{scope}
\path[clip] (  0.00,  0.00) rectangle (505.89,505.89);
\definecolor{drawColor}{RGB}{0,0,0}
\definecolor{fillColor}{RGB}{255,255,255}

\path[draw=drawColor,line width= 0.4pt,line join=round,line cap=round,fill=fillColor] (151.77,281.05) circle ( 12.65);

\node[text=drawColor,anchor=base,inner sep=0pt, outer sep=0pt, scale=  1.80] at (151.77,275) {4};

\path[draw=drawColor,line width= 0.4pt,line join=round,line cap=round] (151.77,337.26) --
	(252.94,168.63);
\definecolor{drawColor}{RGB}{211,211,211}

\path[draw=drawColor,line width= 0.4pt,line join=round,line cap=round] (151.77,337.26) --
	(252.94,337.26);

\definecolor{drawColor}{RGB}{0,0,0}
 \node[text=drawColor,anchor=base west,inner sep=0pt, outer sep=0pt, scale=  1.80] at ( 00.35,333.07) {3. \texttt{bmi}       };
\end{scope}
\begin{scope}
\path[clip] (  0.00,  0.00) rectangle (505.89,505.89);
\definecolor{drawColor}{RGB}{0,0,0}
\definecolor{fillColor}{RGB}{255,255,255}

\path[draw=drawColor,line width= 0.4pt,line join=round,line cap=round,fill=fillColor] (151.77,337.26) circle ( 12.65);

\node[text=drawColor,anchor=base,inner sep=0pt, outer sep=0pt, scale=  1.80] at (151.77,332) {3};
\definecolor{drawColor}{RGB}{211,211,211}

\path[draw=drawColor,line width= 0.4pt,line join=round,line cap=round] (151.77,393.47) --
	(252.94,168.63);
\definecolor{drawColor}{RGB}{0,0,0}

\path[draw=drawColor,line width= 0.4pt,line join=round,line cap=round] (151.77,393.47) --
	(252.94,337.26);

 \node[text=drawColor,anchor=base west,inner sep=0pt, outer sep=0pt, scale=  1.80] at ( 00.35,389.28) {2. \texttt{sex.male}       };
\end{scope}
\begin{scope}
\path[clip] (  0.00,  0.00) rectangle (505.89,505.89);
\definecolor{drawColor}{RGB}{211,211,211}
\definecolor{fillColor}{RGB}{255,255,255}

\path[draw=drawColor,line width= 0.4pt,line join=round,line cap=round,fill=fillColor] (151.77,393.47) circle ( 12.65);
\definecolor{drawColor}{RGB}{0,0,0}

\node[text=drawColor,anchor=base,inner sep=0pt, outer sep=0pt, scale=  1.80] at (151.77,388) {2};
\definecolor{drawColor}{RGB}{211,211,211}

\path[draw=drawColor,line width= 0.4pt,line join=round,line cap=round] (151.77,449.68) --
	(252.94,168.63);
\definecolor{drawColor}{RGB}{0,0,0}

\path[draw=drawColor,line width= 0.4pt,line join=round,line cap=round] (151.77,449.68) --
	(252.94,337.26);

 \node[text=drawColor,anchor=base west,inner sep=0pt, outer sep=0pt, scale=  1.80] at ( 00.35,445.49) {1. \texttt{age}       };
\end{scope}
\begin{scope}
\path[clip] (  0.00,  0.00) rectangle (505.89,505.89);
\definecolor{drawColor}{RGB}{0,0,0}
\definecolor{fillColor}{RGB}{255,255,255}

\path[draw=drawColor,line width= 0.4pt,line join=round,line cap=round,fill=fillColor] (151.77,449.68) circle ( 12.65);

\node[text=drawColor,anchor=base,inner sep=0pt, outer sep=0pt, scale=  1.80] at (151.77,444) {1};

\path[draw=drawColor,line width= 0.4pt,line join=round,line cap=round] (252.94,168.63) --
	(354.12,252.94);

\path[draw=drawColor,line width= 0.4pt,line join=round,line cap=round,fill=fillColor] (252.94,168.63) circle ( 12.65);

\path[draw=drawColor,line width= 0.4pt,line join=round,line cap=round] (252.94,337.26) --
	(354.12,252.94);

\path[draw=drawColor,line width= 0.4pt,line join=round,line cap=round,fill=fillColor] (252.94,337.26) circle ( 12.65);
\end{scope}
\begin{scope}
\path[clip] (  0.00,  0.00) rectangle (505.89,505.89);
\definecolor{drawColor}{RGB}{0,0,0}
\definecolor{fillColor}{RGB}{255,255,255}

\path[draw=drawColor,line width= 0.4pt,line join=round,line cap=round,fill=fillColor] (354.12,252.94) circle ( 12.65);

\node[text=drawColor,anchor=base,inner sep=0pt, outer sep=0pt, scale=  2] at (252.94,480.60) {$q  = 2$};
\end{scope}
\end{tikzpicture}% Created by tikzDevice version 0.12.4 on 2023-10-06 15:59:36
% !TEX encoding = UTF-8 Unicode
\begin{tikzpicture}[x=1pt,y=1pt]
\definecolor{fillColor}{RGB}{255,255,255}
\path[use as bounding box,fill=fillColor,fill opacity=0.00] (130,0) rectangle (375.89,505.89);
\begin{scope}
\path[clip] (  0.00,  0.00) rectangle (505.89,505.89);
\definecolor{drawColor}{RGB}{0,0,0}

\path[draw=drawColor,line width= 0.4pt,line join=round,line cap=round] (151.77, 56.21) --
	(252.94,126.47);
\definecolor{drawColor}{RGB}{211,211,211}

\path[draw=drawColor,line width= 0.4pt,line join=round,line cap=round] (151.77, 56.21) --
	(252.94,252.94);

\path[draw=drawColor,line width= 0.4pt,line join=round,line cap=round] (151.77, 56.21) --
	(252.94,379.42);
\end{scope}
\begin{scope}
\path[clip] (  0.00,  0.00) rectangle (505.89,505.89);
\definecolor{drawColor}{RGB}{0,0,0}
\definecolor{fillColor}{RGB}{255,255,255}

\path[draw=drawColor,line width= 0.4pt,line join=round,line cap=round,fill=fillColor] (151.77, 56.21) circle ( 12.65);

\node[text=drawColor,anchor=base,inner sep=0pt, outer sep=0pt, scale=  1.80] at (151.77, 50) {8};
\definecolor{drawColor}{RGB}{211,211,211}

\path[draw=drawColor,line width= 0.4pt,line join=round,line cap=round] (151.77,112.42) --
	(252.94,126.47);

\path[draw=drawColor,line width= 0.4pt,line join=round,line cap=round] (151.77,112.42) --
	(252.94,252.94);

\path[draw=drawColor,line width= 0.4pt,line join=round,line cap=round] (151.77,112.42) --
	(252.94,379.42);
\end{scope}
\begin{scope}
\path[clip] (  0.00,  0.00) rectangle (505.89,505.89);
\definecolor{drawColor}{RGB}{211,211,211}
\definecolor{fillColor}{RGB}{255,255,255}

\path[draw=drawColor,line width= 0.4pt,line join=round,line cap=round,fill=fillColor] (151.77,112.42) circle ( 12.65);
\definecolor{drawColor}{RGB}{0,0,0}

\node[text=drawColor,anchor=base,inner sep=0pt, outer sep=0pt, scale=  1.8] at (151.77,107) {7};
\definecolor{drawColor}{RGB}{211,211,211}

\path[draw=drawColor,line width= 0.4pt,line join=round,line cap=round] (151.77,168.63) --
	(252.94,126.47);

\path[draw=drawColor,line width= 0.4pt,line join=round,line cap=round] (151.77,168.63) --
	(252.94,252.94);

\path[draw=drawColor,line width= 0.4pt,line join=round,line cap=round] (151.77,168.63) --
	(252.94,379.42);
\end{scope}
\begin{scope}
\path[clip] (  0.00,  0.00) rectangle (505.89,505.89);
\definecolor{drawColor}{RGB}{211,211,211}
\definecolor{fillColor}{RGB}{255,255,255}

\path[draw=drawColor,line width= 0.4pt,line join=round,line cap=round,fill=fillColor] (151.77,168.63) circle ( 12.65);
\definecolor{drawColor}{RGB}{0,0,0}

\node[text=drawColor,anchor=base,inner sep=0pt, outer sep=0pt, scale=  1.80] at (151.77,163) {6};

\path[draw=drawColor,line width= 0.4pt,line join=round,line cap=round] (151.77,224.84) --
	(252.94,126.47);

\path[draw=drawColor,line width= 0.4pt,line join=round,line cap=round] (151.77,224.84) --
	(252.94,252.94);

\path[draw=drawColor,line width= 0.4pt,line join=round,line cap=round] (151.77,224.84) --
	(252.94,379.42);
\end{scope}
\begin{scope}
\path[clip] (  0.00,  0.00) rectangle (505.89,505.89);
\definecolor{drawColor}{RGB}{0,0,0}
\definecolor{fillColor}{RGB}{255,255,255}

\path[draw=drawColor,line width= 0.4pt,line join=round,line cap=round,fill=fillColor] (151.77,224.84) circle ( 12.65);

\node[text=drawColor,anchor=base,inner sep=0pt, outer sep=0pt, scale=  1.80] at (151.77,219.2) {5};

\path[draw=drawColor,line width= 0.4pt,line join=round,line cap=round] (151.77,281.05) --
	(252.94,126.47);
\definecolor{drawColor}{RGB}{211,211,211}

\path[draw=drawColor,line width= 0.4pt,line join=round,line cap=round] (151.77,281.05) --
	(252.94,252.94);
\definecolor{drawColor}{RGB}{0,0,0}

\path[draw=drawColor,line width= 0.4pt,line join=round,line cap=round] (151.77,281.05) --
	(252.94,379.42);
\end{scope}
\begin{scope}
\path[clip] (  0.00,  0.00) rectangle (505.89,505.89);
\definecolor{drawColor}{RGB}{0,0,0}
\definecolor{fillColor}{RGB}{255,255,255}

\path[draw=drawColor,line width= 0.4pt,line join=round,line cap=round,fill=fillColor] (151.77,281.05) circle ( 12.65);

\node[text=drawColor,anchor=base,inner sep=0pt, outer sep=0pt, scale=  1.80] at (151.77,275) {4};

\path[draw=drawColor,line width= 0.4pt,line join=round,line cap=round] (151.77,337.26) --
	(252.94,126.47);

\path[draw=drawColor,line width= 0.4pt,line join=round,line cap=round] (151.77,337.26) --
	(252.94,252.94);

\path[draw=drawColor,line width= 0.4pt,line join=round,line cap=round] (151.77,337.26) --
	(252.94,379.42);
\end{scope}
\begin{scope}
\path[clip] (  0.00,  0.00) rectangle (505.89,505.89);
\definecolor{drawColor}{RGB}{0,0,0}
\definecolor{fillColor}{RGB}{255,255,255}

\path[draw=drawColor,line width= 0.4pt,line join=round,line cap=round,fill=fillColor] (151.77,337.26) circle ( 12.65);

\node[text=drawColor,anchor=base,inner sep=0pt, outer sep=0pt, scale=  1.80] at (151.77,332) {3};

\path[draw=drawColor,line width= 0.4pt,line join=round,line cap=round] (151.77,393.47) --
	(252.94,126.47);
\definecolor{drawColor}{RGB}{211,211,211}

\path[draw=drawColor,line width= 0.4pt,line join=round,line cap=round] (151.77,393.47) --
	(252.94,252.94);

\path[draw=drawColor,line width= 0.4pt,line join=round,line cap=round] (151.77,393.47) --
	(252.94,379.42);
\end{scope}
\begin{scope}
\path[clip] (  0.00,  0.00) rectangle (505.89,505.89);
\definecolor{drawColor}{RGB}{211,211,211}
\definecolor{fillColor}{RGB}{255,255,255}

\path[draw=drawColor,line width= 0.4pt,line join=round,line cap=round,fill=fillColor] (151.77,393.47) circle ( 12.65);
\definecolor{drawColor}{RGB}{0,0,0}

\node[text=drawColor,anchor=base,inner sep=0pt, outer sep=0pt, scale=  1.80] at (151.77,388) {2};

\path[draw=drawColor,line width= 0.4pt,line join=round,line cap=round] (151.77,449.68) --
	(252.94,126.47);
\definecolor{drawColor}{RGB}{211,211,211}

\path[draw=drawColor,line width= 0.4pt,line join=round,line cap=round] (151.77,449.68) --
	(252.94,252.94);

\path[draw=drawColor,line width= 0.4pt,line join=round,line cap=round] (151.77,449.68) --
	(252.94,379.42);
\end{scope}
\begin{scope}
\path[clip] (  0.00,  0.00) rectangle (505.89,505.89);
\definecolor{drawColor}{RGB}{0,0,0}
\definecolor{fillColor}{RGB}{255,255,255}

\path[draw=drawColor,line width= 0.4pt,line join=round,line cap=round,fill=fillColor] (151.77,449.68) circle ( 12.65);

\node[text=drawColor,anchor=base,inner sep=0pt, outer sep=0pt, scale=  1.80] at (151.77,444) {1};

\path[draw=drawColor,line width= 0.4pt,line join=round,line cap=round] (252.94,126.47) --
	(354.12,252.94);

\path[draw=drawColor,line width= 0.4pt,line join=round,line cap=round,fill=fillColor] (252.94,126.47) circle ( 12.65);

\path[draw=drawColor,line width= 0.4pt,line join=round,line cap=round] (252.94,252.94) --
	(354.12,252.94);

\path[draw=drawColor,line width= 0.4pt,line join=round,line cap=round,fill=fillColor] (252.94,252.94) circle ( 12.65);

\path[draw=drawColor,line width= 0.4pt,line join=round,line cap=round] (252.94,379.42) --
	(354.12,252.94);

\path[draw=drawColor,line width= 0.4pt,line join=round,line cap=round,fill=fillColor] (252.94,379.42) circle ( 12.65);
\end{scope}
\begin{scope}
\path[clip] (  0.00,  0.00) rectangle (505.89,505.89);
\definecolor{drawColor}{RGB}{0,0,0}
\definecolor{fillColor}{RGB}{255,255,255}

\path[draw=drawColor,line width= 0.4pt,line join=round,line cap=round,fill=fillColor] (354.12,252.94) circle ( 12.65);

\node[text=drawColor,anchor=base,inner sep=0pt, outer sep=0pt, scale=  2] at (252.94,480.60) {$q  = 3$};
\end{scope}
\end{tikzpicture}% Created by tikzDevice version 0.12.4 on 2023-10-06 15:59:50
% !TEX encoding = UTF-8 Unicode
\begin{tikzpicture}[x=1pt,y=1pt]
\definecolor{fillColor}{RGB}{255,255,255}
\path[use as bounding box,fill=fillColor,fill opacity=0.00] (130,0) rectangle (375.89,505.89);
\begin{scope}
\path[clip] (  0.00,  0.00) rectangle (505.89,505.89);
\definecolor{drawColor}{RGB}{211,211,211}

\path[draw=drawColor,line width= 0.4pt,line join=round,line cap=round] (151.77, 56.21) --
	(252.94,101.18);

\path[draw=drawColor,line width= 0.4pt,line join=round,line cap=round] (151.77, 56.21) --
	(252.94,202.36);

\path[draw=drawColor,line width= 0.4pt,line join=round,line cap=round] (151.77, 56.21) --
	(252.94,303.53);

\path[draw=drawColor,line width= 0.4pt,line join=round,line cap=round] (151.77, 56.21) --
	(252.94,404.71);
\end{scope}
\begin{scope}
\path[clip] (  0.00,  0.00) rectangle (505.89,505.89);
\definecolor{drawColor}{RGB}{211,211,211}
\definecolor{fillColor}{RGB}{255,255,255}

\path[draw=drawColor,line width= 0.4pt,line join=round,line cap=round,fill=fillColor] (151.77, 56.21) circle ( 12.65);
\definecolor{drawColor}{RGB}{0,0,0}

\node[text=drawColor,anchor=base,inner sep=0pt, outer sep=0pt, scale=  1.80] at (151.77, 50) {8};
\definecolor{drawColor}{RGB}{211,211,211}

\path[draw=drawColor,line width= 0.4pt,line join=round,line cap=round] (151.77,112.42) --
	(252.94,101.18);

\path[draw=drawColor,line width= 0.4pt,line join=round,line cap=round] (151.77,112.42) --
	(252.94,202.36);

\path[draw=drawColor,line width= 0.4pt,line join=round,line cap=round] (151.77,112.42) --
	(252.94,303.53);

\path[draw=drawColor,line width= 0.4pt,line join=round,line cap=round] (151.77,112.42) --
	(252.94,404.71);
\end{scope}
\begin{scope}
\path[clip] (  0.00,  0.00) rectangle (505.89,505.89);
\definecolor{drawColor}{RGB}{211,211,211}
\definecolor{fillColor}{RGB}{255,255,255}

\path[draw=drawColor,line width= 0.4pt,line join=round,line cap=round,fill=fillColor] (151.77,112.42) circle ( 12.65);
\definecolor{drawColor}{RGB}{0,0,0}

\node[text=drawColor,anchor=base,inner sep=0pt, outer sep=0pt, scale=  1.8] at (151.77,107) {7};
\definecolor{drawColor}{RGB}{211,211,211}

\path[draw=drawColor,line width= 0.4pt,line join=round,line cap=round] (151.77,168.63) --
	(252.94,101.18);

\path[draw=drawColor,line width= 0.4pt,line join=round,line cap=round] (151.77,168.63) --
	(252.94,202.36);

\path[draw=drawColor,line width= 0.4pt,line join=round,line cap=round] (151.77,168.63) --
	(252.94,303.53);

\path[draw=drawColor,line width= 0.4pt,line join=round,line cap=round] (151.77,168.63) --
	(252.94,404.71);
\end{scope}
\begin{scope}
\path[clip] (  0.00,  0.00) rectangle (505.89,505.89);
\definecolor{drawColor}{RGB}{211,211,211}
\definecolor{fillColor}{RGB}{255,255,255}

\path[draw=drawColor,line width= 0.4pt,line join=round,line cap=round,fill=fillColor] (151.77,168.63) circle ( 12.65);
\definecolor{drawColor}{RGB}{0,0,0}

\node[text=drawColor,anchor=base,inner sep=0pt, outer sep=0pt, scale=  1.80] at (151.77,163) {6};

\path[draw=drawColor,line width= 0.4pt,line join=round,line cap=round] (151.77,224.84) --
	(252.94,101.18);
\definecolor{drawColor}{RGB}{211,211,211}

\path[draw=drawColor,line width= 0.4pt,line join=round,line cap=round] (151.77,224.84) --
	(252.94,202.36);
\definecolor{drawColor}{RGB}{0,0,0}

\path[draw=drawColor,line width= 0.4pt,line join=round,line cap=round] (151.77,224.84) --
	(252.94,303.53);

\path[draw=drawColor,line width= 0.4pt,line join=round,line cap=round] (151.77,224.84) --
	(252.94,404.71);
\end{scope}
\begin{scope}
\path[clip] (  0.00,  0.00) rectangle (505.89,505.89);
\definecolor{drawColor}{RGB}{0,0,0}
\definecolor{fillColor}{RGB}{255,255,255}

\path[draw=drawColor,line width= 0.4pt,line join=round,line cap=round,fill=fillColor] (151.77,224.84) circle ( 12.65);

\node[text=drawColor,anchor=base,inner sep=0pt, outer sep=0pt, scale=  1.80] at (151.77,219.2) {5};
\definecolor{drawColor}{RGB}{211,211,211}

\path[draw=drawColor,line width= 0.4pt,line join=round,line cap=round] (151.77,281.05) --
	(252.94,101.18);

\path[draw=drawColor,line width= 0.4pt,line join=round,line cap=round] (151.77,281.05) --
	(252.94,202.36);
\definecolor{drawColor}{RGB}{0,0,0}

\path[draw=drawColor,line width= 0.4pt,line join=round,line cap=round] (151.77,281.05) --
	(252.94,303.53);
\definecolor{drawColor}{RGB}{211,211,211}

\path[draw=drawColor,line width= 0.4pt,line join=round,line cap=round] (151.77,281.05) --
	(252.94,404.71);
\end{scope}
\begin{scope}
\path[clip] (  0.00,  0.00) rectangle (505.89,505.89);
\definecolor{drawColor}{RGB}{0,0,0}
\definecolor{fillColor}{RGB}{255,255,255}

\path[draw=drawColor,line width= 0.4pt,line join=round,line cap=round,fill=fillColor] (151.77,281.05) circle ( 12.65);

\node[text=drawColor,anchor=base,inner sep=0pt, outer sep=0pt, scale=  1.80] at (151.77,275) {4};
\definecolor{drawColor}{RGB}{211,211,211}

\path[draw=drawColor,line width= 0.4pt,line join=round,line cap=round] (151.77,337.26) --
	(252.94,101.18);

\path[draw=drawColor,line width= 0.4pt,line join=round,line cap=round] (151.77,337.26) --
	(252.94,202.36);
\definecolor{drawColor}{RGB}{0,0,0}

\path[draw=drawColor,line width= 0.4pt,line join=round,line cap=round] (151.77,337.26) --
	(252.94,303.53);

\path[draw=drawColor,line width= 0.4pt,line join=round,line cap=round] (151.77,337.26) --
	(252.94,404.71);
\end{scope}
\begin{scope}
\path[clip] (  0.00,  0.00) rectangle (505.89,505.89);
\definecolor{drawColor}{RGB}{0,0,0}
\definecolor{fillColor}{RGB}{255,255,255}

\path[draw=drawColor,line width= 0.4pt,line join=round,line cap=round,fill=fillColor] (151.77,337.26) circle ( 12.65);

\node[text=drawColor,anchor=base,inner sep=0pt, outer sep=0pt, scale=  1.80] at (151.77,332) {3};
\definecolor{drawColor}{RGB}{211,211,211}

\path[draw=drawColor,line width= 0.4pt,line join=round,line cap=round] (151.77,393.47) --
	(252.94,101.18);

\path[draw=drawColor,line width= 0.4pt,line join=round,line cap=round] (151.77,393.47) --
	(252.94,202.36);

\path[draw=drawColor,line width= 0.4pt,line join=round,line cap=round] (151.77,393.47) --
	(252.94,303.53);

\path[draw=drawColor,line width= 0.4pt,line join=round,line cap=round] (151.77,393.47) --
	(252.94,404.71);
\end{scope}
\begin{scope}
\path[clip] (  0.00,  0.00) rectangle (505.89,505.89);
\definecolor{drawColor}{RGB}{211,211,211}
\definecolor{fillColor}{RGB}{255,255,255}

\path[draw=drawColor,line width= 0.4pt,line join=round,line cap=round,fill=fillColor] (151.77,393.47) circle ( 12.65);
\definecolor{drawColor}{RGB}{0,0,0}

\node[text=drawColor,anchor=base,inner sep=0pt, outer sep=0pt, scale=  1.80] at (151.77,388) {2};
\definecolor{drawColor}{RGB}{211,211,211}

\path[draw=drawColor,line width= 0.4pt,line join=round,line cap=round] (151.77,449.68) --
	(252.94,101.18);

\path[draw=drawColor,line width= 0.4pt,line join=round,line cap=round] (151.77,449.68) --
	(252.94,202.36);

\path[draw=drawColor,line width= 0.4pt,line join=round,line cap=round] (151.77,449.68) --
	(252.94,303.53);

\path[draw=drawColor,line width= 0.4pt,line join=round,line cap=round] (151.77,449.68) --
	(252.94,404.71);
\end{scope}
\begin{scope}
\path[clip] (  0.00,  0.00) rectangle (505.89,505.89);
\definecolor{drawColor}{RGB}{211,211,211}
\definecolor{fillColor}{RGB}{255,255,255}

\path[draw=drawColor,line width= 0.4pt,line join=round,line cap=round,fill=fillColor] (151.77,449.68) circle ( 12.65);
\definecolor{drawColor}{RGB}{0,0,0}

\node[text=drawColor,anchor=base,inner sep=0pt, outer sep=0pt, scale=  1.80] at (151.77,444) {1};

\path[draw=drawColor,line width= 0.4pt,line join=round,line cap=round] (252.94,101.18) --
	(354.12,252.94);

\path[draw=drawColor,line width= 0.4pt,line join=round,line cap=round,fill=fillColor] (252.94,101.18) circle ( 12.65);

\path[draw=drawColor,line width= 0.4pt,line join=round,line cap=round] (252.94,202.36) --
	(354.12,252.94);

\path[draw=drawColor,line width= 0.4pt,line join=round,line cap=round,fill=fillColor] (252.94,202.36) circle ( 12.65);

\path[draw=drawColor,line width= 0.4pt,line join=round,line cap=round] (252.94,303.53) --
	(354.12,252.94);

\path[draw=drawColor,line width= 0.4pt,line join=round,line cap=round,fill=fillColor] (252.94,303.53) circle ( 12.65);

\path[draw=drawColor,line width= 0.4pt,line join=round,line cap=round] (252.94,404.71) --
	(354.12,252.94);

\path[draw=drawColor,line width= 0.4pt,line join=round,line cap=round,fill=fillColor] (252.94,404.71) circle ( 12.65);
\end{scope}
\begin{scope}
\path[clip] (  0.00,  0.00) rectangle (505.89,505.89);
\definecolor{drawColor}{RGB}{0,0,0}
\definecolor{fillColor}{RGB}{255,255,255}

\path[draw=drawColor,line width= 0.4pt,line join=round,line cap=round,fill=fillColor] (354.12,252.94) circle ( 12.65);

\node[text=drawColor,anchor=base,inner sep=0pt, outer sep=0pt, scale=  2] at (252.94,480.60) {$q  = 4$};
\end{scope}
\end{tikzpicture}\input{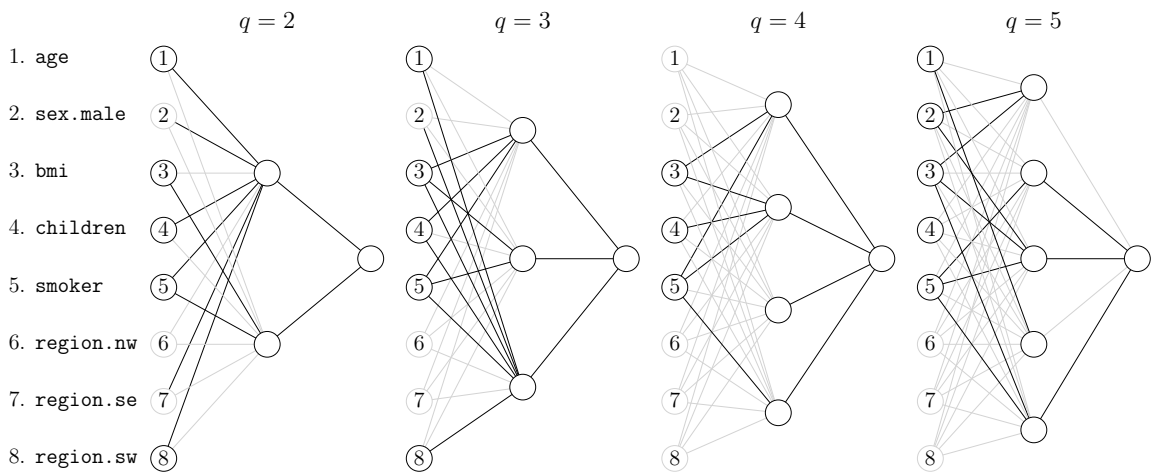}}}
		\caption{Results of the single- and multiple-parameter Wald tests overlaid on the neural network architecture for the \texttt{insurance} data for various hidden layer sizes, where $q$ denotes the number of hidden nodes.}
		\label{fig: plotnn_q}
	\end{figure}

To further investigate the effect of hidden layer size, the results from the Wald tests for values of $q$ between two and five are visualised in Figure \ref{fig: plotnn_q}.
In each model, \texttt{bmi}, \texttt{children} and \texttt{smoker} are found to be statistically significant for the multiple-parameter test at the 5\% significance level, while \texttt{age} is significant in all models except when $q = 4$.
Although we do not test the significance of hidden nodes directly, it is interesting to note that there is no model with more than three hidden nodes that are significant based on the associated single-parameter tests (either all weights entering the node are non-significant or the weight leaving the node is non-significant).
This provides further support for the findings of Figure \ref{fig: insurance_mse} that a simpler neural network is sufficient here (albeit, there, a model with $q = 2$ is suggested based on the RMSE).
It also further highlights the importance of model selection, as it can be misleading to carry out significance testing on input weights that connect to a hidden node that itself is non-significant.
Note that statistically-motivated neural network selection has been considered in \citet{mcinerney2022statistically}.

\begin{figure}[t]
    \centering
    \includegraphics[width = 0.49\textwidth]{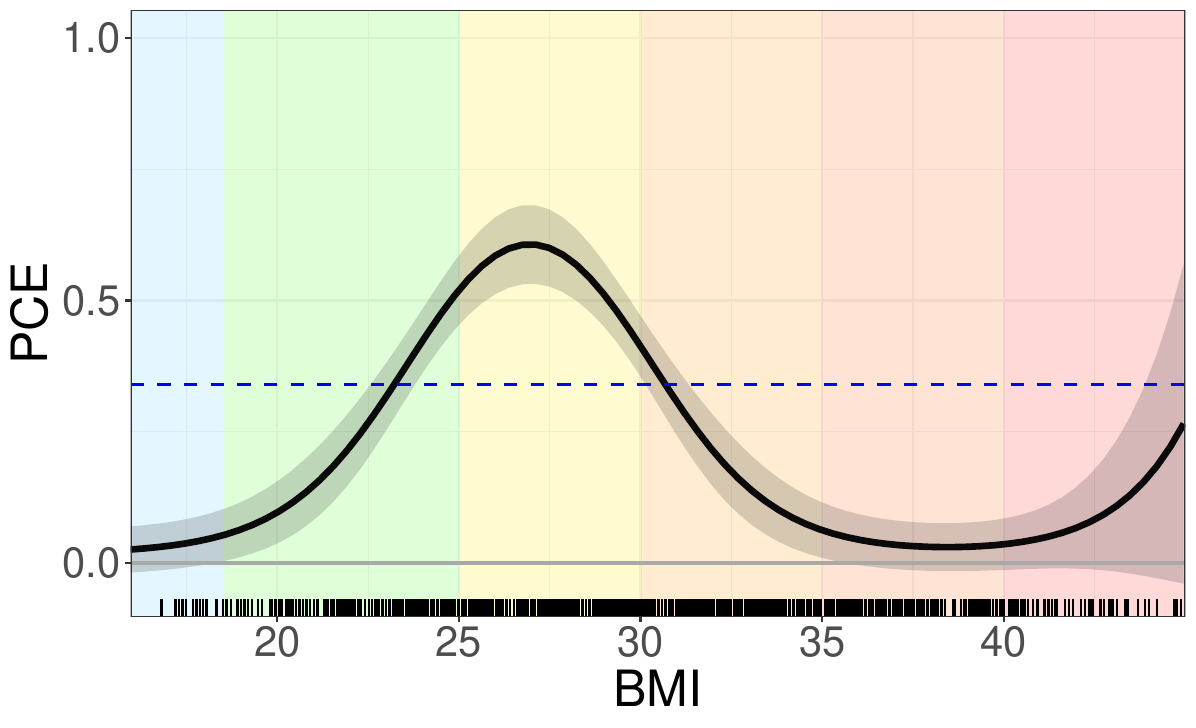}
    \includegraphics[width = 0.49\textwidth]{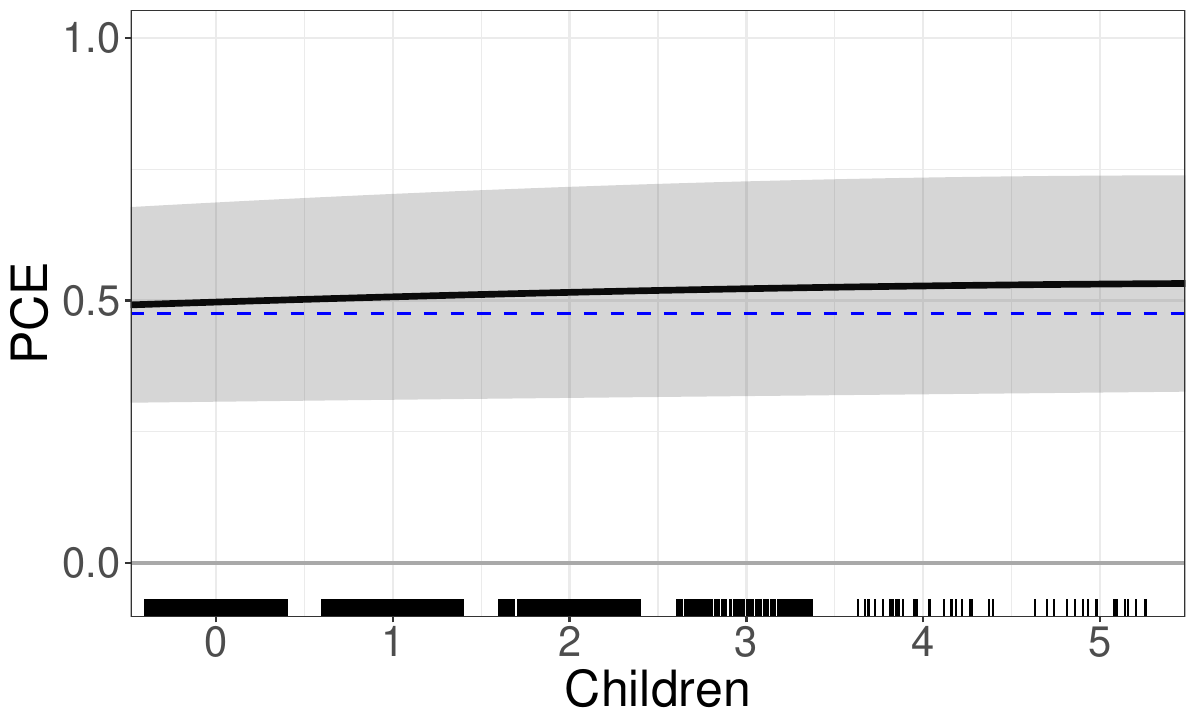}
    \includegraphics[width = 0.49\textwidth]{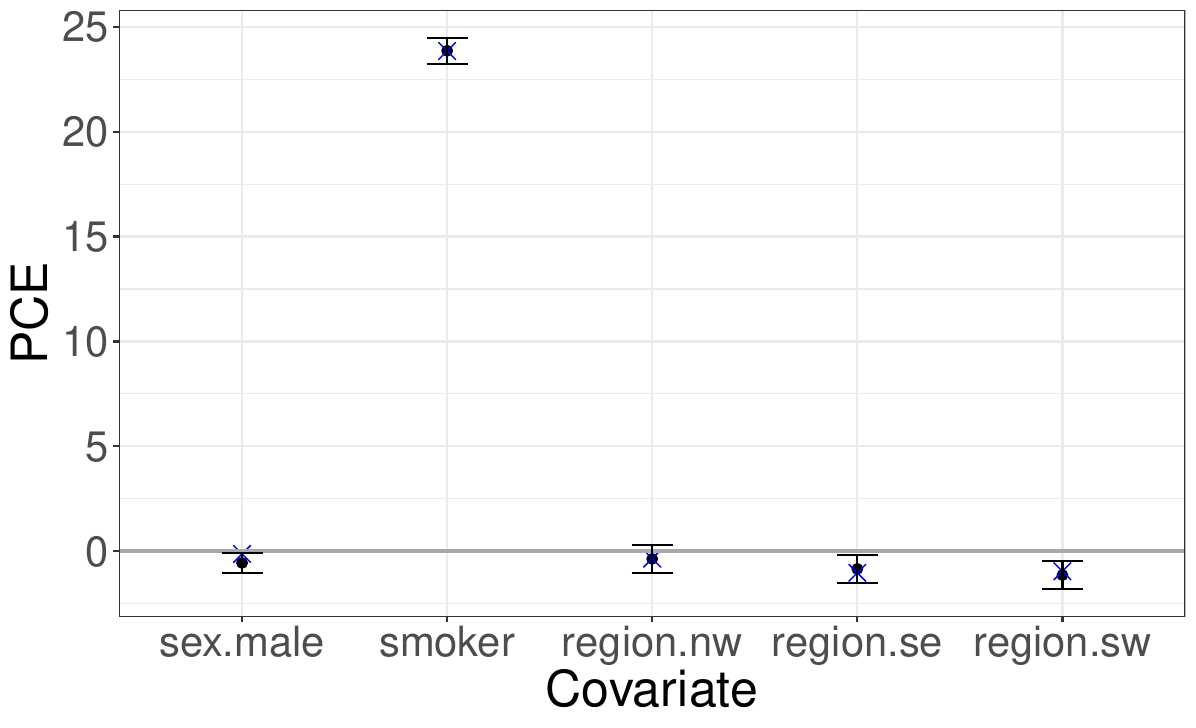}
    \caption{PCE plots for \texttt{bmi} (left), \texttt{children} (right), and the binary covariates in the \texttt{insurance} data (bottom). The blue dashed lines/crosses denote the corresponding effects from the linear model in Table \ref{tab: insurance_lm}. The \texttt{bmi} plot is coloured based on the typical categories: underweight, normal, overweight, obesity class I, obesity class II and obesity class III. For the \texttt{bmi} and \texttt{children} plots, $d$ is set to the respective standard deviation.}
    \label{fig: binary_pce}
\end{figure}

\begin{figure}
    \centering
    \includegraphics[width = 0.49\textwidth]{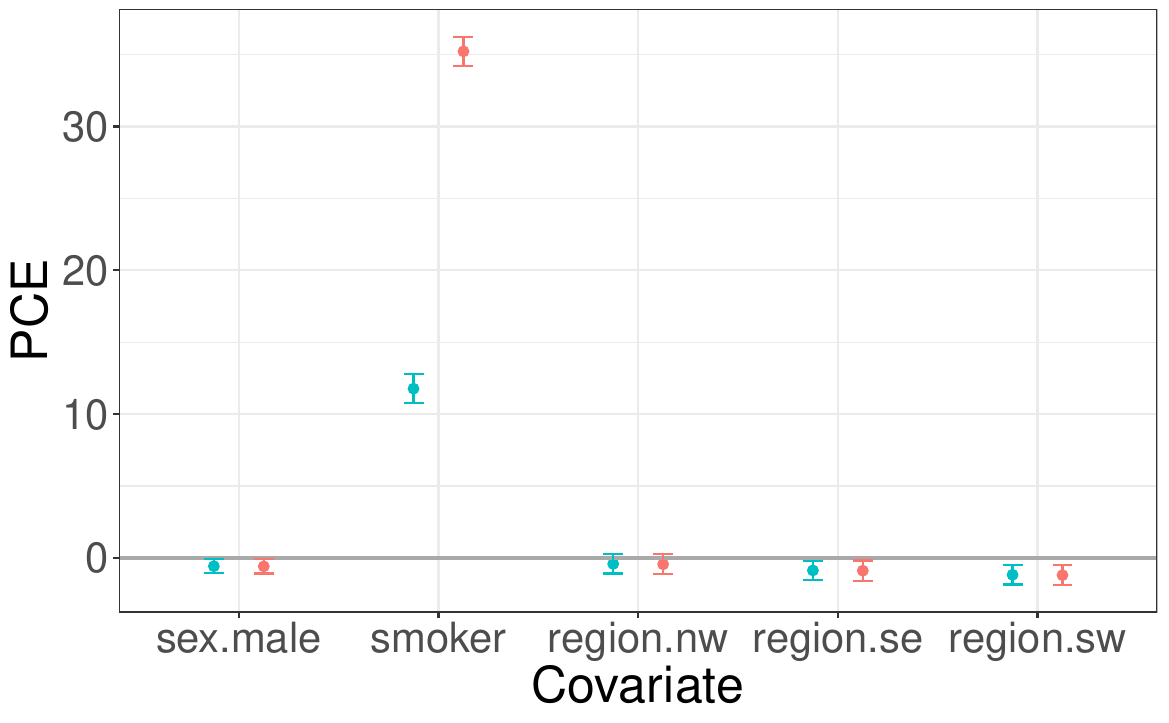}
    \includegraphics[width = 0.49\textwidth]{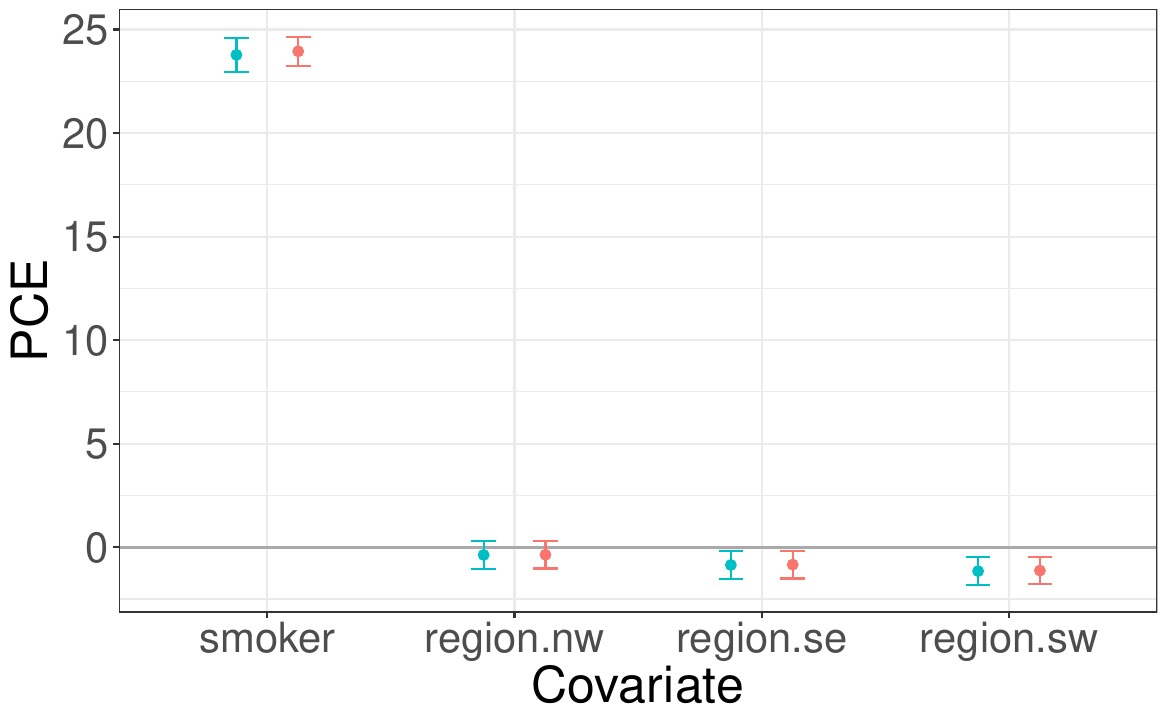}
    \includegraphics[width = 0.49\textwidth]{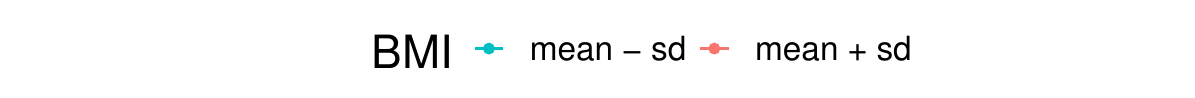}
    \includegraphics[width = 0.49\textwidth]{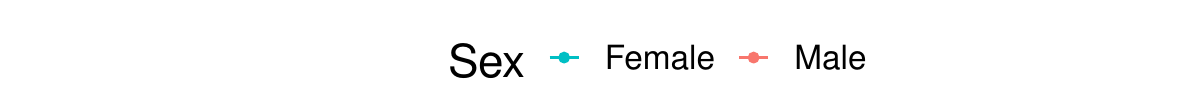}
    \caption{PCE plots investigating possible interaction effects between the binary covariates in the \texttt{insurance} data and \texttt{bmi} (left; set to its mean $\pm$ one standard deviation) and \texttt{sex} (right; set to its two levels).}
    \label{fig: binary_pce_interaction}
\end{figure}

Looking at covariate effects, Figure \ref{fig: binary_pce} contains the PCE plots with their 95\% confidence intervals for the covariates in the $q = 2$ model (the PCE plot for \texttt{age} was displayed earlier in Figure \ref{fig: PCE}).
For \texttt{bmi}, the PCE suggests that increased \texttt{bmi} in the range 20 to 35 leads to increased \texttt{charges}.
The plot has been overlaid with the colours representing the ranges of the different BMI categories \citep{cdc_bmi}.
Interestingly, the effect is only present (in varying strength) for individuals who are currently classified as having a BMI that is normal, overweight, or obese class I.
An increase in \texttt{bmi} has no effect on \texttt{charges} for individuals classed as underweight, or in obesity class II or III.
In other words, underweight individuals are similar to those in the lower range of normal, and obesity classes II and III are similar to those in the upper range of obesity class I.
The coefficient for \texttt{bmi} from the linear model is overlaid in the PCE plot.
It is clear that the effect found by the neural network is highly non-linear, and the linear model yields an effect that somewhat averages the non-linear effect over the different BMI values.
The covariate effect for \texttt{children} is positive, significantly different from zero and appears to be constant, suggesting a linear relationship with \texttt{charges}.
For the binary covariates, the strong effect of smoking status is clear, with a PCE of 23.85 associated with smokers compared to non-smokers (i.e., an increase of \$23,850 in their medical charges).
The other effects are much weaker, and closely align with the results from the multiple-parameter test in Table \ref{tab: insurance_mpwald}, with \texttt{region.sw} having a non-zero negative effect on \texttt{charges}.
Additionally, the PCEs for \texttt{children} and for the binary covariates are very similar to the effects found in the linear model, indicated as blue dashed lines and crosses in Figure \ref{fig: binary_pce}.
Due to the nature of binary covariates (i.e., two-point covariates), the alignment of their PCE values and their corresponding effects from the linear model is to be expected.
However, the neural network additionally captures interactions between covariates, which we can also visualise.
Figure \ref{fig: binary_pce_interaction} explores possible interactions between a continuous covariates (\texttt{bmi}) and the binary covariates, and possible interactions between a binary covariate (\texttt{sex}) and the remaining binary covariates.
For \texttt{bmi}, there is a clear interaction with \texttt{smoker}; the effect of smoking on \texttt{charges} is much stronger for individuals with a higher BMI value.
In contrast, there does not appear to be any interaction between \texttt{sex} and the other binary covariates.

\section{Discussion}\label{sec: disc}
This paper investigates when shallow regularised FNNs can be used for covariate-based Wald testing, and, hence, how they can be made more interpretable and useful for for statistical modellers.
Through an application to insurance data, we have demonstrated how a neural network can be used not only as a predictive model, but also as an inferential one.
This leads to statistically-based outputs that are more familiar in the statistical modelling context, namely, significance tests and estimated covariate effects.
This is critical to ensure the model is intelligible, so that it can be used to provide relevant explanations.

Through extensive simulation studies, we have explored the inferential performance of both single- and multiple-parameter Wald tests in the neural network context, and highlight the circumstances where they perform as expected.
This leads to practical recommendations for the neural network architecture when taking a statistical-modelling perspective.
We find that the ability to estimate uncertainty comes at a direct trade-off with model flexibility.
We recommend that (i) the hidden layer should not be larger than the input layer and (ii) a small ridge penalty should be used to make the estimation more stable.
We have used a ridge penalty for the purpose of estimation stability only, but note that it could be varied and selected through cross-validation or an information criterion.

To accompany the statistical tests, we have proposed covariate-effect plots that mimic regression coefficients.
These plots visualise the effect of a $d$-unit increase in the covariate on the response, and they also allow for the detection of possible interaction effects, which, in the future, could also be complimented with more formal statistical testing.
Our focus is on simpler single-hidden-layer neural networks, which are underutilised in the statistical community despite the fact that they extend more familiar (linear-type) models.
We believe that our work (available within the \texttt{interpretnn} R package \citep{mcinerney2023interpretnn}) makes these models more accessible within the broader statistical-modelling context, and can increase their utility to complement the current statistical toolbox.

\section*{Acknowledgement}
This publication has emanated from research conducted with the financial support of Research Ireland under Grant number 18/CRT/6049.
The second author was supported by the Confirm Smart Manufacturing Centre (https://confirm.ie/) funded by Research Ireland (Grant Number: 16/RC/3918).
For the purpose of Open Access, the authors have applied a CC BY public copyright licence to any Author Accepted Manuscript version arising from this submission.

\bibliographystyle{apalike}
\bibliography{refs}

\appendix

\section{Practical Issues of FNNs}\label{app: practicalities}
There are some practical considerations that arise when fitting neural networks.
Two models may appear different upon initial inspection, but may, in fact, be the same model. 
In particular, we consider model \emph{reducibility} and weight \emph{uniqueness} below, which may mask two equivalent models.

\subsection{Reducibilty}
An FNN is said to be \emph{reducible} if its input-to-output mapping is equivalent to another FNN that has fewer hidden nodes \citep{sussmann1992uniqueness}.
If an FNN is reducible, then at least one of the hidden nodes is \emph{redundant} (i.e., it provides no contribution in the prediction of the response).
There are three conditions that, if any are true, result in a reducible FNN:
\begin{itemize}
  \item One of the $\gamma_k$ output weights, for $k=1,...,q$, equals zero.
  ($\exists ~ k \in \{1, 2, \dotsc, q\} \text{~s.t.~} \gamma_k = 0$.)
  \item The net input, $s_k(x_{i}) = \sum_{j=0}^p \omega_{jk}x_{ij}$, for two hidden nodes, $k_1, k_2 \in \{1,2,...,q\}$, are sign-equivalent.
  ($\exists ~ k_1, k_2 \in \{1, 2, \dotsc, q\} \text{~s.t.~} \forall i \in 1, \dotsc n, ~ |s_{k_1}(x_{i})| = |s_{k_2}(x_{i})|$.)
  \item A hidden node has a net input that is constant.
  ($\exists ~ k\in \{1, 2, \dotsc, q\} \text{~s.t.~} \forall i \in 1, \dotsc n, ~  s_{k}(x_{i}) = c$.)
\end{itemize}
An FNN that does not contain any redundant hidden nodes is called \textit{irreducible}.

\subsection{Uniqueness}
If an FNN is irreducible, \citet{sussmann1992uniqueness} showed that the neural network weights are unique, up to a finite group of symmetries.
A weight vector, $\Tilde{\theta}$, is a symmetry of $\theta$ if one of the following is true (assuming a sigmoid activation on the hidden nodes):
\begin{itemize}
  \item $\Tilde{\omega}_k=-\omega_k$, $\Tilde{\gamma}_k=-\gamma_k$, and $\Tilde{\omega}_0=\gamma_0+\gamma_k$, for any $k=1,2,...,q$.
  \item If the order of the hidden nodes is permuted.
\end{itemize}
Therefore, $2^q(q!)$ symmetries exist, all of which are equivalent and can be considered the global solution.

\clearpage

\section{Simulation: Estimation of $\Sigma$}\label{app: sim_valid}
This simulation study aims to determine what conditions are required to ensure a high probability that the estimate of variance-covariance matrix, $\hat{\Sigma}$, is positive definite.
The variance-covariance matrix is positive definite if its eigenvalues are all positive real numbers.
This is a necessary requirement for any method of uncertainty quantification that uses $\hat{\Sigma}$, such as the Wald hypothesis tests (Section \ref{sec: hypo_test}), and the delta method.
Factors that can affect the positive definiteness of $\hat{\Sigma}$ include the architecture (the size of $q$ relative to $p$), the parameter values, the number of parameters relative to the sample size, and the presence or absence of a ridge penalty (commonly used in neural network optimisation).

The simulation setup is similar to the setup of Section \ref{sec: sim}.
However, we have run simulations where there is no ridge penalty ($\lambda = 0$) and where there is a small ridge penalty ($\lambda = 0.01$) to determine its effect; additional sample sizes are also investigated (100, 250, 500, 1000, 2000 and 5000).
For each simulation scenario, the percentage of variance-covariance matrices that are positive definite is reported (PD).
The results are summarised in Table \ref{tab: sim_valid_vc}.

For the scenarios where  $\lambda = 0$, when the number of non-zero covariates (N) is greater than or equal to the number of hidden nodes ($N \geq q$), and the sample size is relatively large ($n \geq 500$), the percentage of variance-covariance matrices that are positive definite is very high ($\geq 99\%$).
However, when there are more hidden nodes than non-zero covariates, the estimation of the variance-covariance matrix is quite unstable, i.e., it cannot be reliably estimated or requires large sample sizes.
The addition of a ridge penalty clearly allows for a more stable estimation of $\Sigma$ across the full range of settings considered here.
Therefore, we recommend the use of the ridge penalty in practice to stabilise the estimation of the variance-covariance matrix so that statistical inference can at least proceed.
Nevertheless, to attain acceptable inferential performance, it is still important not to have an overly complex hidden layer relative to the number of inputs (as shown in Section \ref{sec: sim}).
It is also worth highlighting that the user themselves can assess the positive-definiteness of the information matrix in practice, and seek to simplify the model when this matrix is not positive definite.
This justifies the need for neural network model selection methods that focus on producing parsimonious models \citep{mcinerney2022statistically}.

\begin{table}
\centering
\caption{Simulation: percentage of variance-covariance matrices that are positive definite.}
\label{tab: sim_valid_vc}
\begin{tabular}{@{}l@{~~}l@{~~} c@{~~} c@{~~}c@{~~}  c@{~~}c@{~~}c@{~~}c@{~~}c@{~~}c@{~~} }
\toprule
{$\lambda$} & {$q$} & {} & {N-Z} &  & $n = 100$ & $n = 250$ & $n = 500$ & $n = 1000$ & $n = 2000$ & $n = 5000$ \\
  \midrule

0 & 2 && 5-1 &  & 99.1 & 100.0 &  100.0 &  100.0  &  100.0 &  100.0    \\ 
  &&& 3-3 &  & 99.4 & 100.0 &  100.0 &  100.0  &  100.0 &  100.0 \\ [0.2cm]

& 4 && 5-1 &  & 60.1 & 81.5 & 99.6 &  100.0  &  100.0 &  100.0  \\ 
  &&& 3-3 &  & 76.6 & 85.3 & 98.3 &  99.6  &  99.9 &  100.0   \\ [0.2cm]

& 6 && 5-1 &  & 53.2 & 42.4 & 82.2 &  85.3 &  99.4 &  99.3   \\ 
  &&& 3-3 &  & 45.8 & 30.0 & 37.8 &  44.5 &  44.8 &  34.1   \\  [0.2cm]

  \midrule 

  0.01 & 2 && 5-1 && 99.6 & 100.0 & 100.0 & 100.0 & 100.0 & 100.0    \\ 
  &&& 3-3 && 99.9 & 100.0 & 100.0 & 100.0 & 100.0 & 100.0   \\  [0.2cm]

& 4 && 5-1 &&  98.3 & 99.8 & 100.0 & 100.0 & 100.0 & 100.0  \\ 
  &&& 3-3 && 97.6 & 99.1 & 99.8 & 100.0 & 100.0 & 100.0 \\ [0.2cm]

& 6 && 5-1 && 96.4 & 98.6 & 99.9 & 100.0 & 100.0 & 100.0  \\ 
  &&& 3-3 && 96.3 & 96.2 & 97.7 & 98.0 & 98.7 & 98.7   \\ 
  \bottomrule
  \\
  \end{tabular}
\end{table}

\begin{figure}
    \centering
    \includegraphics[width=\textwidth]{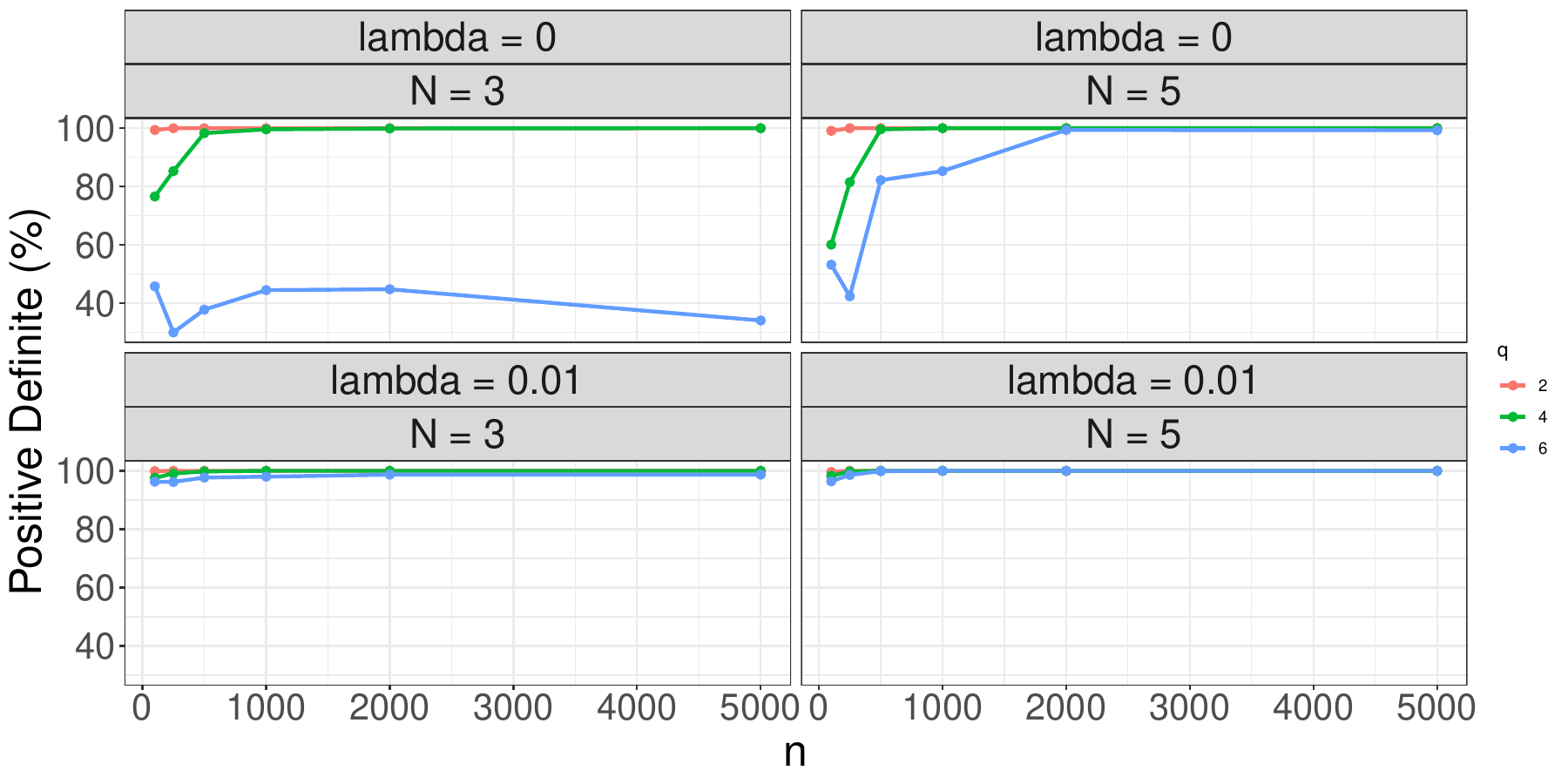}
    \caption{Plot of the percentage of variance-covariance matrices that are positive definite for different values of $N$, $q$ and $\lambda$.}
    \label{fig: positivedefinite}
\end{figure}

\clearpage
\section{Estimation and inference metrics}\label{app: full_est_results}
This section contains the full simulation results from Section \ref{sec: sim} of the main paper.
Tables \ref{tab: sim_mp_wald_full_5-1} and \ref{tab: sim_mp_wald_full_3-3} contain the rejection rate results for the multiple-parameter Wald test for the 5-1 and 3-3 non-zero-to-zero covariates (N-Z) scenarios, respectively.
Tables \ref{tab: est_and_inf_5-1} and \ref{tab: est_and_inf_3-3} contain the results for the additional estimation and inference metrics, including the rejection rate results for the single-parameter Wald test, for the 5-1 and 3-3 non-zero-to-zero covariates (N-Z) scenarios, respectively.

\begin{table}[h]
\footnotesize
\centering
\caption{Simulation: estimates of rejection rates for the multiple-parameter Wald test for the 5-1 scenario.}
\label{tab: sim_mp_wald_full_5-1}
\begin{tabular}{@{} c@{~~}c@{~~}  c@{~~}c@{~~}c@{~~} c@{~~} c@{~~}c@{~~}c@{~~} c@{~~} c@{~~}c@{~~}c@{~~}}
\toprule
{N-Z}  & {5-1} & \multicolumn{3}{c}{$q = 2$} & & \multicolumn{3}{c}{$q = 4$} & & \multicolumn{3}{c}{$q = 6$} \\
\cmidrule(r){3-5} \cmidrule(r){7-9} \cmidrule(){11-13}
 {} & & $n = 500$ & $n = 1000$ & $n = 2000$ & & $n = 500$ & $n = 1000$ & $n = 2000$ & & $n = 500$ & $n = 1000$ & $n = 2000$  \\
  \midrule

   \boldsymbol{$\omega_1$} &   &  0.050  &  0.054 &  0.053 && 0.038 &    0.048 & 0.057 && 0.053 & 0.064 & 0.035\\ 
   $\omega_2$ &   &  0.999  &  1.000 &  1.000 && 0.767   & 1.000 & 1.000 && 0.990 & 1.000 & 1.000\\ 
   $\omega_3$ &   &  1.000  &  1.000 &  1.000 && 1.000    & 1.000 & 1.000 && 1.000 & 1.000 & 1.000\\ 
   $\omega_4$ &   &  1.000  &  1.000 &  1.000 && 1.000   & 1.000 & 1.000 && 1.000 & 1.000 & 1.000\\ 
   $\omega_5$ &   &  1.000  &  1.000 &  1.000 && 1.000   & 1.000 & 1.000 && 1.000 & 1.000 & 1.000\\ 
   $\omega_6$ &   &  1.000  &  1.000 &  1.000 && 1.000  & 1.000 & 1.000 && 1.000 & 1.000 & 1.000\\
  \bottomrule
 \multicolumn{13}{p{\textwidth}}{\scriptsize
  True zero weight vectors are highlighted in \textbf{bold}.}\\
  \end{tabular}
\end{table}

\begin{table}[h]
\footnotesize
\centering
\caption{Simulation: estimates of rejection rates for the multiple-parameter Wald test for the 3-3 scenario.}
\label{tab: sim_mp_wald_full_3-3}
\begin{tabular}{@{} c@{~~}c@{~~}  c@{~~}c@{~~}c@{~~} c@{~~} c@{~~}c@{~~}c@{~~} c@{~~} c@{~~}c@{~~}c@{~~}}
\toprule
{N-Z}  & {3-3} & \multicolumn{3}{c}{$q = 2$} & & \multicolumn{3}{c}{$q = 4$} & & \multicolumn{3}{c}{$q = 6$} \\
\cmidrule(r){3-5} \cmidrule(r){7-9} \cmidrule(){11-13}
 {} & & $n = 500$ & $n = 1000$ & $n = 2000$ & & $n = 500$ & $n = 1000$ & $n = 2000$ & & $n = 500$ & $n = 1000$ & $n = 2000$  \\
  \midrule

   \boldsymbol{$\omega_1$} &   &  0.055  &  0.039 &  0.042 && 0.035 & 0.035 & 0.037 && 0.489 & 0.320 & 0.299    \\ 
   $\omega_2$ &   &  1.000  &  1.000 &  1.000 && 0.851 & 0.996 & 1.000 && 0.979 & 0.998 & 1.000    \\ 
    \boldsymbol{$\omega_3$} &   &  0.049  &  0.052 &  0.055 && 0.044 & 0.044 & 0.037  && 0.453 & 0.328 & 0.365  \\ 
    \boldsymbol{$\omega_4$} &   &  0.051  &  0.040 &  0.048 && 0.039 & 0.039 & 0.039 && 0.454 & 0.350 & 0.340    \\ 
   $\omega_5$ &   &  1.000  &  1.000 &  1.000 && 1.000 & 1.000 & 1.000 && 1.000 & 1.000 & 1.000 \\ 
   $\omega_6$ &   &  1.000  &  1.000 &  1.000 && 1.000 & 1.000 & 1.000 && 1.000 & 1.000 & 1.000 \\ 
  \bottomrule
  \multicolumn{13}{p{\textwidth}}{\scriptsize
  True zero weight vectors are highlighted in \textbf{bold}.}\\
  \end{tabular}
\end{table}

\clearpage

\footnotesize
\setlength\LTleft{0pt}  
\setlength\LTright{0pt} 
\begin{longtable}{@{}l@{~~} c@{~~}c@{~~}  c@{~~}c@{~~}c@{~~}c@{~~}c@{~~}  c@{~~}  c@{~~}c@{~~}c@{~~}c@{~~}c@{~~}  c@{~~}  c@{~~}c@{~~}c@{~~}c@{~~}c@{~~}}
\caption{Simulation: estimation and inference metrics for the 5-1 non-zero-to-zero (N-Z) scenario.}\\ 
\toprule
{N-Z}  & {5-1} & {} & \multicolumn{5}{c}{$n = 500$} && \multicolumn{5}{c}{$n = 1000$} && \multicolumn{5}{c}{$n = 2000$} \\
\cmidrule(){4-8} \cmidrule(){10-14} \cmidrule(){16-20}
{$q$} &  {} & $\theta$ & $\hat{\theta}$ & SE & SEE & CP & RR && $\hat{\theta}$ & SE & SEE & CP & RR && $\hat{\theta}$ & SE & SEE & CP & RR \\
  \midrule
2   & $\omega_{01}$ & -2.77 & -2.77 & 0.43 & 0.42 & 0.94 & 1.00 &  & -2.75 & 0.29 & 0.28 & 0.94 & 1.00 &  & -2.77 & 0.20 & 0.20 & 0.96 & 1.00\\
 & \boldsymbol{$\omega_{11}$} & ~0.00 & ~0.00 & 0.13 & 0.12 & 0.94 & 0.06 &  & ~0.00 & 0.09 & 0.09 & 0.93 & 0.07 &  & ~0.00 & 0.06 & 0.06 & 0.93 & 0.07\\
 & $\omega_{21}$ & -0.14 & -0.14 & 0.14 & 0.13 & 0.95 & 0.60 &  & -0.14 & 0.09 & 0.09 & 0.95 & 0.69 &  & -0.15 & 0.06 & 0.06 & 0.94 & 0.82\\
 & $\omega_{31}$ & -0.56 & -0.56 & 0.17 & 0.17 & 0.95 & 0.98 &  & -0.56 & 0.11 & 0.11 & 0.95 & 1.00 &  & -0.57 & 0.07 & 0.07 & 0.96 & 1.00\\
 & $\omega_{41}$ & ~2.24 & ~2.24 & 0.38 & 0.36 & 0.93 & 1.00 &  & ~2.23 & 0.25 & 0.24 & 0.94 & 1.00 &  & ~2.24 & 0.17 & 0.17 & 0.96 & 1.00\\
 & $\omega_{51}$ & -2.52 & -2.51 & 0.41 & 0.40 & 0.94 & 1.00 &  & -2.51 & 0.28 & 0.28 & 0.94 & 1.00 &  & -2.52 & 0.19 & 0.19 & 0.96 & 1.00\\
 & $\omega_{61}$ & -2.52 & -2.52 & 0.41 & 0.40 & 0.94 & 1.00 &  & -2.51 & 0.27 & 0.27 & 0.93 & 1.00 &  & -2.52 & 0.19 & 0.19 & 0.95 & 1.00\\
 & $\omega_{02}$ & ~1.70 & ~1.70 & 0.14 & 0.13 & 0.94 & 1.00 &  & ~1.70 & 0.09 & 0.09 & 0.94 & 1.00 &  & ~1.70 & 0.06 & 0.06 & 0.94 & 1.00\\
 & \boldsymbol{$\omega_{12}$} & ~0.00 & ~0.00 & 0.05 & 0.05 & 0.95 & 0.05 &  & ~0.00 & 0.04 & 0.04 & 0.94 & 0.06 &  & ~0.00 & 0.02 & 0.02 & 0.95 & 0.05\\
 & $\omega_{22}$ & -0.27 & -0.27 & 0.06 & 0.06 & 0.95 & 0.58 &  & -0.27 & 0.04 & 0.04 & 0.94 & 0.70 &  & -0.27 & 0.03 & 0.03 & 0.95 & 0.83\\
 & $\omega_{32}$ & ~0.56 & ~0.56 & 0.07 & 0.06 & 0.95 & 0.99 &  & ~0.56 & 0.05 & 0.05 & 0.96 & 1.00 &  & ~0.56 & 0.03 & 0.03 & 0.94 & 1.00\\
 & $\omega_{42}$ & ~1.71 & ~1.71 & 0.13 & 0.13 & 0.95 & 1.00 &  & ~1.71 & 0.09 & 0.09 & 0.94 & 1.00 &  & ~1.71 & 0.06 & 0.07 & 0.95 & 1.00\\
 & $\omega_{52}$ & ~1.05 & ~1.06 & 0.10 & 0.10 & 0.94 & 1.00 &  & ~1.06 & 0.07 & 0.07 & 0.95 & 1.00 &  & ~1.06 & 0.05 & 0.05 & 0.95 & 1.00\\
 & $\omega_{62}$ & -2.06 & -2.07 & 0.16 & 0.15 & 0.93 & 1.00 &  & -2.07 & 0.12 & 0.11 & 0.94 & 1.00 &  & -2.06 & 0.07 & 0.08 & 0.96 & 1.00\\
 & $\gamma_{0}$ & -1.54 & -1.55 & 0.06 & 0.07 & 0.94 & 0.34 &  & -1.55 & 0.05 & 0.05 & 0.94 & 0.52 &  & -1.55 & 0.04 & 0.04 & 0.95 & 0.85\\
 & $\gamma_{1}$ & ~1.37 & ~1.39 & 0.09 & 0.09 & 0.94 & 1.00 &  & ~1.38 & 0.07 & 0.07 & 0.95 & 1.00 &  & ~1.37 & 0.04 & 0.04 & 0.95 & 1.00\\
 & $\gamma_{2}$ & ~2.73 & ~2.73 & 0.10 & 0.10 & 0.95 & 1.00 &  & ~2.73 & 0.07 & 0.07 & 0.95 & 1.00 &  & ~2.73 & 0.05 & 0.05 & 0.96 & 1.00\\[0.2cm]

4  & $\omega_{01}$ & -2.77 & -2.79 & 0.50 & 0.49 & 0.94 & 1.00 &  & -2.79 & 0.32 & 0.33 & 0.95 & 1.00 &  & -2.78 & 0.25 & 0.25 & 0.94 & 1.00\\
 & \boldsymbol{$\omega_{11}$} & ~0.00 & ~0.01 & 0.17 & 0.16 & 0.95 & 0.05 &  & ~0.00 & 0.11 & 0.11 & 0.95 & 0.05 &  & ~0.00 & 0.08 & 0.08 & 0.94 & 0.06\\
 & $\omega_{21}$ & -0.14 & -0.14 & 0.17 & 0.17 & 0.95 & 0.54 &  & -0.15 & 0.10 & 0.10 & 0.95 & 0.77 &  & -0.15 & 0.09 & 0.08 & 0.95 & 0.84\\
 & $\omega_{31}$ & -0.56 & -0.57 & 0.20 & 0.19 & 0.94 & 0.93 &  & -0.57 & 0.13 & 0.12 & 0.94 & 0.99 &  & -0.56 & 0.10 & 0.10 & 0.96 & 1.00\\
 & $\omega_{41}$ & ~2.24 & ~2.26 & 0.43 & 0.41 & 0.94 & 1.00 &  & ~2.26 & 0.28 & 0.28 & 0.96 & 1.00 &  & ~2.26 & 0.21 & 0.21 & 0.94 & 1.00\\
 & $\omega_{51}$ & -2.52 & -2.53 & 0.49 & 0.48 & 0.94 & 1.00 &  & -2.54 & 0.33 & 0.33 & 0.95 & 1.00 &  & -2.53 & 0.26 & 0.25 & 0.94 & 1.00\\
 & $\omega_{61}$ & -2.52 & -2.55 & 0.46 & 0.47 & 0.95 & 1.00 &  & -2.54 & 0.32 & 0.32 & 0.95 & 1.00 &  & -2.53 & 0.24 & 0.23 & 0.94 & 1.00\\
 & $\omega_{02}$ & ~1.70 & ~1.69 & 0.32 & 0.30 & 0.92 & 1.00 &  & ~1.71 & 0.23 & 0.21 & 0.94 & 1.00 &  & ~1.71 & 0.15 & 0.15 & 0.94 & 1.00\\
 & \boldsymbol{$\omega_{12}$} & ~0.00 & ~0.00 & 0.09 & 0.08 & 0.96 & 0.04 &  & ~0.00 & 0.06 & 0.06 & 0.94 & 0.06 &  & ~0.00 & 0.05 & 0.04 & 0.95 & 0.05\\
 & $\omega_{22}$ & -0.27 & -0.28 & 0.11 & 0.10 & 0.94 & 0.56 &  & -0.27 & 0.06 & 0.06 & 0.95 & 0.78 &  & -0.27 & 0.05 & 0.04 & 0.94 & 0.82\\
 & $\omega_{32}$ & ~0.56 & ~0.56 & 0.14 & 0.14 & 0.95 & 0.94 &  & ~0.57 & 0.11 & 0.10 & 0.94 & 1.00 &  & ~0.56 & 0.07 & 0.07 & 0.95 & 1.00\\
 & $\omega_{42}$ & ~1.71 & ~1.70 & 0.27 & 0.26 & 0.93 & 1.00 &  & ~1.72 & 0.19 & 0.18 & 0.94 & 1.00 &  & ~1.71 & 0.13 & 0.12 & 0.95 & 1.00\\
 & $\omega_{52}$ & ~1.05 & ~1.06 & 0.17 & 0.16 & 0.95 & 1.00 &  & ~1.06 & 0.12 & 0.12 & 0.95 & 1.00 &  & ~1.06 & 0.08 & 0.08 & 0.96 & 1.00\\
 & $\omega_{62}$ & -2.06 & -2.07 & 0.31 & 0.30 & 0.94 & 1.00 &  & -2.08 & 0.22 & 0.21 & 0.94 & 1.00 &  & -2.08 & 0.14 & 0.14 & 0.95 & 1.00\\
 & $\omega_{03}$ & ~2.08 & ~2.12 & 0.46 & 0.45 & 0.94 & 1.00 &  & ~2.10 & 0.32 & 0.30 & 0.94 & 1.00 &  & ~2.08 & 0.22 & 0.22 & 0.94 & 1.00\\
 & \boldsymbol{$\omega_{13}$} & ~0.00 & ~0.01 & 0.12 & 0.11 & 0.95 & 0.05 &  & ~0.00 & 0.08 & 0.07 & 0.94 & 0.06 &  & ~0.00 & 0.05 & 0.05 & 0.94 & 0.06\\
 & $\omega_{23}$ & -0.20 & -0.22 & 0.14 & 0.12 & 0.95 & 0.51 &  & -0.21 & 0.08 & 0.08 & 0.95 & 0.77 &  & -0.20 & 0.06 & 0.06 & 0.93 & 0.85\\
 & $\omega_{33}$ & -0.60 & -0.61 & 0.16 & 0.15 & 0.96 & 0.94 &  & -0.61 & 0.11 & 0.11 & 0.95 & 1.00 &  & -0.60 & 0.07 & 0.07 & 0.94 & 1.00\\
 & $\omega_{43}$ & -2.89 & -2.96 & 0.51 & 0.50 & 0.93 & 1.00 &  & -2.92 & 0.36 & 0.34 & 0.94 & 1.00 &  & -2.89 & 0.25 & 0.24 & 0.93 & 1.00\\
 & $\omega_{53}$ & -1.43 & -1.49 & 0.29 & 0.29 & 0.94 & 1.00 &  & -1.44 & 0.20 & 0.19 & 0.94 & 1.00 &  & -1.43 & 0.14 & 0.14 & 0.94 & 1.00\\
 & $\omega_{63}$ & ~1.48 & ~1.48 & 0.29 & 0.28 & 0.92 & 1.00 &  & ~1.49 & 0.21 & 0.19 & 0.95 & 1.00 &  & ~1.48 & 0.13 & 0.13 & 0.94 & 1.00\\
 & $\omega_{04}$ & ~3.00 & ~3.01 & 0.40 & 0.41 & 0.94 & 1.00 &  & ~3.02 & 0.29 & 0.29 & 0.96 & 1.00 &  & ~3.01 & 0.20 & 0.20 & 0.96 & 1.00\\
 & \boldsymbol{$\omega_{14}$} & ~0.00 & ~0.00 & 0.10 & 0.10 & 0.95 & 0.05 &  & ~0.00 & 0.08 & 0.07 & 0.96 & 0.04 &  & ~0.00 & 0.06 & 0.05 & 0.94 & 0.06\\
 & $\omega_{24}$ & -0.29 & -0.30 & 0.12 & 0.11 & 0.96 & 0.53 &  & -0.29 & 0.07 & 0.07 & 0.94 & 0.77 &  & -0.29 & 0.05 & 0.05 & 0.95 & 0.86\\
 & $\omega_{34}$ & -0.42 & -0.42 & 0.15 & 0.16 & 0.94 & 0.93 &  & -0.42 & 0.11 & 0.11 & 0.95 & 0.99 &  & -0.42 & 0.07 & 0.07 & 0.94 & 1.00\\
 & $\omega_{44}$ & ~1.40 & ~1.40 & 0.20 & 0.20 & 0.94 & 1.00 &  & ~1.41 & 0.13 & 0.13 & 0.95 & 1.00 &  & ~1.40 & 0.10 & 0.09 & 0.95 & 1.00\\
 & $\omega_{54}$ & ~2.86 & ~2.88 & 0.40 & 0.43 & 0.94 & 1.00 &  & ~2.89 & 0.31 & 0.31 & 0.95 & 1.00 &  & ~2.87 & 0.20 & 0.20 & 0.96 & 1.00\\
 & $\omega_{64}$ & -2.08 & -2.09 & 0.27 & 0.27 & 0.93 & 1.00 &  & -2.09 & 0.18 & 0.19 & 0.94 & 1.00 &  & -2.08 & 0.13 & 0.13 & 0.95 & 1.00\\
 & $\gamma_{0}$ & -1.54 & -1.55 & 0.27 & 0.27 & 0.95 & 0.14 &  & -1.54 & 0.20 & 0.18 & 0.94 & 0.24 &  & -1.53 & 0.14 & 0.14 & 0.95 & 0.22\\
 & $\gamma_{1}$ & ~1.37 & ~1.41 & 0.13 & 0.13 & 0.96 & 1.00 &  & ~1.38 & 0.10 & 0.09 & 0.93 & 1.00 &  & ~1.38 & 0.07 & 0.07 & 0.94 & 1.00\\
 & $\gamma_{2}$ & ~2.73 & ~2.84 & 0.50 & 0.53 & 0.94 & 1.00 &  & ~2.77 & 0.40 & 0.38 & 0.95 & 1.00 &  & ~2.74 & 0.23 & 0.23 & 0.95 & 1.00\\
 & $\gamma_{3}$ & -1.81 & -1.81 & 0.22 & 0.22 & 0.94 & 1.00 &  & -1.82 & 0.17 & 0.16 & 0.94 & 1.00 &  & -1.82 & 0.12 & 0.12 & 0.95 & 1.00\\
 & $\gamma_{4}$ & -2.56 & -2.66 & 0.45 & 0.47 & 0.94 & 1.00 &  & -2.60 & 0.35 & 0.34 & 0.96 & 1.00 &  & -2.57 & 0.20 & 0.20 & 0.94 & 1.00\\[0.2cm]

6  & $\omega_{01}$ & -2.77 & -2.68 & 1.12 & 0.84 & 0.91 & 0.98 &  & -2.76 & 0.67 & 0.60 & 0.92 & 1.00 &  & -2.75 & 0.42 & 0.42 & 0.95 & 1.00\\
 & \boldsymbol{$\omega_{11}$} & ~0.00 & ~0.02 & 0.24 & 0.20 & 0.94 & 0.06 &  & ~0.00 & 0.14 & 0.13 & 0.95 & 0.05 &  & ~0.00 & 0.10 & 0.10 & 0.95 & 0.05\\
 & $\omega_{21}$ & -0.14 & -0.14 & 0.24 & 0.20 & 0.93 & 0.50 &  & -0.14 & 0.15 & 0.14 & 0.94 & 0.76 &  & -0.15 & 0.10 & 0.10 & 0.95 & 0.86\\
 & $\omega_{31}$ & -0.56 & -0.56 & 0.33 & 0.25 & 0.92 & 0.91 &  & -0.57 & 0.20 & 0.18 & 0.93 & 0.99 &  & -0.57 & 0.13 & 0.12 & 0.95 & 1.00\\
 & $\omega_{41}$ & ~2.24 & ~2.26 & 0.68 & 0.61 & 0.91 & 1.00 &  & ~2.27 & 0.47 & 0.44 & 0.93 & 1.00 &  & ~2.24 & 0.31 & 0.31 & 0.96 & 1.00\\
 & $\omega_{51}$ & -2.52 & -2.50 & 0.87 & 0.69 & 0.91 & 1.00 &  & -2.56 & 0.54 & 0.49 & 0.92 & 1.00 &  & -2.52 & 0.33 & 0.34 & 0.95 & 1.00\\
 & $\omega_{61}$ & -2.52 & -2.55 & 0.75 & 0.66 & 0.90 & 1.00 &  & -2.55 & 0.50 & 0.47 & 0.94 & 1.00 &  & -2.52 & 0.33 & 0.33 & 0.95 & 1.00\\
 & $\omega_{02}$ & ~1.70 & ~1.75 & 0.58 & 0.36 & 0.91 & 0.97 &  & ~1.72 & 0.28 & 0.24 & 0.92 & 1.00 &  & ~1.71 & 0.17 & 0.16 & 0.94 & 1.00\\
 & \boldsymbol{$\omega_{12}$} & ~0.00 & ~0.00 & 0.14 & 0.11 & 0.94 & 0.06 &  & ~0.00 & 0.07 & 0.07 & 0.93 & 0.07 &  & ~0.00 & 0.05 & 0.05 & 0.96 & 0.04\\
 & $\omega_{22}$ & -0.27 & -0.28 & 0.14 & 0.12 & 0.93 & 0.54 &  & -0.28 & 0.09 & 0.08 & 0.94 & 0.78 &  & -0.27 & 0.05 & 0.06 & 0.94 & 0.88\\
 & $\omega_{32}$ & ~0.56 & ~0.57 & 0.23 & 0.17 & 0.90 & 0.92 &  & ~0.57 & 0.13 & 0.13 & 0.94 & 0.99 &  & ~0.56 & 0.09 & 0.08 & 0.94 & 1.00\\
 & $\omega_{42}$ & ~1.71 & ~1.75 & 0.43 & 0.31 & 0.91 & 1.00 &  & ~1.73 & 0.27 & 0.22 & 0.93 & 1.00 &  & ~1.72 & 0.15 & 0.15 & 0.93 & 1.00\\
 & $\omega_{52}$ & ~1.05 & ~1.04 & 0.39 & 0.22 & 0.91 & 1.00 &  & ~1.07 & 0.16 & 0.15 & 0.93 & 1.00 &  & ~1.06 & 0.10 & 0.10 & 0.94 & 1.00\\
 & $\omega_{62}$ & -2.06 & -2.14 & 0.46 & 0.36 & 0.93 & 1.00 &  & -2.11 & 0.29 & 0.25 & 0.93 & 1.00 &  & -2.08 & 0.17 & 0.17 & 0.95 & 1.00\\
 & $\omega_{03}$ & ~2.08 & ~2.14 & 0.63 & 0.52 & 0.91 & 0.98 &  & ~2.09 & 0.41 & 0.34 & 0.94 & 1.00 &  & ~2.09 & 0.23 & 0.24 & 0.95 & 1.00\\
 & \boldsymbol{$\omega_{13}$} & ~0.00 & ~0.01 & 0.17 & 0.13 & 0.95 & 0.05 &  & -0.01 & 0.10 & 0.09 & 0.94 & 0.06 &  & ~0.00 & 0.06 & 0.06 & 0.95 & 0.05\\
 & $\omega_{23}$ & -0.20 & -0.22 & 0.19 & 0.15 & 0.92 & 0.54 &  & -0.21 & 0.11 & 0.10 & 0.93 & 0.75 &  & -0.20 & 0.07 & 0.07 & 0.95 & 0.87\\
 & $\omega_{33}$ & -0.60 & -0.62 & 0.23 & 0.18 & 0.90 & 0.90 &  & -0.59 & 0.14 & 0.13 & 0.95 & 0.99 &  & -0.60 & 0.09 & 0.09 & 0.94 & 1.00\\
 & $\omega_{43}$ & -2.89 & -2.97 & 0.84 & 0.60 & 0.91 & 1.00 &  & -2.91 & 0.52 & 0.39 & 0.94 & 1.00 &  & -2.91 & 0.26 & 0.27 & 0.94 & 1.00\\
 & $\omega_{53}$ & -1.43 & -1.48 & 0.48 & 0.33 & 0.92 & 1.00 &  & -1.43 & 0.32 & 0.22 & 0.93 & 1.00 &  & -1.44 & 0.15 & 0.15 & 0.94 & 1.00\\
 & $\omega_{63}$ & ~1.48 & ~1.49 & 0.62 & 0.34 & 0.90 & 1.00 &  & ~1.47 & 0.33 & 0.22 & 0.94 & 1.00 &  & ~1.49 & 0.15 & 0.15 & 0.94 & 1.00\\
 & $\omega_{04}$ & ~3.00 & ~3.08 & 0.54 & 0.47 & 0.88 & 0.98 &  & ~3.02 & 0.36 & 0.34 & 0.94 & 1.00 &  & ~3.00 & 0.24 & 0.23 & 0.95 & 1.00\\
 & \boldsymbol{$\omega_{14}$} & ~0.00 & ~0.00 & 0.13 & 0.12 & 0.95 & 0.05 &  & ~0.00 & 0.08 & 0.07 & 0.94 & 0.06 &  & ~0.00 & 0.06 & 0.06 & 0.96 & 0.04\\
 & $\omega_{24}$ & -0.29 & -0.30 & 0.13 & 0.12 & 0.92 & 0.53 &  & -0.30 & 0.09 & 0.08 & 0.94 & 0.77 &  & -0.29 & 0.06 & 0.06 & 0.95 & 0.88\\
 & $\omega_{34}$ & -0.42 & -0.43 & 0.18 & 0.15 & 0.91 & 0.92 &  & -0.43 & 0.13 & 0.12 & 0.94 & 0.99 &  & -0.42 & 0.08 & 0.09 & 0.94 & 1.00\\
 & $\omega_{44}$ & ~1.40 & ~1.43 & 0.27 & 0.21 & 0.90 & 1.00 &  & ~1.40 & 0.16 & 0.15 & 0.94 & 1.00 &  & ~1.40 & 0.11 & 0.10 & 0.94 & 1.00\\
 & $\omega_{54}$ & ~2.86 & ~2.93 & 0.57 & 0.48 & 0.90 & 1.00 &  & ~2.88 & 0.41 & 0.35 & 0.94 & 1.00 &  & ~2.87 & 0.26 & 0.25 & 0.94 & 1.00\\
 & $\omega_{64}$ & -2.08 & -2.13 & 0.39 & 0.33 & 0.90 & 1.00 &  & -2.10 & 0.22 & 0.21 & 0.93 & 1.00 &  & -2.08 & 0.16 & 0.15 & 0.94 & 1.00\\
 & $\omega_{05}$ & ~2.79 & ~2.69 & 0.46 & 0.40 & 0.92 & 0.99 &  & ~2.74 & 0.30 & 0.28 & 0.94 & 1.00 &  & ~2.77 & 0.20 & 0.20 & 0.94 & 1.00\\
 & \boldsymbol{$\omega_{15}$} & ~0.00 & ~0.00 & 0.09 & 0.08 & 0.94 & 0.06 &  & ~0.00 & 0.07 & 0.06 & 0.95 & 0.05 &  & ~0.00 & 0.04 & 0.04 & 0.96 & 0.04\\
 & $\omega_{25}$ & ~0.27 & ~0.26 & 0.11 & 0.10 & 0.91 & 0.55 &  & ~0.27 & 0.07 & 0.06 & 0.94 & 0.77 &  & ~0.27 & 0.05 & 0.04 & 0.95 & 0.85\\
 & $\omega_{35}$ & -0.56 & -0.54 & 0.11 & 0.10 & 0.93 & 0.92 &  & -0.55 & 0.08 & 0.08 & 0.94 & 0.99 &  & -0.55 & 0.05 & 0.05 & 0.94 & 1.00\\
 & $\omega_{45}$ & ~2.03 & ~1.99 & 0.28 & 0.26 & 0.91 & 1.00 &  & ~2.02 & 0.19 & 0.18 & 0.93 & 1.00 &  & ~2.02 & 0.13 & 0.12 & 0.94 & 1.00\\
 & $\omega_{55}$ & -2.83 & -2.76 & 0.41 & 0.36 & 0.90 & 1.00 &  & -2.80 & 0.27 & 0.25 & 0.94 & 1.00 &  & -2.82 & 0.18 & 0.17 & 0.94 & 1.00\\
 & $\omega_{65}$ & -1.93 & -1.90 & 0.27 & 0.25 & 0.92 & 1.00 &  & -1.91 & 0.18 & 0.16 & 0.94 & 1.00 &  & -1.92 & 0.12 & 0.12 & 0.94 & 1.00\\
 & $\omega_{06}$ & -1.82 & -1.85 & 0.21 & 0.19 & 0.91 & 0.97 &  & -1.83 & 0.14 & 0.13 & 0.93 & 1.00 &  & -1.82 & 0.09 & 0.09 & 0.95 & 1.00\\
 & \boldsymbol{$\omega_{16}$} & ~0.00 & ~0.00 & 0.09 & 0.08 & 0.94 & 0.06 &  & ~0.00 & 0.05 & 0.05 & 0.94 & 0.06 &  & ~0.00 & 0.04 & 0.04 & 0.94 & 0.06\\
 & $\omega_{26}$ & ~0.20 & ~0.20 & 0.09 & 0.08 & 0.91 & 0.52 &  & ~0.20 & 0.06 & 0.06 & 0.93 & 0.75 &  & ~0.20 & 0.04 & 0.04 & 0.94 & 0.86\\
 & $\omega_{36}$ & -0.39 & -0.41 & 0.09 & 0.09 & 0.91 & 0.91 &  & -0.39 & 0.06 & 0.06 & 0.92 & 0.98 &  & -0.39 & 0.04 & 0.04 & 0.97 & 1.00\\
 & $\omega_{46}$ & -1.92 & -1.95 & 0.25 & 0.23 & 0.92 & 1.00 &  & -1.94 & 0.15 & 0.15 & 0.93 & 1.00 &  & -1.92 & 0.11 & 0.10 & 0.95 & 1.00\\
 & $\omega_{56}$ & -2.78 & -2.83 & 0.30 & 0.28 & 0.91 & 1.00 &  & -2.80 & 0.20 & 0.19 & 0.94 & 1.00 &  & -2.79 & 0.13 & 0.13 & 0.95 & 1.00\\
 & $\omega_{66}$ & -2.38 & -2.42 & 0.28 & 0.26 & 0.91 & 1.00 &  & -2.40 & 0.17 & 0.17 & 0.94 & 1.00 &  & -2.39 & 0.12 & 0.12 & 0.94 & 1.00\\
 & $\gamma_{0}$ & -1.54 & -1.50 & 0.44 & 0.35 & 0.92 & 0.38 &  & -1.52 & 0.28 & 0.23 & 0.94 & 0.49 &  & -1.53 & 0.16 & 0.16 & 0.96 & 0.37\\
 & $\gamma_{1}$ & ~1.37 & ~1.57 & 0.49 & 0.33 & 0.92 & 1.00 &  & ~1.44 & 0.23 & 0.19 & 0.94 & 1.00 &  & ~1.40 & 0.13 & 0.13 & 0.96 & 1.00\\
 & $\gamma_{2}$ & ~2.73 & ~2.83 & 0.66 & 0.58 & 0.93 & 1.00 &  & ~2.80 & 0.59 & 0.45 & 0.94 & 1.00 &  & ~2.77 & 0.32 & 0.31 & 0.95 & 1.00\\
 & $\gamma_{3}$ & -1.81 & -1.82 & 0.49 & 0.28 & 0.93 & 1.00 &  & -1.83 & 0.22 & 0.19 & 0.94 & 1.00 &  & -1.82 & 0.13 & 0.13 & 0.94 & 1.00\\
 & $\gamma_{4}$ & -2.56 & -2.67 & 0.54 & 0.49 & 0.93 & 1.00 &  & -2.62 & 0.39 & 0.39 & 0.93 & 1.00 &  & -2.60 & 0.28 & 0.27 & 0.94 & 1.00\\
 & $\gamma_{5}$ & -2.87 & -3.08 & 0.49 & 0.33 & 0.93 & 1.00 &  & -2.94 & 0.20 & 0.18 & 0.95 & 1.00 &  & -2.91 & 0.11 & 0.11 & 0.95 & 1.00\\
 & $\gamma_{6}$ & ~2.98 & ~2.98 & 0.23 & 0.21 & 0.93 & 1.00 &  & ~2.98 & 0.14 & 0.13 & 0.94 & 1.00 &  & ~2.98 & 0.10 & 0.09 & 0.95 & 1.00\\

  \bottomrule
  \multicolumn{20}{p{\textwidth}}{\scriptsize
  SE, standard deviation of estimates over 1000 replications; SEE, average of estimated standard errors over 1000 replications; CP, the empirical coverage probability of a nominal 95\% confidence interval; RR, the rejection rate for the single-parameter Wald test with a 5\% significance level.
  True zero parameters are highlighted in \textbf{bold}.}\\
  \label{tab: est_and_inf_5-1}
  \end{longtable}

  %%%%%%%%%%%%%%%%%%
  % 3-3
  %%%%%%%%%%%%%%%%%%

\footnotesize
\setlength\LTleft{0pt}  
\setlength\LTright{0pt} 
\begin{longtable}{@{}l@{~~} c@{~~}c@{~~}  c@{~~}c@{~~}c@{~~}c@{~~}c@{~~}  c@{~~}  c@{~~}c@{~~}c@{~~}c@{~~}c@{~~}  c@{~~}  c@{~~}c@{~~}c@{~~}c@{~~}c@{~~}}
  \caption{Simulation: estimation and inference metrics for the 3-3 non-zero-to-zero (N-Z) scenario.} \\

\toprule
{N-Z}  & {3-3} & {} & \multicolumn{5}{c}{$n = 500$} && \multicolumn{5}{c}{$n = 1000$} && \multicolumn{5}{c}{$n = 2000$} \\
\cmidrule(){4-8} \cmidrule(){10-14} \cmidrule(){16-20}
{$q$} &  {} & $\theta$ & $\hat{\theta}$ & SE & SEE & CP & RR && $\hat{\theta}$ & SE & SEE & CP & RR && $\hat{\theta}$ & SE & SEE & CP & RR \\
  \midrule
2  & $\omega_{01}$ & -2.77 & -2.75 & 0.35 & 0.34 & 0.94 & 1.00 &  & -2.76 & 0.25 & 0.24 & 0.94 & 1.00 &  & -2.76 & 0.17 & 0.18 & 0.96 & 1.00\\
& \boldsymbol{$\omega_{11}$} & ~0.00 & ~0.00 & 0.11 & 0.10 & 0.93 & 0.07 &  & ~0.00 & 0.07 & 0.07 & 0.96 & 0.04 &  & ~0.00 & 0.05 & 0.05 & 0.96 & 0.04\\
& $\omega_{21}$ & -0.14 & -0.14 & 0.11 & 0.10 & 0.94 & 0.65 &  & -0.14 & 0.07 & 0.07 & 0.95 & 0.75 &  & -0.14 & 0.05 & 0.05 & 0.95 & 0.89\\
& \boldsymbol{$\omega_{31}$} & ~0.00 & -0.01 & 0.11 & 0.10 & 0.94 & 0.06 &  & ~0.00 & 0.07 & 0.07 & 0.95 & 0.05 &  & ~0.00 & 0.05 & 0.05 & 0.95 & 0.05\\
& \boldsymbol{$\omega_{41}$} & ~0.00 & ~0.00 & 0.10 & 0.10 & 0.94 & 0.06 &  & ~0.00 & 0.07 & 0.07 & 0.96 & 0.04 &  & ~0.00 & 0.05 & 0.05 & 0.96 & 0.04\\
& $\omega_{51}$ & -2.52 & -2.49 & 0.35 & 0.35 & 0.94 & 1.00 &  & -2.51 & 0.24 & 0.24 & 0.95 & 1.00 &  & -2.51 & 0.17 & 0.17 & 0.95 & 1.00\\
& $\omega_{61}$ & -2.52 & -2.50 & 0.36 & 0.37 & 0.94 & 1.00 &  & -2.51 & 0.25 & 0.25 & 0.94 & 1.00 &  & -2.52 & 0.17 & 0.18 & 0.95 & 1.00\\
& $\omega_{02}$ & ~1.70 & ~1.70 & 0.11 & 0.11 & 0.94 & 1.00 &  & ~1.70 & 0.08 & 0.08 & 0.94 & 1.00 &  & ~1.70 & 0.06 & 0.06 & 0.94 & 1.00\\
& \boldsymbol{$\omega_{12}$} & ~0.00 & ~0.00 & 0.04 & 0.04 & 0.95 & 0.05 &  & ~0.00 & 0.03 & 0.03 & 0.95 & 0.05 &  & ~0.00 & 0.02 & 0.02 & 0.95 & 0.05\\
& $\omega_{22}$ & -0.27 & -0.27 & 0.04 & 0.04 & 0.95 & 0.65 &  & -0.27 & 0.03 & 0.03 & 0.94 & 0.76 &  & -0.27 & 0.02 & 0.02 & 0.95 & 0.89\\
& \boldsymbol{$\omega_{32}$} & ~0.00 & ~0.00 & 0.04 & 0.04 & 0.95 & 0.05 &  & ~0.00 & 0.03 & 0.03 & 0.95 & 0.05 &  & ~0.00 & 0.02 & 0.02 & 0.96 & 0.04\\
& \boldsymbol{$\omega_{42}$} & ~0.00 & 0.00 & 0.04 & 0.04 & 0.96 & 0.04 &  & ~0.00 & 0.03 & 0.03 & 0.96 & 0.04 &  & ~0.00 & 0.02 & 0.02 & 0.95 & 0.05\\
& $\omega_{52}$ & ~1.05 & ~1.07 & 0.09 & 0.09 & 0.94 & 1.00 &  & ~1.06 & 0.06 & 0.06 & 0.93 & 1.00 &  & ~1.06 & 0.04 & 0.04 & 0.95 & 1.00\\
& $\omega_{62}$ & -2.06 & -2.08 & 0.16 & 0.15 & 0.94 & 1.00 &  & -2.06 & 0.10 & 0.10 & 0.94 & 1.00 &  & -2.06 & 0.08 & 0.07 & 0.95 & 1.00\\
& $\gamma_{0}$ & -1.54 & -1.54 & 0.10 & 0.09 & 0.94 & 0.22 &  & -1.55 & 0.06 & 0.06 & 0.95 & 0.55 &  & -1.55 & 0.04 & 0.04 & 0.95 & 0.83\\
& $\gamma_{1}$ & ~1.37 & ~1.39 & 0.10 & 0.10 & 0.96 & 1.00 &  & ~1.38 & 0.06 & 0.06 & 0.94 & 1.00 &  & ~1.38 & 0.05 & 0.05 & 0.95 & 1.00\\
& $\gamma_{2}$ & ~2.73 & ~2.72 & 0.13 & 0.12 & 0.94 & 1.00 &  & ~2.73 & 0.08 & 0.08 & 0.94 & 1.00 &  & ~2.73 & 0.06 & 0.06 & 0.95 & 1.00\\[0.2cm]

4 & $\omega_{01}$ & -2.77 & -2.73 & 0.38 & 0.37 & 0.92 & 0.96 &  & -2.74 & 0.29 & 0.28 & 0.93 & 1.00 &  & -2.76 & 0.21 & 0.20 & 0.96 & 1.00\\
& \boldsymbol{$\omega_{11}$} & ~0.00 & ~0.00 & 0.12 & 0.12 & 0.94 & 0.06 &  & ~0.00 & 0.08 & 0.08 & 0.95 & 0.05 &  & ~0.00 & 0.06 & 0.06 & 0.94 & 0.06\\
& $\omega_{21}$ & -0.14 & -0.14 & 0.12 & 0.12 & 0.93 & 0.62 &  & -0.14 & 0.09 & 0.09 & 0.94 & 0.81 &  & -0.14 & 0.06 & 0.06 & 0.95 & 0.92\\
& \boldsymbol{$\omega_{31}$} & ~0.00 & ~0.00 & 0.12 & 0.11 & 0.95 & 0.05 &  & ~0.00 & 0.08 & 0.08 & 0.96 & 0.04 &  & ~0.00 & 0.06 & 0.06 & 0.95 & 0.05\\
& \boldsymbol{$\omega_{41}$} & ~0.00 & 0.00 & 0.12 & 0.11 & 0.95 & 0.05 &  & ~0.00 & 0.09 & 0.08 & 0.96 & 0.04 &  & ~0.00 & 0.06 & 0.06 & 0.95 & 0.05\\
& $\omega_{51}$ & -2.52 & -2.45 & 0.38 & 0.37 & 0.93 & 1.00 &  & -2.48 & 0.28 & 0.28 & 0.94 & 1.00 &  & -2.51 & 0.20 & 0.20 & 0.95 & 1.00\\
& $\omega_{61}$ & -2.52 & -2.48 & 0.39 & 0.39 & 0.92 & 1.00 &  & -2.50 & 0.29 & 0.29 & 0.94 & 1.00 &  & -2.52 & 0.20 & 0.20 & 0.95 & 1.00\\
& $\omega_{02}$ & ~1.70 & ~1.71 & 0.37 & 0.27 & 0.91 & 0.97 &  & ~1.73 & 0.21 & 0.19 & 0.94 & 1.00 &  & ~1.72 & 0.13 & 0.13 & 0.96 & 1.00\\
& \boldsymbol{$\omega_{12}$} & ~0.00 & ~0.01 & 0.08 & 0.08 & 0.96 & 0.04 &  & ~0.00 & 0.06 & 0.06 & 0.95 & 0.05 &  & ~0.00 & 0.04 & 0.04 & 0.95 & 0.05\\
& $\omega_{22}$ & -0.27 & -0.30 & 0.11 & 0.10 & 0.94 & 0.61 &  & -0.29 & 0.07 & 0.07 & 0.94 & 0.82 &  & -0.28 & 0.05 & 0.05 & 0.95 & 0.93\\
& \boldsymbol{$\omega_{32}$} & ~0.00 & ~0.00 & 0.08 & 0.07 & 0.95 & 0.05 &  & ~0.00 & 0.05 & 0.05 & 0.96 & 0.04 &  & ~0.00 & 0.04 & 0.04 & 0.94 & 0.06\\
& \boldsymbol{$\omega_{42}$} & ~0.00 & -0.01 & 0.08 & 0.08 & 0.95 & 0.05 &  & ~0.00 & 0.05 & 0.06 & 0.95 & 0.05 &  & ~0.00 & 0.04 & 0.04 & 0.95 & 0.05\\
& $\omega_{52}$ & ~1.05 & ~1.05 & 0.20 & 0.18 & 0.92 & 1.00 &  & ~1.05 & 0.14 & 0.14 & 0.95 & 1.00 &  & ~1.05 & 0.10 & 0.10 & 0.95 & 1.00\\
& $\omega_{62}$ & -2.06 & -2.11 & 0.36 & 0.30 & 0.93 & 1.00 &  & -2.11 & 0.22 & 0.21 & 0.96 & 1.00 &  & -2.09 & 0.15 & 0.15 & 0.95 & 1.00\\
& $\omega_{03}$ & ~2.08 & ~1.96 & 0.63 & 0.58 & 0.93 & 0.98 &  & ~2.01 & 0.46 & 0.43 & 0.94 & 1.00 &  & ~2.05 & 0.29 & 0.30 & 0.95 & 1.00\\
& \boldsymbol{$\omega_{13}$} & ~0.00 & ~0.01 & 0.08 & 0.08 & 0.96 & 0.04 &  & ~0.00 & 0.05 & 0.05 & 0.94 & 0.06 &  & 0.00 & 0.04 & 0.03 & 0.95 & 0.05\\
& $\omega_{23}$ & -0.20 & -0.21 & 0.11 & 0.10 & 0.93 & 0.61 &  & -0.20 & 0.08 & 0.07 & 0.94 & 0.81 &  & -0.20 & 0.05 & 0.05 & 0.94 & 0.91\\
& \boldsymbol{$\omega_{33}$} & ~0.00 & ~0.00 & 0.08 & 0.07 & 0.95 & 0.05 &  & ~0.00 & 0.05 & 0.05 & 0.95 & 0.05 &  & ~0.00 & 0.04 & 0.04 & 0.94 & 0.06\\
& \boldsymbol{$\omega_{43}$} & ~0.00 & ~0.00 & 0.08 & 0.07 & 0.96 & 0.04 &  & ~0.00 & 0.05 & 0.05 & 0.95 & 0.05 &  & ~0.00 & 0.03 & 0.03 & 0.94 & 0.06\\
& $\omega_{53}$ & -1.43 & -1.38 & 0.49 & 0.34 & 0.94 & 1.00 &  & -1.41 & 0.31 & 0.25 & 0.94 & 1.00 &  & -1.42 & 0.17 & 0.17 & 0.96 & 1.00\\
& $\omega_{63}$ & ~1.48 & ~1.35 & 0.60 & 0.35 & 0.93 & 1.00 &  & ~1.44 & 0.34 & 0.26 & 0.95 & 1.00 &  & ~1.47 & 0.17 & 0.18 & 0.95 & 1.00\\
& $\omega_{04}$ & ~3.00 & ~3.00 & 0.36 & 0.38 & 0.93 & 0.98 &  & ~3.00 & 0.29 & 0.30 & 0.95 & 0.99 &  & ~3.02 & 0.22 & 0.23 & 0.96 & 1.00\\
& \boldsymbol{$\omega_{14}$} & ~0.00 & ~0.01 & 0.11 & 0.10 & 0.95 & 0.05 &  & ~0.00 & 0.08 & 0.07 & 0.94 & 0.06 &  & ~0.00 & 0.05 & 0.05 & 0.95 & 0.05\\
& $\omega_{24}$ & -0.29 & -0.29 & 0.12 & 0.11 & 0.93 & 0.61 &  & -0.29 & 0.07 & 0.07 & 0.94 & 0.83 &  & -0.29 & 0.05 & 0.05 & 0.94 & 0.92\\
& \boldsymbol{$\omega_{34}$} & ~0.00 & 0.00 & 0.10 & 0.09 & 0.94 & 0.06 &  & ~0.00 & 0.06 & 0.06 & 0.95 & 0.05 &  & ~0.00 & 0.05 & 0.05 & 0.95 & 0.05\\
& \boldsymbol{$\omega_{44}$} & ~0.00 & -0.01 & 0.10 & 0.09 & 0.96 & 0.04 &  & ~0.00 & 0.07 & 0.07 & 0.95 & 0.05 &  & ~0.00 & 0.05 & 0.05 & 0.95 & 0.05\\
& $\omega_{54}$ & ~2.86 & ~2.88 & 0.35 & 0.38 & 0.94 & 1.00 &  & ~2.88 & 0.30 & 0.32 & 0.95 & 1.00 &  & ~2.90 & 0.24 & 0.25 & 0.96 & 1.00\\
& $\omega_{64}$ & -2.08 & -2.08 & 0.25 & 0.26 & 0.94 & 1.00 &  & -2.08 & 0.17 & 0.18 & 0.95 & 1.00 &  & -2.09 & 0.13 & 0.13 & 0.96 & 1.00\\
& $\gamma_{0}$ & -1.54 & -1.47 & 0.54 & 0.44 & 0.95 & 0.10 &  & -1.47 & 0.36 & 0.31 & 0.94 & 0.18 &  & -1.50 & 0.21 & 0.21 & 0.96 & 0.14\\
& $\gamma_{1}$ & ~1.37 & ~1.42 & 0.15 & 0.15 & 0.95 & 1.00 &  & ~1.40 & 0.10 & 0.10 & 0.94 & 1.00 &  & ~1.38 & 0.07 & 0.07 & 0.95 & 1.00\\
& $\gamma_{2}$ & ~2.73 & ~2.68 & 0.52 & 0.54 & 0.94 & 1.00 &  & ~2.70 & 0.48 & 0.53 & 0.94 & 1.00 &  & ~2.70 & 0.38 & 0.41 & 0.94 & 1.00\\
& $\gamma_{3}$ & -1.81 & -1.84 & 0.85 & 0.43 & 0.94 & 1.00 &  & -1.87 & 0.52 & 0.29 & 0.95 & 1.00 &  & -1.85 & 0.19 & 0.19 & 0.95 & 1.00\\
& $\gamma_{4}$ & -2.56 & -2.62 & 0.44 & 0.49 & 0.94 & 1.00 &  & -2.60 & 0.42 & 0.46 & 0.94 & 1.00 &  & -2.56 & 0.34 & 0.36 & 0.94 & 1.00\\[0.2cm]

6 & $\omega_{01}$ & -2.77 & ~0.04 & 2.81 & 0.76 & 0.62 & 0.88 &  & ~0.24 & 2.90 & 0.69 & 0.66 & 0.92 &  & ~0.48 & 2.41 & 0.51 & 0.69 & 0.94\\
& \boldsymbol{$\omega_{11}$} & ~0.00 & ~0.03 & 1.42 & 0.40 & 0.81 & 0.19 &  & ~0.01 & 1.23 & 0.33 & 0.85 & 0.15 &  & -0.01 & 0.96 & 0.24 & 0.87 & 0.13\\
& $\omega_{21}$ & -0.14 & -0.26 & 1.43 & 0.42 & 0.59 & 0.43 &  & -0.25 & 1.25 & 0.35 & 0.67 & 0.55 &  & -0.26 & 1.05 & 0.26 & 0.68 & 0.75\\
& \boldsymbol{$\omega_{31}$} & ~0.00 & -0.03 & 1.43 & 0.42 & 0.80 & 0.20 &  & ~0.01 & 1.16 & 0.32 & 0.87 & 0.13 &  & -0.02 & 1.01 & 0.25 & 0.86 & 0.14\\
& \boldsymbol{$\omega_{41}$} & ~0.00 & -0.04 & 1.35 & 0.39 & 0.79 & 0.21 &  & ~0.00 & 1.17 & 0.32 & 0.84 & 0.16 &  & -0.04 & 1.13 & 0.26 & 0.87 & 0.13\\
& $\omega_{51}$ & -2.52 & -0.27 & 2.15 & 0.59 & 0.70 & 0.91 &  & -0.15 & 2.15 & 0.52 & 0.80 & 0.94 &  & -0.07 & 2.09 & 0.44 & 0.83 & 0.95\\
& $\omega_{61}$ & -2.52 & -2.53 & 1.61 & 0.79 & 0.65 & 0.93 &  & -2.35 & 1.33 & 0.63 & 0.76 & 0.95 &  & -2.38 & 1.13 & 0.52 & 0.77 & 0.95\\
& $\omega_{02}$ & ~1.70 & ~1.87 & 1.80 & 0.73 & 0.50 & 0.88 &  & ~2.02 & 1.40 & 0.64 & 0.54 & 0.91 &  & ~2.06 & 1.24 & 0.54 & 0.53 & 0.92\\
& \boldsymbol{$\omega_{12}$} & ~0.00 & ~0.04 & 1.22 & 0.32 & 0.80 & 0.20 &  & ~0.00 & 0.76 & 0.22 & 0.88 & 0.12 &  & ~0.02 & 0.75 & 0.19 & 0.87 & 0.13\\
& $\omega_{22}$ & -0.27 & -0.31 & 1.09 & 0.33 & 0.55 & 0.42 &  & -0.29 & 1.02 & 0.27 & 0.60 & 0.57 &  & -0.23 & 0.87 & 0.22 & 0.60 & 0.73\\
& \boldsymbol{$\omega_{32}$} & ~0.00 & ~0.07 & 1.12 & 0.32 & 0.79 & 0.21 &  & -0.03 & 0.80 & 0.23 & 0.88 & 0.12 &  & ~0.05 & 0.90 & 0.20 & 0.86 & 0.14\\
& \boldsymbol{$\omega_{42}$} & ~0.00 & ~0.01 & 1.13 & 0.31 & 0.81 & 0.19 &  & ~0.02 & 0.74 & 0.22 & 0.88 & 0.12 &  & -0.02 & 0.91 & 0.21 & 0.87 & 0.13\\
& $\omega_{52}$ & ~1.05 & -0.40 & 1.92 & 0.53 & 0.52 & 0.90 &  & -0.35 & 1.60 & 0.42 & 0.62 & 0.95 &  & -0.53 & 1.51 & 0.36 & 0.64 & 0.95\\
& $\omega_{62}$ & -2.06 & -0.46 & 2.43 & 0.64 & 0.56 & 0.94 &  & -0.20 & 2.21 & 0.52 & 0.62 & 0.96 &  & ~0.05 & 2.17 & 0.45 & 0.62 & 0.96\\
& $\omega_{03}$ & ~2.08 & ~2.58 & 1.60 & 0.87 & 0.49 & 0.87 &  & ~2.65 & 1.43 & 0.78 & 0.48 & 0.90 &  & ~2.48 & 1.17 & 0.62 & 0.49 & 0.94\\
& \boldsymbol{$\omega_{13}$} & ~0.00 & ~0.01 & 1.25 & 0.37 & 0.82 & 0.18 &  & ~0.03 & 0.83 & 0.25 & 0.85 & 0.15 &  & ~0.03 & 0.93 & 0.22 & 0.88 & 0.12\\
& $\omega_{23}$ & -0.20 & -0.35 & 1.22 & 0.37 & 0.54 & 0.43 &  & -0.30 & 1.07 & 0.29 & 0.53 & 0.55 &  & -0.23 & 0.93 & 0.24 & 0.53 & 0.76\\
& \boldsymbol{$\omega_{33}$} & ~0.00 & ~0.10 & 1.19 & 0.36 & 0.81 & 0.19 &  & -0.01 & 0.87 & 0.25 & 0.86 & 0.14 &  & ~0.04 & 0.88 & 0.21 & 0.86 & 0.14\\
& \boldsymbol{$\omega_{43}$} & ~0.00 & ~0.09 & 1.22 & 0.36 & 0.81 & 0.19 &  & ~0.02 & 0.85 & 0.25 & 0.86 & 0.14 &  & ~0.00 & 0.88 & 0.21 & 0.87 & 0.13\\
& $\omega_{53}$ & -1.43 & -0.17 & 2.58 & 0.69 & 0.51 & 0.92 &  & -0.18 & 2.39 & 0.57 & 0.52 & 0.95 &  & -0.12 & 2.35 & 0.48 & 0.52 & 0.96\\
& $\omega_{63}$ & ~1.48 & ~0.22 & 2.50 & 0.67 & 0.52 & 0.94 &  & ~0.37 & 2.27 & 0.56 & 0.55 & 0.94 &  & ~0.40 & 2.17 & 0.47 & 0.55 & 0.96\\
& $\omega_{04}$ & ~3.00 & ~3.05 & 1.20 & 0.74 & 0.47 & 0.88 &  & ~3.02 & 1.05 & 0.59 & 0.49 & 0.91 &  & ~3.03 & 0.95 & 0.49 & 0.50 & 0.95\\
& \boldsymbol{$\omega_{14}$} & ~0.00 & -0.01 & 0.85 & 0.24 & 0.79 & 0.21 &  & -0.01 & 0.70 & 0.19 & 0.86 & 0.14 &  & ~0.01 & 0.63 & 0.15 & 0.90 & 0.10\\
& $\omega_{24}$ & -0.29 & -0.16 & 0.92 & 0.25 & 0.56 & 0.41 &  & -0.22 & 0.85 & 0.22 & 0.54 & 0.52 &  & -0.22 & 0.78 & 0.18 & 0.54 & 0.74\\
& \boldsymbol{$\omega_{34}$} & ~0.00 & -0.01 & 0.94 & 0.24 & 0.81 & 0.19 &  & -0.01 & 0.63 & 0.19 & 0.86 & 0.14 &  & ~0.01 & 0.76 & 0.16 & 0.88 & 0.12\\
& \boldsymbol{$\omega_{44}$} & ~0.00 & ~0.03 & 0.84 & 0.24 & 0.81 & 0.19 &  & -0.01 & 0.71 & 0.19 & 0.86 & 0.14 &  & -0.01 & 0.85 & 0.17 & 0.87 & 0.13\\
& $\omega_{54}$ & ~2.86 & ~1.87 & 2.59 & 0.72 & 0.50 & 0.90 &  & ~2.05 & 2.34 & 0.60 & 0.49 & 0.94 &  & ~1.99 & 2.29 & 0.48 & 0.51 & 0.96\\
& $\omega_{64}$ & -2.08 & -2.13 & 0.96 & 0.51 & 0.51 & 0.94 &  & -2.15 & 0.77 & 0.40 & 0.54 & 0.94 &  & -2.14 & 0.76 & 0.33 & 0.55 & 0.95\\
& $\omega_{05}$ & ~2.79 & ~2.69 & 1.39 & 0.61 & 0.54 & 0.86 &  & ~2.81 & 1.03 & 0.48 & 0.58 & 0.92 &  & ~2.72 & 1.01 & 0.34 & 0.58 & 0.94\\
& \boldsymbol{$\omega_{15}$} & ~0.00 & ~0.00 & 0.59 & 0.16 & 0.80 & 0.20 &  & ~0.00 & 0.53 & 0.13 & 0.86 & 0.14 &  & -0.01 & 0.50 & 0.09 & 0.89 & 0.11\\
& $\omega_{25}$ & ~0.27 & ~0.28 & 0.80 & 0.18 & 0.58 & 0.41 &  & ~0.30 & 0.67 & 0.15 & 0.59 & 0.53 &  & ~0.29 & 0.54 & 0.11 & 0.59 & 0.76\\
& \boldsymbol{$\omega_{35}$} & ~0.00 & -0.03 & 0.70 & 0.18 & 0.80 & 0.20 &  & ~0.03 & 0.63 & 0.13 & 0.86 & 0.14 &  & ~0.01 & 0.57 & 0.10 & 0.85 & 0.15\\
& \boldsymbol{$\omega_{45}$} & ~0.00 & -0.01 & 0.52 & 0.15 & 0.82 & 0.18 &  & ~0.05 & 0.56 & 0.13 & 0.86 & 0.14 &  & ~0.01 & 0.53 & 0.10 & 0.87 & 0.13\\
& $\omega_{55}$ & -2.83 & -2.91 & 0.98 & 0.57 & 0.53 & 0.90 &  & -2.93 & 0.64 & 0.42 & 0.59 & 0.95 &  & -2.93 & 0.71 & 0.32 & 0.57 & 0.96\\
& $\omega_{65}$ & -1.93 & -2.11 & 0.64 & 0.41 & 0.57 & 0.92 &  & -2.03 & 0.54 & 0.31 & 0.60 & 0.93 &  & -2.05 & 0.53 & 0.24 & 0.62 & 0.96\\
& $\omega_{06}$ & -1.82 & -1.88 & 0.77 & 0.32 & 0.71 & 0.88 &  & -1.96 & 0.32 & 0.22 & 0.79 & 0.90 &  & -1.95 & 0.44 & 0.17 & 0.79 & 0.92\\
& \boldsymbol{$\omega_{16}$} & ~0.00 & ~0.03 & 0.42 & 0.10 & 0.79 & 0.21 &  & ~0.00 & 0.10 & 0.05 & 0.88 & 0.12 &  & ~0.00 & 0.10 & 0.04 & 0.87 & 0.13\\
& $\omega_{26}$ & ~0.20 & ~0.14 & 0.27 & 0.10 & 0.68 & 0.41 &  & ~0.11 & 0.29 & 0.06 & 0.74 & 0.56 &  & ~0.13 & 0.39 & 0.05 & 0.69 & 0.74\\
& \boldsymbol{$\omega_{36}$} & ~0.00 & ~0.02 & 0.33 & 0.10 & 0.81 & 0.19 &  & ~0.00 & 0.21 & 0.06 & 0.86 & 0.14 &  & ~0.00 & 0.29 & 0.05 & 0.86 & 0.14\\
& \boldsymbol{$\omega_{46}$} & ~0.00 & ~0.01 & 0.28 & 0.09 & 0.81 & 0.19 &  & ~0.00 & 0.23 & 0.06 & 0.86 & 0.14 &  & -0.01 & 0.24 & 0.05 & 0.86 & 0.14\\
& $\omega_{56}$ & -2.78 & -2.64 & 0.54 & 0.33 & 0.70 & 0.91 &  & -2.59 & 0.38 & 0.22 & 0.76 & 0.95 &  & -2.62 & 0.38 & 0.18 & 0.76 & 0.94\\
& $\omega_{66}$ & -2.38 & -2.29 & 0.49 & 0.29 & 0.72 & 0.93 &  & -2.29 & 0.27 & 0.20 & 0.78 & 0.94 &  & -2.31 & 0.25 & 0.16 & 0.81 & 0.94\\
& $\gamma_{0}$ & -1.54 & -2.19 & 1.47 & 0.61 & 0.70 & 0.60 &  & -1.79 & 1.13 & 0.54 & 0.79 & 0.69 &  & -1.55 & 0.72 & 0.42 & 0.86 & 0.66\\
& $\gamma_{1}$ & ~1.37 & ~2.19 & 2.02 & 0.77 & 0.53 & 0.94 &  & ~2.11 & 1.66 & 0.72 & 0.61 & 0.94 &  & ~2.04 & 1.35 & 0.59 & 0.64 & 0.96\\
& $\gamma_{2}$ & ~2.73 & ~1.29 & 2.90 & 0.79 & 0.39 & 0.95 &  & ~0.65 & 2.49 & 0.69 & 0.50 & 0.95 &  & ~0.22 & 2.24 & 0.51 & 0.51 & 0.94\\
& $\gamma_{3}$ & -1.81 & -1.35 & 2.20 & 0.66 & 0.38 & 0.96 &  & -1.55 & 1.54 & 0.55 & 0.39 & 0.94 &  & -1.60 & 1.29 & 0.43 & 0.43 & 0.96\\
& $\gamma_{4}$ & -2.56 & -2.68 & 1.42 & 0.81 & 0.38 & 0.96 &  & -2.56 & 1.26 & 0.72 & 0.39 & 0.94 &  & -2.42 & 1.12 & 0.56 & 0.41 & 0.96\\
& $\gamma_{5}$ & -2.87 & -2.57 & 1.81 & 0.50 & 0.42 & 0.94 &  & -2.62 & 1.25 & 0.33 & 0.45 & 0.95 &  & -2.55 & 1.27 & 0.25 & 0.46 & 0.94\\
& $\gamma_{6}$ & ~2.98 & ~4.19 & 1.07 & 0.49 & 0.59 & 0.94 &  & ~4.29 & 0.64 & 0.34 & 0.60 & 0.94 &  & ~4.19 & 0.77 & 0.27 & 0.63 & 0.94\\

  \bottomrule
  \multicolumn{20}{p{\textwidth}}{\scriptsize
  SE, standard deviation of estimates over 1000 replications; SEE, average of estimated standard errors over 1000 replications; CP, the empirical coverage probability of a nominal 95\% confidence interval; RR, the rejection rate for the single-parameter Wald test with a 5\% significance level.
  True zero parameters are highlighted in \textbf{bold}.}\\
    \label{tab: est_and_inf_3-3}
  \end{longtable}

\end{document}